\documentclass[11pt]{article}

\usepackage[]{inputenc}
\usepackage[T1]{fontenc}
\usepackage[hidelinks]{hyperref} 
\usepackage{mathrsfs}
\usepackage{color}
\usepackage{authblk}

\usepackage{fullpage}

\usepackage{amsmath}
\usepackage{amssymb}
\usepackage[toc,page]{appendix}

\usepackage{amsthm}
\usepackage{float}
\usepackage{placeins}
\usepackage{wrapfig}
\usepackage{caption}
\usepackage{graphicx}
\usepackage{pgfplots}
\usepackage{subcaption}
\usepackage{wasysym}

\def\sgn{\pm_o}
\def \be {\begin{equation}}
\def \ee {\end{equation}}
\def \bea {\begin{eqnarray}}
\def \eea {\end{eqnarray}}

\def \rmin r_{\rm min}

\title{Evolution of creases on the event horizon of a black hole merger}

\date{}                                           

\def\be{\begin{equation}}
\def\ee{\end{equation}}
\def\ba{\begin{eqnarray}}
\def\ea{\end{eqnarray}}

\newcommand{\mc}[1]{{\mathcal{#1}}}
\def\cH{{\cal H}}

\def\scri{{\cal I}}

\begin{document}

\author[1]{Maxime Gadioux\thanks{mjhg2@cam.ac.uk}}
\author[2]{Robie A. Hennigar\thanks{robie.hennigar@icc.ub.edu}}
\author[1]{Harvey S. Reall\thanks{hsr1000@cam.ac.uk}}

\affil[1]{\small Department of Applied Mathematics and Theoretical Physics, University of Cambridge, Wilberforce Road, Cambridge CB3 0WA, United Kingdom}
\affil[2]{\small Departament de F\'isica Qu\`antica i Astrof\'isica, Institut de Ci\`encies del Cosmos,
  Universitat de Barcelona, Mart\'i i Franqu\`es 1, E-08028 Barcelona, Spain}

\maketitle

\begin{abstract}

A generic black hole merger occurs through a restructuring of creases (sharp edges) on the event horizon. This process is studied for a black hole merger in the limit of infinite mass ratio, for which constructing the event horizon reduces to finding a null hypersurface that asymptotes to a Rindler horizon in the Kerr spacetime. Geometrical properties of the creases on this horizon are determined and the results are compared with the predictions of an exact local description of the event horizon in a generic merger. The crease set is shown to have finite area. A recently proposed expression for the gravitational entropy of a crease is shown to diverge at the instant of merger. Caustics on (and off) the event horizon are determined by exploiting the correspondence with the problem of gravitational lensing by a Kerr black hole. Caustics form an ``astroid tube'' in spacetime, two edges of which lie on the event horizon of the merger. Perturbative expressions for the location of this tube are presented, extending previous work to much higher perturbative order.
\end{abstract}

\newpage 

\section{Introduction}

\begin{figure}[t]
    \centering
    \includegraphics[width=0.5\textwidth]{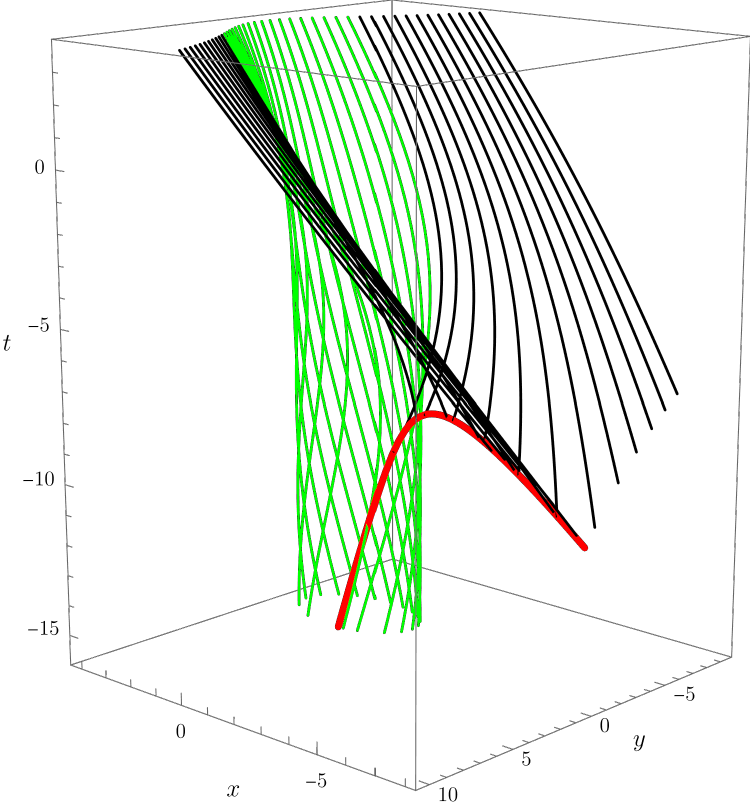}
    \caption{Event horizon for an orthogonal merger (small black hole spin perpendicular to its velocity). The  small black hole has $a/M=0.5$. The vertical axis is the Boyer-Lindquist time coordinate. The quasi-Cartesian spatial coordinates $(x,y,z)$ are defined below. The black hole spin lies along the $z$-axis, and the figure shows a $z=0$ cross-section. Generators that belong to the small black hole at $t=-\infty$ are shown in green and those that originate from the crease set are shown in black. The crease set is shown in red. This figure is similar to Fig.~12 of Emparan \textit{et al.}~\cite{Emparan:2017vyp}. The difference between their results and ours is too small to be seen at this scale. In all of our plots we use units such that $M=1$.}
    \label{fig:horizon_generators}
\end{figure}

Determining the event horizon of a black hole merger is very difficult. The spacetime must be found via numerical simulation and finding the event horizon is rarely attempted because it is computationally expensive. However, there is a physically interesting situation in which the problem is considerably simpler. This is the case of an extreme mass ratio merger, where one black hole is much larger than the other. Specifically, in the limit of {\it infinite} mass ratio, Emparan, Mart\'inez and Zilh\~ao have presented a beautiful exact description of the event horizon \cite{Emparan:2016ylg,Emparan:2017vyp}. If one focuses on length scales comparable to the size of the small black hole then, for infinite mass ratio, the curvature of the large black hole is negligible and so the spacetime is described {\it exactly} by the Kerr solution associated with the small black hole. The event horizon of the large black hole at late time after the merger is described by an infinite planar {\it acceleration horizon} in the Kerr spacetime. Hence the problem of finding the event horizon of the merger reduces to finding the null hypersurface in the Kerr spacetime that asymptotes to this acceleration horizon at late time. Since the Kerr metric is known explicitly, this a much simpler problem than determining the event horizon in a numerically-determined black hole merger spacetime. The result of this calculation is a surface such as the one shown in Fig.~\ref{fig:horizon_generators}. 
When regarded as an approximation to the event horizon of a merger with large but finite mass ratio, this model describes only the final ``plunge'' part of the merger process, it does not describe the ``inspiral'' phase.

The event horizon of a black hole merger cannot be smooth. Various types of non-smooth features can arise \cite{Siino:2004xe,Gadioux:2023pmw} and these correspond to past endpoints of horizon generators (i.e.~points at which generators enter the horizon). Generically, the largest (highest dimensional) structures are {\it creases}. On a constant time cross-section of the horizon, a crease is a sharp edge, at which two smooth sections of the horizon meet, resembling the edge of a chisel. Creases terminate at {\it caustic points}. Recently, two of us have argued that a generic (non-axisymmetric) black hole merger always occurs via a ``crease perestroika'' as follows \cite{Gadioux:2023pmw}. Before the merger, each black hole possesses a crease and, at the ``instant of merger'' (w.r.t.~a given time foliation), these creases join at a point and then split in a transverse direction to form a connected horizon ($\supset \, \subset \rightarrow \text{\Large  $\times$} \rightarrow\text{\rotatebox[origin=c]{90}{$\supset \, \subset$}}$). After the merger the horizon has an almost flat ``bridge'' with a pair of creases along its edges. These creases subsequently shrink and disappear via pair-annihilation of their caustic endpoints (a ``caustic perestroika'' \cite{Gadioux:2023pmw}). Creases on the event horizon can be seen clearly in the figures of Ref.~\cite{Bohn:2016soe} showing the event horizon for numerical simulations of black hole mergers.\footnote{
See also the numerical simulations of \cite{Cohen:2011cf} and the flat space toy model of \cite{Husa:1999nm}.
}

\begin{figure}
    \centering
    \includegraphics[width=\textwidth]{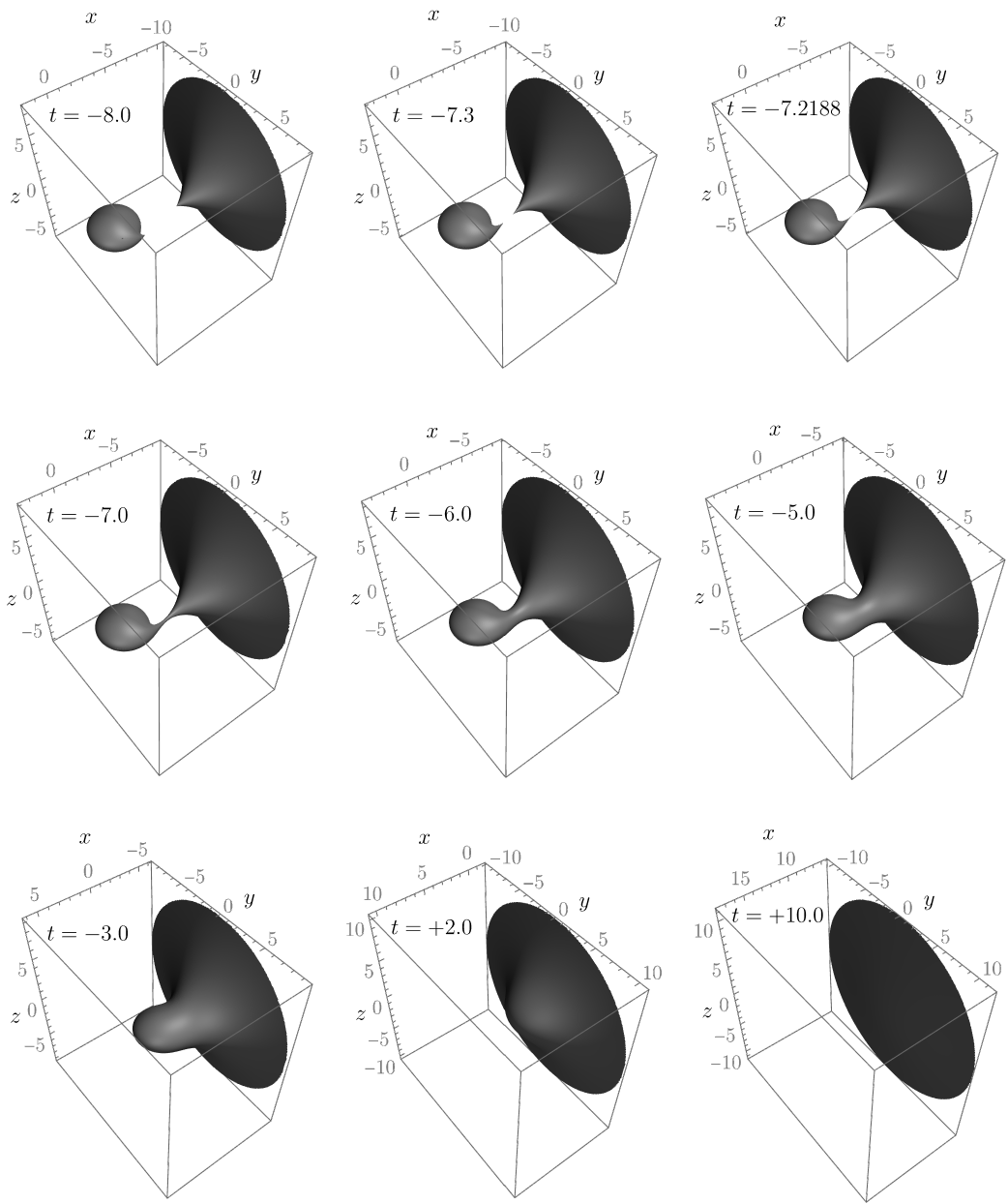}
    \caption{Constant Boyer-Lindquist time cross-sections of the event horizon for an extremal small black hole in an orthogonal merger. The small black hole spin axis is in the $z$-direction. Creases are not visible at the scale of this plot. (Units: $M=1$.)}
    \label{fig:shiny_black_extremal}
\end{figure}

\begin{figure}
    \centering
    \includegraphics[width=\textwidth]{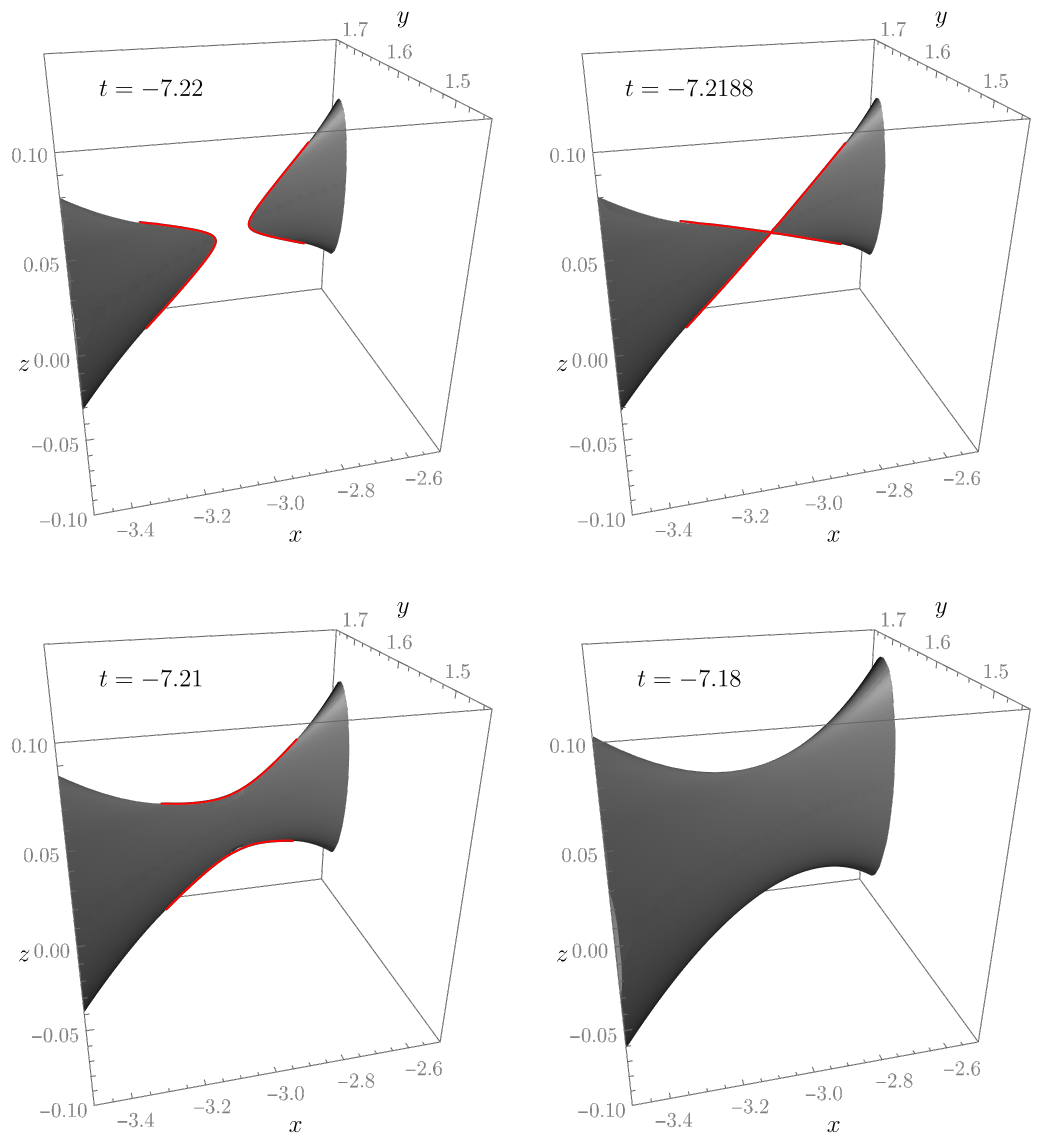}
    \caption{Behaviour near the instant of merger of Fig.~\ref{fig:shiny_black_extremal}. The creases are shown in red and exist before and just after the merger. Now one sees the chisel-like features on the two horizons before the merger. Near the instant of merger, the creases have a hyperbolic shape, and the horizon becomes very flat in the direction transverse to the plane of the creases, exactly as described in \cite{Gadioux:2023pmw}. The end points of the creases are caustic points. The creases shrink and vanish (via a caustic perestroika) between the third and fourth images of this figure.}
    \label{fig:shiny_black_extremal_zoom}
\end{figure}

\begin{figure}
    \centering
    \includegraphics[width=0.45\textwidth]{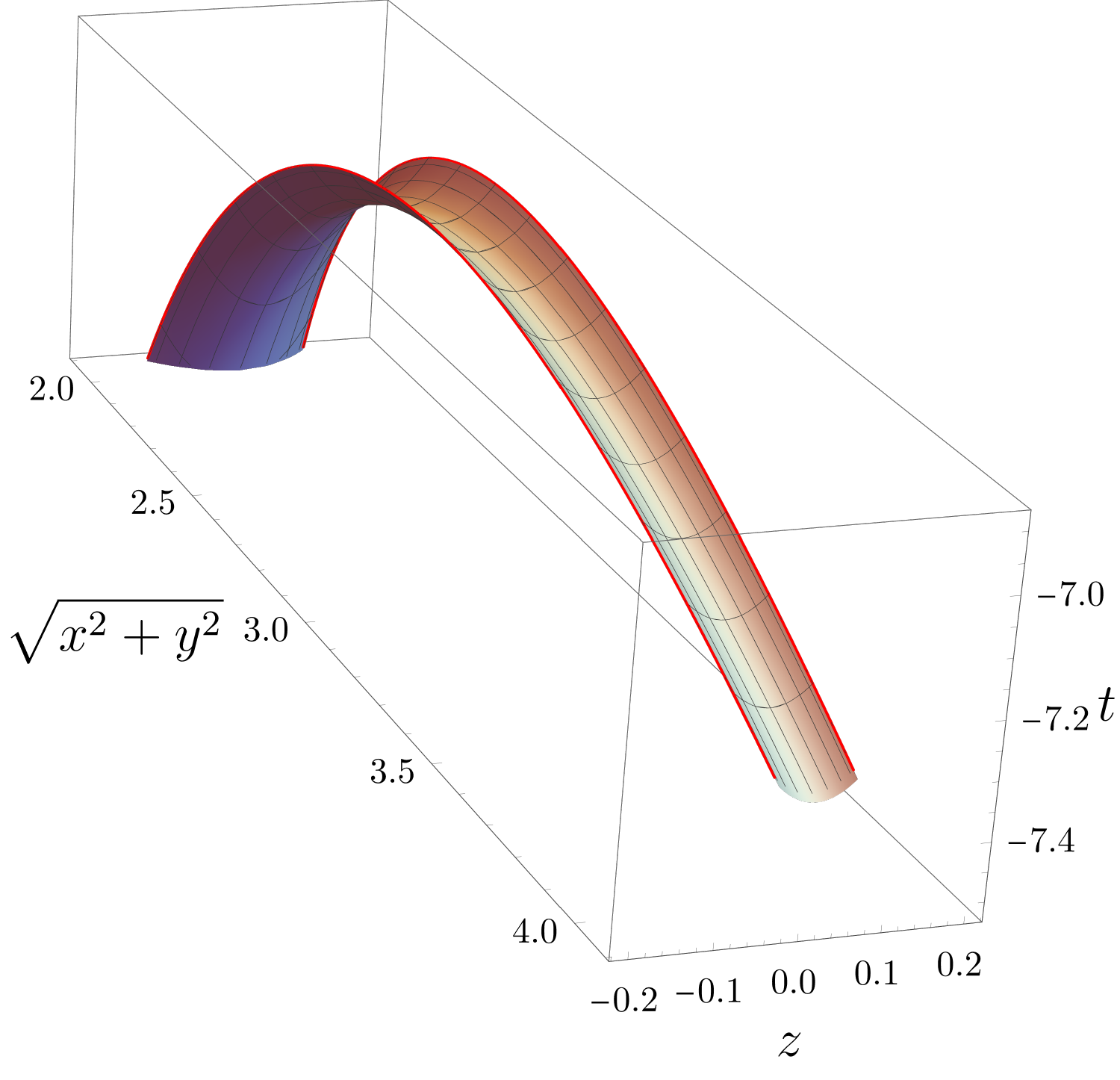}
    \caption{The crease set for an orthogonal merger with an extremal small black hole. The red lines bounding the crease set are lines of caustic points. The ``thick end'' of the crease set lies on small black hole horizon, the ``thin end'' lies on the large black hole horizon. This figure shows the $z$-direction, which is suppressed in Fig.~\ref{fig:horizon_generators}. This figure is generated from our perturbative results, which are accurate near the large black hole but only qualitatively correct near the small black hole. (Units such that $M=1$.)}
    \label{fig:crease_set_example}
\end{figure}

In this paper we shall use the model of Emparan {\it et al.}~to demonstrate the important role played by these non-smooth features of the event horizon during a black hole merger. Following Emparan {\it et al.}, we shall study how the merger evolves in time, using the time foliation defined by Boyer-Lindquist coordinates. An example of our results is shown in Fig.~\ref{fig:shiny_black_extremal} (see also Fig.~9 of \cite{Emparan:2017vyp}). Just before the merger, each black hole develops a spike. However, if we zoom in on the tips of these spikes, as shown in Fig.~\ref{fig:shiny_black_extremal_zoom}, then we discover that they are resolved into chisel-like structures, where the sharp edge of each chisel is a crease. This zoomed-in view confirms that the merger proceeds according to the crease perestroika of \cite{Gadioux:2023pmw}. We shall determine the {\it crease set}, the set of all crease points in spacetime. This is a spacelike 2-dimensional submanifold bounded by two lines of caustic points. An example is shown in Fig.~\ref{fig:crease_set_example}. We shall show that it has finite area, even though the caustic lines have infinite length. Ref.~\cite{Gadioux:2023pmw} used properties of entanglement entropy to suggest that creases might contribute to black hole entropy in a specific way. However, we shall show that this contribution diverges at the instant of merger and so this suggestion seems unlikely to be correct.

Our study was partly motivated by the fact that the results of Emparan {\it et al.}~for the behaviour of creases in a non-axisymmetric merger \cite{Emparan:2017vyp} {\it disagree} with the general results of \cite{Gadioux:2023pmw}. For example, the results of Emparan {\it et al.}~indicate that the ``instant of merger'' is a merger of the caustic endpoints of the creases on each black hole, rather than the crease perestroika predicted by \cite{Gadioux:2023pmw}. We shall show that the reason for this disagreement is that Emparan {\it et al.}~determined the crease set incorrectly. The surface they identified as the crease set actually lies {\it outside} the event horizon.
When this mistake is corrected, the behaviour of creases is rather different than stated in \cite{Emparan:2017vyp}. For example, in the case of an orthogonal merger (small black hole spin orthogonal to its velocity) we find that the creases are aligned roughly parallel to the spin of the black hole, in contrast with \cite{Emparan:2017vyp}, which predicts they are perpendicular to the spin. Another difference is that Ref.~\cite{Emparan:2017vyp} found that, in the Boyer-Lindquist time foliation, a hole forms in the horizon just after the merger. This would correspond to a horizon of toroidal topology if the large black hole were of finite size. However, we find that no hole in the horizon forms in this time foliation (e.g.~Fig.~\ref{fig:shiny_black_extremal_zoom} differs significantly from the corresponding Fig.~10 of \cite{Emparan:2017vyp}). 

Constructing the event horizon involves studying null geodesics of the Kerr spacetime. Starting with boundary conditions corresponding to a late time Rindler horizon, the geodesic equation must be solved backwards in time along each horizon generator until one reaches a point where the generator intersects another generator (a crease point), or a caustic forms. This has significant overlap with studies of gravitational lensing by a Kerr black hole. Similarly to that work, we use two techniques. First, we use perturbation theory to study geodesics with large impact parameter in the Kerr spacetime. These correspond to horizon generators of the large black hole that remain far from the instant of merger. We have extended these perturbative calculations to many orders higher than considered previously. This enables us to get fairly close to the instant of merger (where the impact parameter is a few times the mass of the small black hole). Second, we use numerical integration of the geodesic equation to obtain horizon generators near to the instant of merger or to the small black hole.

In gravitational lensing, caustics play an important role since they lead to an enhancement of the brightness of a source. In lensing by a Kerr black hole, the set of caustic points forms a two-dimensional tube in spacetime \cite{rauch}. At large distance this tube lies roughly along a line on the opposite side of the black hole to the observer. A cross-section of the tube has the form of an astroid (4-pointed star). Our results extend previous studies of this tube \cite{Sereno:2006ss,Sereno:2007gd} to much higher perturbative order. In our description of a black hole merger, two edges of this tube lie on the event horizon with the rest of the tube lying outside the horizon. 

This paper is organized as follows. In Section \ref{sec:prelim} we introduce the model of Emparan {\it et al.}~and deduce certain properties of the event horizon in this model. We then explain the relation of this problem to the study of gravitational lensing by a Kerr black hole. In Section \ref{sec:results} we give a summary of our main results. Later sections explain how these results are obtained. Section \ref{sec:local_model} uses the local description of \cite{Gadioux:2023pmw} to explain certain geometrical properties of the horizon near the instant of merger. Section \ref{sec:numerics} describes our numerical method for determining the event horizon. Finally, Section \ref{sec:perturbative} describes our perturbative approach. 

\section{Preliminaries}

\label{sec:prelim}

\subsection{General properties of the horizon}

\label{sec:gen_horizon}

\begin{figure}[t]
    \centering
    \includegraphics[width=0.75\textwidth]{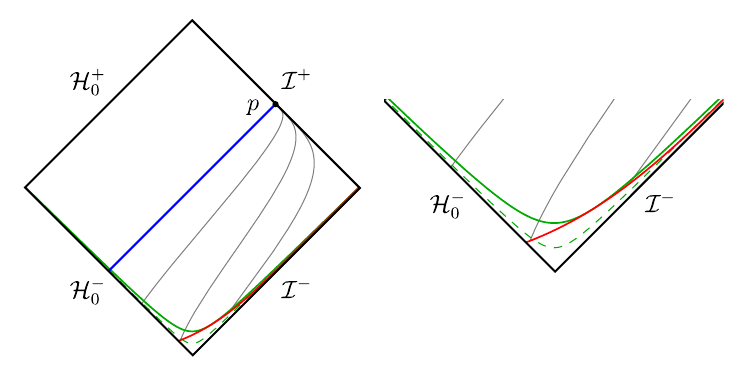}
    \caption{Penrose diagram illustrating the construction of Emparan {\it et al.}~for the case of a non-rotating small black hole \cite{Emparan:2016ylg}. The figure on the right zooms in on the region near past timelike infinity. $p$ is the point at $\scri^+$ used to define $\cH^+$. $\cH_0^\pm$ are the horizons of the Schwarzschild spacetime ($r=2M$). By symmetry we can choose $\theta_o=\pi/2$ and $\phi_o=0$ and consider only geodesics in the equatorial plane. The blue line shows the special ``central'' generator of $\cH^+$ with $\phi \equiv 0$, a radial null geodesic. Other generators of $\cH^+$ are non-radial and their projections onto the Penrose diagram are shown in grey: they appear as timelike curves ending at $p$. ``Creaseless'' horizon generators originate from $\cH_0^-$, all other generators enter $\cH^+$ at a caustic point. The set of caustic points is shown in red: it is a spacelike curve with $\phi=\pi$, extending to spatial infinity. (The location of this line is shown incorrectly in Fig.~6 of \cite{Emparan:2016ylg}.) Two constant $t$ surfaces are shown in green: the solid line is the ``instant of merger'' $t=t_\star$; the dashed line has $t<t_\star$, which interesects the caustic line in two points, one on the large black hole and one on the small black hole. In outgoing Eddington-Finkelstein coordinates, the central generator has $u \equiv 0$ and we find that the caustic line intersects $\cH_0^-$ at $u = -8.265M$. A $\phi=0$ ($\phi=\pi$) cross-section of the black hole region covers the part of the diagram to the left of the blue (red) curve.}
    \label{fig:penrose}
\end{figure}

As explained in \cite{Emparan:2016ylg,Emparan:2017vyp}, to describe the event horizon of an infinite mass ratio merger, we need to find a null hypersurface in the Kerr spacetime that, near future null infinity $\scri^+$, approaches an outgoing planar null hypersurface, i.e.~an acceleration horizon. In Minkowski spacetime, an acceleration horizon can be defined as the boundary of the causal past of a point $p \in \scri^+$ \cite{Jacobson:1999mi,Jacobson:2003wv}. (One can think of $p$ as the intersection of an accelerating worldline with $\scri^+$). Thus in the model of Emparan {\it et al}.~we can define the event horizon of the black hole merger as the boundary of the causal past of a point $p \in \scri^+$ in the Kerr spacetime:\footnote{
This is called an ``asymptotic Rindler horizon'' in \cite{Jacobson:2003wv}.} $\cH^+ = \dot{J}^{-}(p)$. We shall denote the usual future and past event horizons of the Kerr solution as ${\cal H}_0^\pm$. The generators of $\cH^+$ reach $\scri^+$ at $p$ hence $\cH^+$ lies outside $\cH_0^+$. Fig.~\ref{fig:penrose} illustrates this construction for the special case of a non-rotating small black hole.

Following Emparan {\it et al.}, we shall use the Boyer-Lindquist time coordinate $t$ as a time function. We use $\Sigma_t$ to denote a surface of constant $t$. The structure of $\cH^+$ can be seen in Fig.~\ref{fig:horizon_generators}. A late time cross-section of $\cH^+$ has $\mathbb{R}^2$ topology, corresponding to the final black hole after the merger. An early time cross-section of $\cH^+$ is disconnected, with a component of $S^2$ topology and a component of $\mathbb{R}^2$ topology, corresponding to the small and large black holes long before the merger. 

A generator of $\cH^+$ can have a past endpoint, which is where the null geodesic enters $\cH^+$. Any point lying on two (or more) generators is necessarily an endpoint of these generators \cite{Beem:1997uv} and the set of such points is called the {\it crease set}. For an event horizon that is smooth at late time (as we expect here), the following results hold \cite{Gadioux:2023pmw}. An endpoint that lies on only one generator is necessarily a caustic point (i.e.~a point conjugate to a late-time cross-section of $\cH^+$). Non-caustic endpoints lying on exactly two generators form a 2d spacelike submanifold called the {\it crease submanifold}. This corresponds to a self-intersection of $\cH^+$. 

In the special case where the small black hole is non-rotating, the Kerr spacetime reduces to the Schwarzschild spacetime. In this case, the merger preserves axisymmetry and \cite{Emparan:2016ylg} showed that all endpoints belong to the crease set, which is a spacelike line of caustic points shown in Fig.~\ref{fig:penrose} (the crease submanifold is empty). Each caustic point lies on infinitely many generators, related by the axisymmetry. For a rotating small black hole, an axisymmetric merger gives the same structure  \cite{Emparan:2017vyp}. In the rotating case, for a non-axisymmetric merger the analysis of \cite{Emparan:2017vyp} obtained a set of endpoints consisting of a crease submanifold with the topology of an infinite strip, bounded by two lines of caustic points. Although we disagree with the analysis that led to this conclusion, we will show that this structure for the set of endpoints is correct, as shown in Fig.~\ref{fig:crease_set_example}. In this case, two generators enter the horizon at each point of the crease submanifold and one generator enters at each caustic point. The crease set is the same as the crease submanifold.

We define the {\it time of merger} by the condition that $\Sigma_t \cap \cH^+$ is disconnected for $t<t_{\star}$ and connected for $t>t_{\star}$. For $t<t_\star$ each component of $\Sigma_t \cap \cH^+$ exhibits a 1-dimensional crease, corresponding to the intersection of $\Sigma_t$ with the crease submanifold. The endpoints of each crease are caustic points. 

If a generator is extended as a null geodesic beyond its past endpoint (if present), then this geodesic must originate either from the white hole region of the Kerr spacetime, or from past null infinity $\scri^-$ (or non-generically from past timelike infinity). Roughly speaking, one can view these two types of generator as belonging to the small or large black hole before the merger. The Schwarzschild  \cite{penrose1980} and Kerr \cite{Cameron:2020itp} spacetimes have the ``Penrose property'' that $\scri^- \subset  J^-(p)$ for any $p \in \scri^+$. Hence $\cH^+$ cannot intersect $\scri^-$, so any generator of $\cH^+$ that can be extended as a null geodesic to $\scri^-$ must have a past endpoint (i.e. leave $\cH^+$) before reaching $\scri^-$. These generators belong to the large black hole at early time. Hence every generator that belongs to the large black hole before the merger necessarily has a past endpoint. This can be seen in Fig \ref{fig:horizon_generators}. The crease set belongs to $\cH^+$ so it cannot intersect $\scri^-$, instead it extends to spatial infinity as shown in Fig.~\ref{fig:penrose}.
  
A generator belonging to the small black hole at early time can be extended to the past, as a null geodesic, into the white hole region of the Kerr spacetime. Such a generator might have an endpoint outside this region, inside this region, or originate from the white hole inner horizon.\footnote{The latter two possibilities correspond to the generators referred to as ``creaseless'' by Emparan {\it et al}. Note that Raychaudhuri's equation implies that all generators are necessarily incomplete to the past.} In particular, the crease set extends into the white hole region and hence intersects $\cH_0^-$. As $t \rightarrow -\infty$, $\Sigma_t$ approaches $\cH_0^- \cup \scri^-$. Hence at $t=-\infty$ the small black hole horizon cross-section is $\cH^+ \cap \cH_0^-$, i.e. a cross-section of the Kerr {\it white} hole horizon. The presence of a crease implies that this is a non-smooth cross-section, given by a non-smooth function $u=u(\theta,\phi)$ in outgoing Kerr coordinates. The non-smoothness occurs at the crease. However, since all cross-sections of $\cH_0^-$ are isometric, the induced metric on this cross-section is smooth, and isometric to the induced metric on a smooth cross-section of $\cH_0^-$. But this is isometric to a cross-section of $\cH_0^+$. Hence, in the far past, the induced metric on the small black hole horizon is isometric to a cross-section of a Kerr black hole. However, at $t=-\infty$ the expansion and shear of the generators of the horizon of the small black hole differ from those of a Kerr black hole: the latter are zero whereas the former are non-zero, discontinuous at the crease, and divergent at its caustic endpoints. 
 
The white hole region should be regarded as unphysical. In a more physical description it would be replaced by a region describing gravitational collapse of a star to form the small black hole in the far past. 

To determine $\cH^+$ it is convenient to use outgoing Kerr coordinates $(u,r,\theta,\phi)$. As explained above, $\cH^+$ is the boundary of the causal past of a point $p \in \scri^+$. The symmetries of Kerr imply that there is no loss of generality in  assuming $p$ to have coordinates $u=0$, $\phi=0$ and $\theta=\theta_o$ for some $\theta_o$. The parameter $\theta_o$ is the angle between the small black hole spin and the collision axis (see Fig 1. of \cite{Emparan:2017vyp}). Equivalently, $\theta_o$ is the angle between the small black hole spin and the outward normal to a constant time cross-section of the large black hole horizon long before the merger, i.e.~the angle between the small black hole spin and the large black hole's (pre-merger) velocity. In the rest frame of the large black hole, $\pi-\theta_o$ is the angle between the small black hole's spin and its velocity. 

At late time, the generators of $\cH^+$ are outgoing null geodesics in the Kerr spacetime, far from $\cH_0^+$. These geodesics are labelled uniquely using a pair of impact parameters $(\alpha,\beta)$ \cite{Bardeen:1973tla} which can be used as coordinates on a late time (planar) cross-section of $\cH^+$. (The definition of $(\alpha,\beta)$ is given in equation \eqref{alpha_beta_def} of Appendix \ref{app:kerr_geos}.) The generators of $\cH^+$ are the null geodesics which have $(u,r,\theta,\phi) \rightarrow (0,\infty, \theta_o,0)$ as $\lambda \rightarrow \infty$, where $\lambda$ is an affine parameter. To determine $\cH^+$ we need to follow each of these geodesics to the past until we reach its past endpoint. The past endpoint is either a point at which it intersects another such geodesic (corresponding to an endpoint in the crease set) or a point at which the expansion of the generators diverges (a caustic endpoint). A generic past endpoint will correspond to an intersection of geodesics. To determine $\cH^+$, for each $(\alpha,\beta)$ we will need to identify $(\alpha',\beta')$ such that, when evolved to the past starting from $\scri^+$, the geodesic with parameters $(\alpha,\beta)$ intersects the geodesic with parameters $(\alpha',\beta')$, and this is the {\it first} such intersection along the geodesic.

Based on numerical evidence, Emparan {\it et al}.~observed that, outside of a {\it creaseless disc} in the $(\alpha,\beta)$ plane, the geodesics with parameters $(\alpha,\beta)$ and $(\alpha,-\beta)$ ($\beta \ne 0$) intersect each other. They also observed that $\theta = \pi-\theta_o$ at the intersection. We give an analytical explanation of these observations in Appendix \ref{app:false_crease}. Emparan {\it et al}.~interpreted this intersection as the endpoint along each horizon generator, with the points of intersection determining the crease set. They found that this set has the structure of an infinite spacelike strip, bounded by two lines of caustic points. The latter lie on generators with parameters $(\alpha,0)$ with $\alpha>0$ on one line and $\alpha<0$ on the other line. 

The mistake in the analysis of Emparan {\it et al}.~is that the intersection found by them is not the {\it first} intersection (in a backwards-in-time sense) along these generators: we shall show that there is an earlier intersection involving a different pair of generators. Hence the intersection of Emparan {\it et al}.~lies outside $\cH^+$ so the set they identified as the crease set is not actually the crease set. We shall refer to it as the {\it false crease set} and the corresponding disc in the $(\alpha,\beta)$ plane as the {\it false creaseless disc}. We shall see that the true crease set gives a creaseless disc which lies slightly inside the false creaseless disc. This disc corresponds to generators which emerge from the white hole region. 

As just explained, the pairs of generators that meet at the false crease set are related by the map $\beta \rightarrow -\beta$. We have been unable to find an analogous simple characterization of the pairs of generators that meet at the true crease set, which is therefore more difficult to determine. Knowing the location of the false crease set is helpful because it gives an upper bound on how far we have to evolve each generator backwards in time until we reach the true crease set. 

\subsection{Relation to gravitational lensing}

\label{sec:lensing}

The problem of determining $\cH^+$ is closely related to the study of gravitational lensing by a Kerr black hole. In the latter problem one considers an observer far from the black hole, with spatial coordinates $(r_o,\theta_o,\phi_o=0)$ (with $r_0 \gg M$) where $\theta_o$ is the angle between the line of sight to the observer and the spin axis of the black hole (the subscript ``o'' denotes ``observer''). In the limit $r_o \rightarrow \infty$ we can view this idealized observer as living at infinity and making an observation at retarded time $u_o=0$, corresponding to our event $p \in \scri^+$. The null geodesics reaching the observer are precisely the generators of our $\cH^+$. The impact parameters $(\alpha,\beta)$ can be regarded as coordinates on the ``screen'' of the observer \cite{Bardeen:1973tla}. 

A typical gravitational lensing problem is to consider a source at fixed spatial position $(r_s,\theta_s,\phi_s)$ (``s'' for ``source'') and determine how this appears to the observer. To do this one needs to determine which values of $(\alpha,\beta)$ give a null geodesic which intersects the worldline of the source. This is done by solving the geodesic equation backwards in time from $p$. Each such pair $(\alpha,\beta)$ corresponds to a distinct image of the source. Generically, distinct images arise from geodesics which intersect the worldline of the source at different times $u_s$, i.e.~there is a time delay between different images. 
In our problem, we need to determine the crease set, i.e.~points of intersection of horizon generators. If we imagine a source at such a point of intersection then it will give rise to distinct images with {\it vanishing} time delay. Conversely, if a source at $(r_s,\theta_s,\phi_s)$ gives rise to a pair of images with vanishing time delay then $(u_s,r_s,\theta_s,\phi_s)$ is a potential crease point (with $u_s$ fixed by $u \rightarrow 0$ as $\lambda \rightarrow \infty$). For it to be an actual crease point (rather than e.g.~a false crease point), this intersection needs to be the first such intersection (when evolved backwards in time from $p$) along each geodesic. 

\begin{figure}
    \centering
    \includegraphics[width=0.45\textwidth]{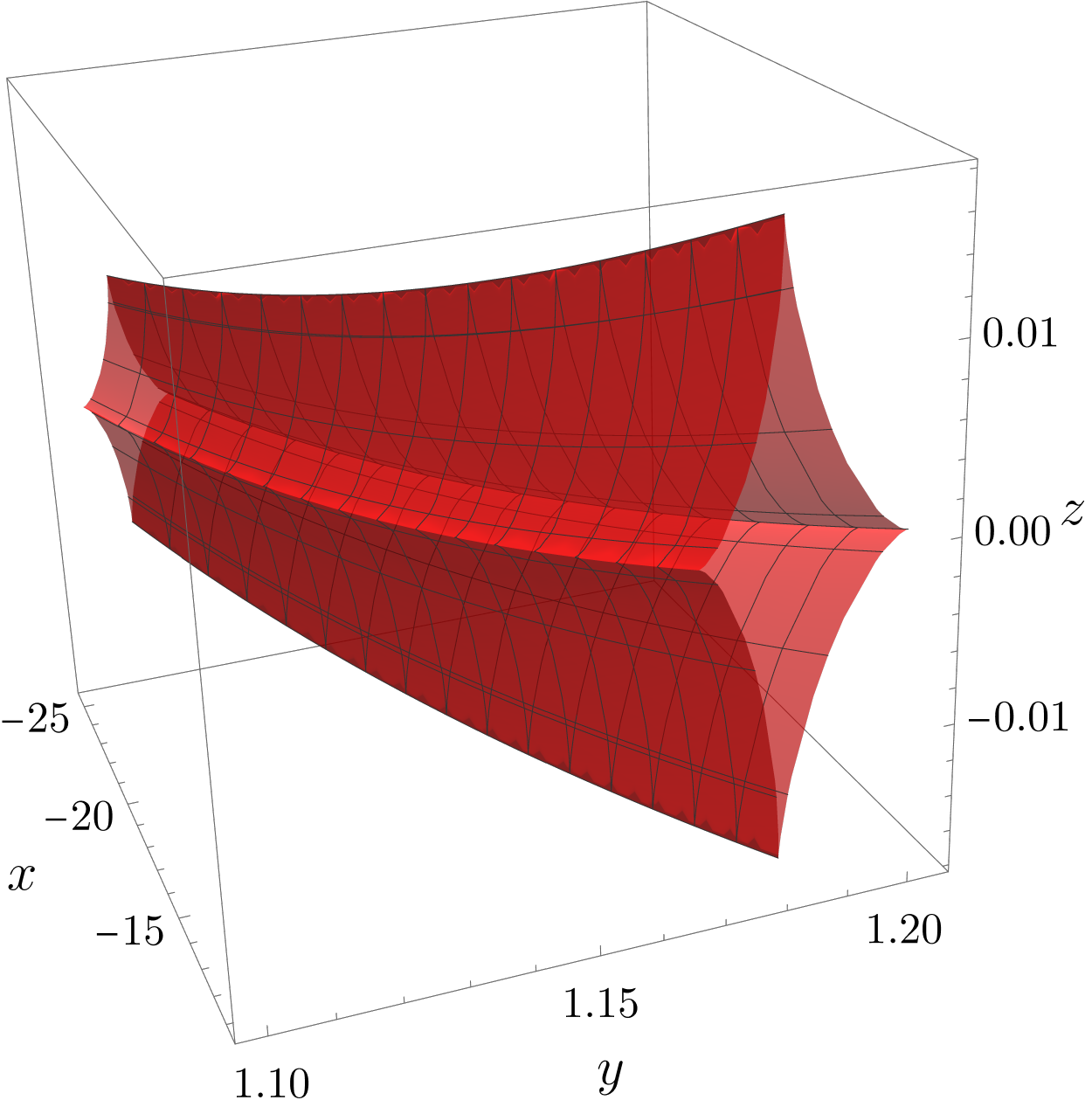}
    \quad \includegraphics[width=0.45\textwidth]{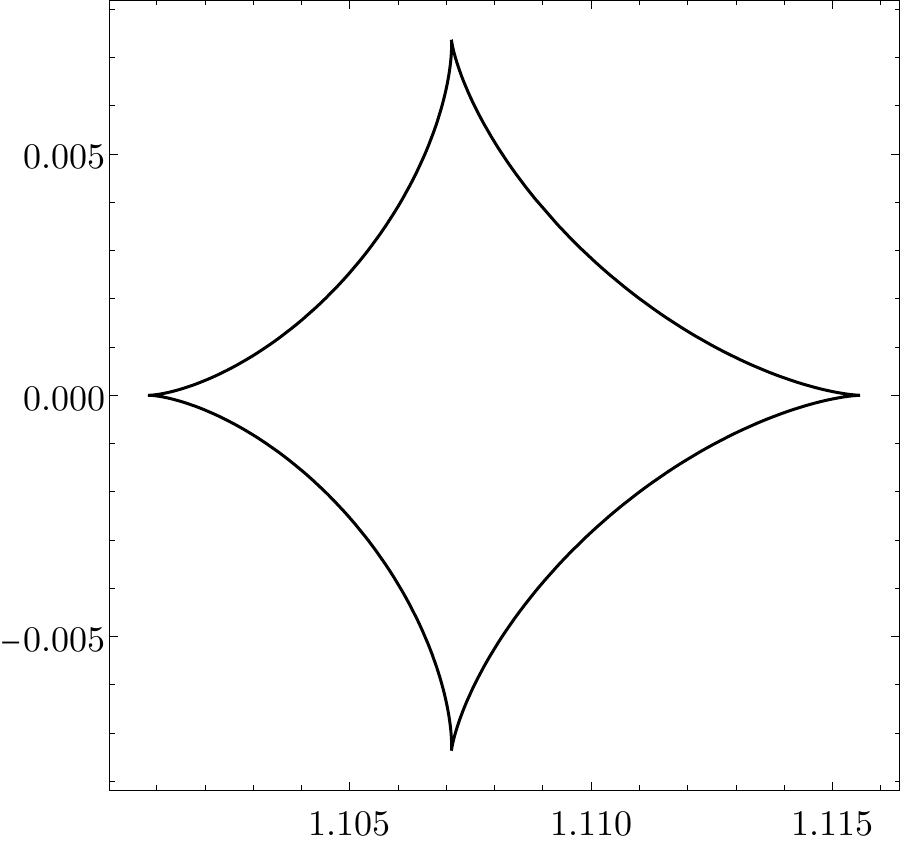}
    \caption{The caustic tube arising from lensing by an extremal Kerr black hole with $\theta_o=\pi/2$ (observer in the equatorial plane). The plot shows a portion of the tube far away from the black hole, where a perturbative approach is valid \cite{Sereno:2006ss,Sereno:2007gd}. The right panel shows a constant $x = -25$ cross-section of the tube. Note the different scales on the axes. (In terms of the parameter $\varepsilon$ defined below, this section of tube has $\varepsilon \in [0.1, 0.15]$ and the cross-section has $\varepsilon = 0.1$.)}
    \label{fig:caustic_tube}
\end{figure}

The problem of searching for multiple images with vanishing time delay does not seem to have been studied in the lensing literature. However, the problem of determining caustic points has received considerable attention \cite{rauch,Sereno:2006ss,Sereno:2007gd}. To determine caustic points, one evolves null geodesics backwards from $p$ until the expansion of the geodesics diverges. Caustic points are important in lensing because if a source passes through a caustic point then its brightness is enhanced. For a Schwarzschild black hole, there is a line of caustic points on the ``optical axis'' $\theta = \pi-\theta_o$, $\phi=\pi$ (i.e.~antipodal to the observer).\footnote{This is the ``primary caustic''. Higher order caustics arise from geodesics that orbit the black hole one or more times. We shall not discuss these higher order caustics.} For a Kerr black hole, the set of caustic points has the structure of a two-dimensional {\it tube} \cite{rauch}. Far from the black hole, this is parallel to the optical axis and displaced from it by a distance proportional to $a$. 
The tube can be plotted by projecting onto the spatial Boyer-Lindquist coordinates, i.e.~suppressing the time coordinate, as shown in Fig.~\ref{fig:caustic_tube}. A cross-section through this tube gives an {\it astroid} ($4$-pointed star) also shown in Fig.~\ref{fig:caustic_tube}. The vertices of the astroid form $4$ spacelike curves in spacetime. These points correspond to caustic points of type $A_3$ (``swallowtail'') in the classification of generic caustics. The four ``faces'' of the caustic tube are $2$-dimensional null surfaces in spacetime \cite{Gadioux:2023pmw}. Points on these surfaces are caustic points of type $A_2$ (``cusp''). 

\begin{figure}[h]
    \centering
    \includegraphics[width=0.45\textwidth]{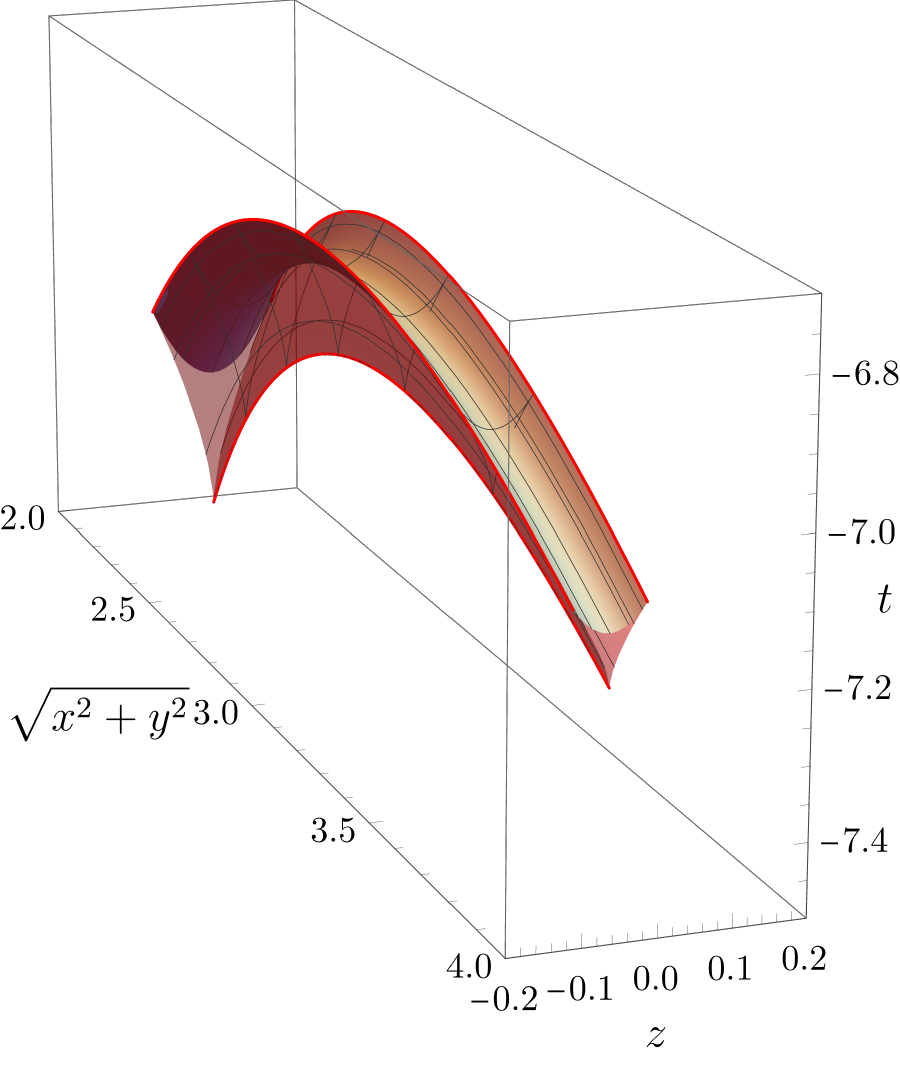}
    \caption{Crease set and caustic tube for an orthogonal merger with an extremal small black hole. The crease set is the same as in Fig.~\ref{fig:crease_set_example} and the caustic tube ``hangs down'' from it. 
    A spatial coordinate is necessarily suppressed in this figure, so the astroid structure of the tube is not apparent: the bottom edge of the figure corresponds to the pair of $A_3$ lines along the edges of the false crease set (i.e.~the caustics of \cite{Emparan:2017vyp}). This figure is generated from our perturbative results which are only qualitatively correct near the small black hole.}
    \label{fig:creasencaustic}
\end{figure}

It is known that, generically, only caustics of type $A_3$ can occur as endpoints of horizon generators \cite{Siino:2004xe,Gadioux:2023pmw} and that such endpoints should occur at the edges of the crease set. We already know that the false crease set is a strip bounded by a pair of caustic lines \cite{Emparan:2017vyp}. The latter must correspond to two of the four lines of $A_3$ points (since $A_3$ points always lie adjacent to an intersection whereas $A_2$ points do not \cite{Gadioux:2023pmw}). It is natural to guess that the true crease set is also a strip, bounded by the other pair of $A_3$ lines. The crease set together with this pair of lines is the set of endpoints of the horizon generators. $A_2$ points cannot occur on $\cH^+$ \cite{Siino:2004xe,Gadioux:2023pmw} so, except for two of the $A_3$ lines, all of the caustic tube must lie {\it outside} $\cH^+$. When a generator is evolved backwards in time it reaches the crease set before reaching the caustic tube (except for the special generators with $A_3$ endpoints). Thus, along each null geodesic, the crease set ``lies to the future'' of the caustic tube. This is shown in Fig.~\ref{fig:creasencaustic}.

\begin{figure}[h]
    \centering
    \includegraphics[width=0.6\textwidth]{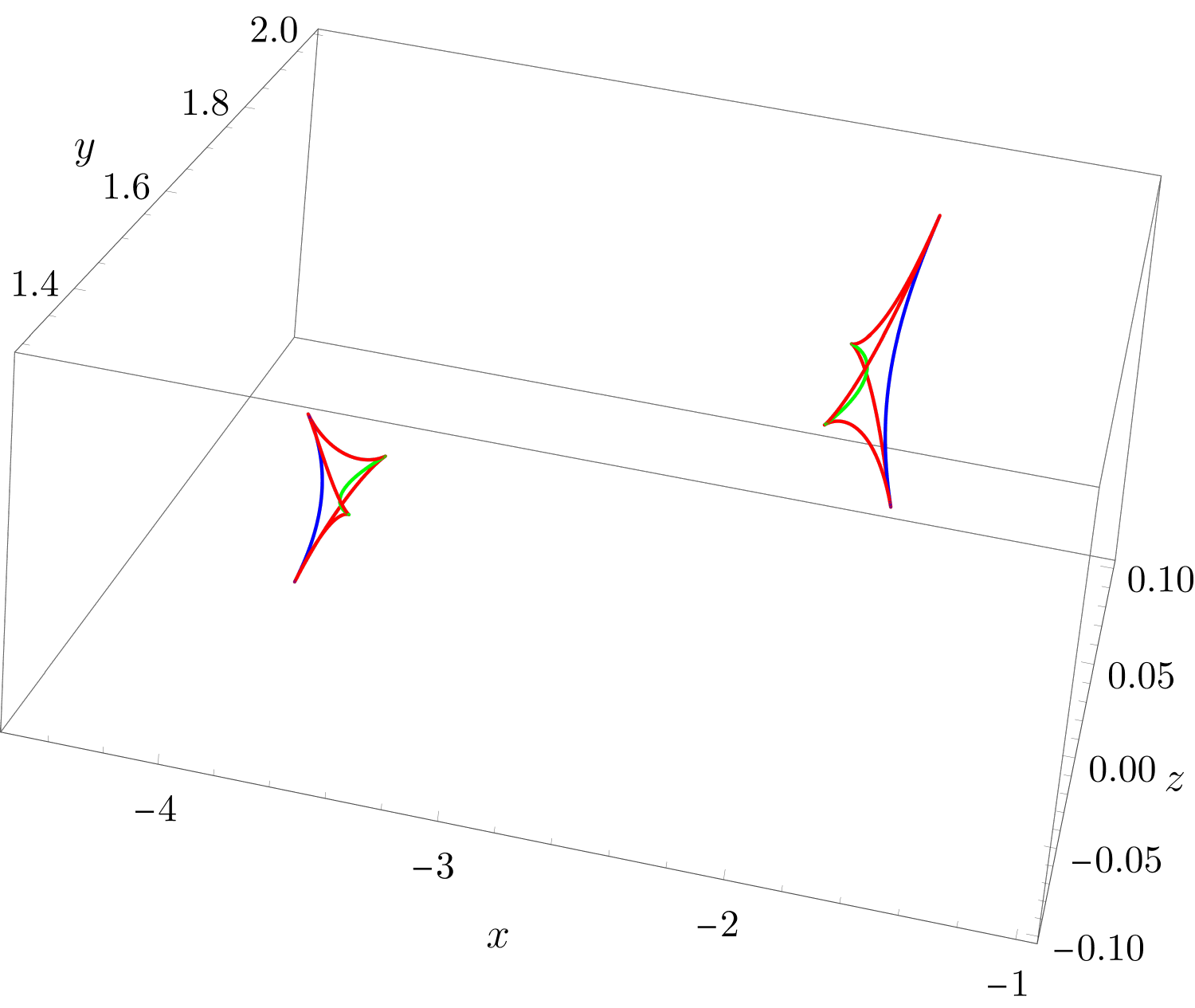}
    \caption{Constant time cross-section of the caustic tube (red), crease (blue) and false crease (green) for an orthogonal merger with an extremal small black hole. The figure is constructed from our perturbative results (at time $t=-7.2$), which give a qualitatively correct description.}
    \label{fig:crease_caustic_section}
\end{figure}

When projected onto the spatial coordinates, both the crease set and the false crease set lie inside the caustic tube, connecting opposite vertices of the astroid. As explained above, the lensing interpretation of these sets is as points at which a source produces a pair of images with vanishing time delay. Taking an {\it intersection} with a surface of constant $t$ gives a rather different picture shown in Fig.~\ref{fig:crease_caustic_section}. At early time, a surface of constant $t$ intersects the caustic tube in two places, near to the small and large black hole horizons. Each intersection has the form of an astroid. The creases and false creases are lines connecting opposite vertices of these astroids, with the overall structure resembling a deformed tetrahedron. The only parts of this structure that lie on $\cH^+$ are the (blue) crease lines and their $A_3$-caustic endpoints. 

A well-studied object in lensing is the {\it black hole shadow}. This is the disc in the $(\alpha,\beta)$ plane corresponding to null geodesics which originate from the white hole region and reach the observer at infinity. This is not the same as the creaseless disc because the latter corresponds to horizon {\it generators}, not just null geodesics, that emerge from the white hole region. The creaseless disc is a subset of the black hole shadow \cite{Emparan:2017vyp}. A point belonging to the shadow but not the creaseless disc corresponds to a null geodesic that emerges from the white hole region and subsequently enters the event horizon as a generator. 
For the special case of a Schwarzschild black hole, all endpoints are caustic points so the creaseless disc corresponds to null geodesics without a caustic outside the white hole region. This disc has radius $4.46M$ \cite{rauch,Emparan:2016ylg} whereas the shadow has radius $\sqrt{27}M=5.20M$, with both discs centred on the origin. For Kerr, the shadow is not exactly circular and it is shifted to the right (for $a>0$) \cite{Bardeen:1973tla}. The same is true of the false creaseless disc  \cite{Emparan:2017vyp} and, as we shall show, the true creaseless disc.

\section{Results}

\label{sec:results}

In this section we shall present our main results, obtained via the perturbative and numerical approaches. The explanations of how these results are obtained will be given in later sections. Recall that the merger is defined by two parameters: $\chi \equiv a/M$ and $\theta_o$ (the merger angle). We shall sometimes use $\mu_o \equiv \cos \theta_o$. All of our figures use units such that $M=1$. 

\subsection{Horizon plots}

For visualization purposes, it is convenient to convert the spatial Boyer-Linquist coordinates to quasi-Cartesian coordinates defined as follows~\cite{Emparan:2017vyp}: 
\begin{align}
x' &= \sqrt{r^2 + a^2 + \frac{2 M a^2 r \sin^2\theta}{\Sigma}} \sin \theta \cos \phi \, ,
\nonumber \\ \label{Emp_Cart_Primed}
y' &= \sqrt{r^2 + a^2 + \frac{2 M a^2 r \sin^2\theta}{\Sigma}} \sin \theta \sin \phi
\, ,
\\
z' &= r \cos \theta \, .\nonumber
\end{align}
These coordinates have the property that circles of constant $z'$ and ${x'}^2+{y'}^2$ (i.e.~orbits of $\partial/\partial \phi$) have circumference $2\pi \sqrt{{x'}^2+{y'}^2}$, as in Euclidean space. We now perform a rotation of the axes:
\be 
\label{Emp_Cart}
x = x' \sin \theta_o + z' \cos \theta_o \, , \quad y = y' \, , \quad z = z' \sin \theta_o - x' \cos \theta_o \, .
\ee
In the gravitational lensing interpretation, the observer lies on the $x$-axis (with large positive $x$). For the black hole merger interpretation, these are coordinates in which the small black hole is at rest at the origin with spin axis lying in the $xz$-plane, at angle $\theta_0$ to the positive $x$-axis. For large negative $t$, the large black hole has large negative $x$. The large black hole moves in the positive $x$-direction.   

Figs \ref{fig:horizon_generators}, \ref{fig:shiny_black_extremal} and \ref{fig:shiny_black_extremal_zoom} are obtained by determining the horizon generators, and their intersections, numerically. Figs \ref{fig:shiny_black_extremal} and \ref{fig:shiny_black_extremal_zoom} show the case of an orthogonal ($\theta_o=\pi/2$) merger with an extremal small black hole. This is the case for which the creases are largest. The creases have a hyperbolic shape before and after the merger, in agreement with the exact local results of \cite{Gadioux:2023pmw}.

\subsection{Crease set}

Our results suggest that, for a non-axisymmetric merger, the crease set and the crease submanifold coincide in the spacetimes we are considering. In general, the crease set may include a one-dimensional ``corner submanifold'' at which three leaves of the crease submanifold meet \cite{Gadioux:2023pmw} (giving rise to a corner on a horizon cross-section) but we do not find any evidence for this. Generically, caustic points are of type $A_3$, and lie on exactly one horizon generator and so do not belong to the crease set \cite{Siino:2004xe,Gadioux:2023pmw}. They belong to the closure of the crease set.

We determine the crease submanifold by looking for pairwise intersections of generators. Using perturbative calculations, valid for generators with large impact parameter ($\sqrt{\alpha^2 + \beta^2} \gg M$), we find that this submanifold is given by the following parametric equations in Boyer-Lindquist coordinates:\footnote{
Recall that we often follow the gravitational lensing notation where a subscript ``s'' indicates the position of a possible ``source'' and ``o'' to the ``observer''. For us, ``o'' refers to the point $p \in \scri^+$ used to define $\cH^+$ and ``s'' labels any point on a null geodesic extending to $p$. We also use $\mu \equiv \cos \theta$.
}
\begin{align}
\label{eps_def}
    r_s \equiv &\,  \frac{M} {4 \varepsilon^2} \, ,
    \\
\label{eqn:t_crease}    
\frac{t_s}{M} =& \, - \frac{1}{4 \varepsilon^2} - \frac{15 \pi }{4} \varepsilon +  \left(\frac{225 \pi^2}{256} - 8 \right)\varepsilon^2 +  \left\{-\frac{15  \pi \left( 375 \pi^2 + 1664 \right)}{8192}  \right. 
\nonumber\\
&\left.+ \frac{5 \pi \chi ^2}{32}  \left(13 - 6 \mu_o^2-3 (1-\mu_o^2) \cos^2 \varphi  \right)\right\} \varepsilon^3 + \left\{ \frac{3 \left(16875 \pi^{4}-60000 \pi^{2}-1048576\right) }{65536} 
\right. 
\nonumber \\
&\left. 
+ \frac{\left(16384 - 225 \pi^{2} \left(1-\mu_{o}^{2}\right) \cos^{2}\varphi - 225 \pi^2 \left(2 \mu_{o}^{2}+5\right)\right) \chi^{2}}{1024} \right\} \varepsilon^4  + \mathcal{O}\left(\varepsilon^5 \right)
\\
\label{eqn:mu_crease}
\mu_s =& \, - \mu_o - \left\{ \frac{15 \pi \chi^2}{16} \left(1-\mu_o^2\right)^{3/2} \sin \varphi \right\} \varepsilon^4 + \mathcal{O}\left(\varepsilon^5\right)\, ,
\\
\label{eqn:phi_crease}
\phi_s =& \, \pi \, {\rm sign}(\cos \varphi) - 4 \chi \varepsilon^2 - \frac{5 \pi \chi }{4 } \varepsilon^3  + \left(\frac{225 \pi^2}{128} - 16 \right) \chi \varepsilon^4  + \mathcal{O}\left(\varepsilon^5\right)
\end{align}
where $\varphi$ and $0 < \varepsilon \ll 1$ parameterize the submanifold and $\chi = a/M$. Equation \eqref{eps_def} is exact (it is the definition of $\varepsilon$) and we have truncated the other equations to the lowest order for which the crease set is non-trivial. Our perturbative results extend to much higher order than displayed above: including terms up to $\mathcal{O}(\varepsilon^8)$ for mergers of generic orientation and to $\mathcal{O}(\varepsilon^{13})$ for orthogonal mergers ($\mu_o = 0$).  
We have defined $\varphi$ to be the polar angle in the $(\alpha,\beta)$ plane. The mapping from this plane to the crease submanifold is two-to-one, and restricting $\varphi$ to the range $\varphi \in (\pi/2,3\pi/2)$ (or its complement) gives a one-to-one map. The limits $\varphi \rightarrow \pi/2,3\pi/2$ correspond to the lines of caustic points along the edges of the crease submanifold. (At higher order in $\varepsilon$ than shown above the location of these lines is not exactly at $\varphi = \pi/2,3\pi/2$.)


Converting to the quasi-Cartesian coordinates \eqref{Emp_Cart}, an example of the crease set determined using the above equations is shown in Fig.~\ref{fig:crease_set_example}. The most important difference between the crease set and the false crease set is that the crease set ``hangs down'' from two lines of caustic points whereas the false crease set ``bulges up'' from two lines of caustic points (Fig.~14 of \cite{Emparan:2017vyp}).\footnote{
This is the reason why the results of \cite{Emparan:2017vyp} disagree with those of \cite{Gadioux:2023pmw}. The latter proves that if a surface of constant time is tangent to a line of caustics then, locally, a slightly later surface of constant time does not intersect the crease submanifold. This would not be true if the crease submanifold bulges up. 
}
This is the reason why no hole forms in the horizon w.r.t.~the Boyer-Lindquist time slicing. As the parameters $a/M$ and $\theta_o$ are varied, the qualitative shape of the crease set is the same as in Fig.~\ref{fig:crease_set_example} but its width varies. It is largest for the case of an orthogonal merger involving an extremal small black hole, as shown in Fig.~\ref{fig:crease_set_example}. In the limits $\theta_o \rightarrow 0,\pi$ the merger becomes axisymmetric and the crease set shrinks to a line of caustic points \cite{Emparan:2017vyp}.
  
\begin{figure}
    \centering
    \includegraphics[width=0.5\textwidth]{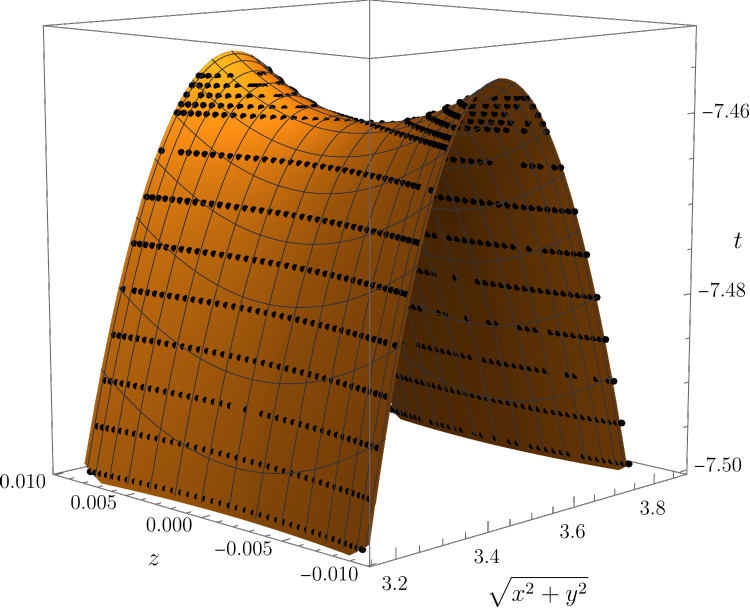}
    \caption{Numerical data for the crease submanifold (black points) for an orthogonal merger with parameters $M=1$, $a=1/2$. The numerical data are obtained for several different fixed values of $t$, which is why they lie in horizontal planes. The surface is a fit to the numerical data. The merger occurs at $t=-7.46091$, and the caustic perestroikas (at which the creases disappear) occur at $t=-7.45453$.}
    \label{fig:prelim_crease}
\end{figure}

Our perturbative calculations are valid for $\varepsilon \ll 1$, which corresponds to $r_s \gg M$, which describes the part of the crease set that belongs to the large black hole before the merger. However, in Fig.~\ref{fig:crease_set_example}, we have extrapolated $\varepsilon$ to $\mathcal{O}(1)$ values, which lie outside the regime of validity of our perturbative calculations. Near to the small black hole we expect these results to be unreliable, and the crease submanifold must be determined numerically. The results of such a numerical calculation are shown in Fig.~\ref{fig:prelim_crease}.  

\begin{figure}
    \centering
    \includegraphics[width=0.60\textwidth]{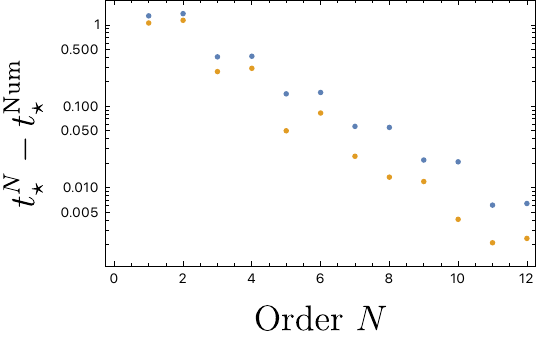}
    \caption{Comparison of the time of merger $t_\star^N$ as computed from the perturbative calculations including terms up to order $\varepsilon^N$ in the expansion of the time function with the time of merger determined numerically $t_\star^{\rm Num}$. Here blue dots correspond to an orthogonal merger with $a = 1/2$ and the orange dots correspond to an orthogonal merger with $a = 1$. When a series including terms up to $\varepsilon^{13}$ is used, the analytical and numerical results agree up to three decimal places. The units are such that $M=1$.}
    \label{fig:merger_error}
\end{figure}

The time of merger $t_\star$ is determined by the condition that the surface $t=t_\star$ is tangent to the crease submanifold (this is the defining condition of a crease perestroika). The instant of merger in these figures has $r \sim 3M$ in these figures and hence $\varepsilon \approx 0.3$. We can test this accuracy of our perturbative results for these values of $\varepsilon$ by comparing our perturbative prediction for the time of merger $t_\star$ with our numerical results for this quantity. If the perturbative results are extended to sufficiently high order then we obtain good agreement, as shown in Fig.~\ref{fig:merger_error}.

\begin{figure}[h]
    \centering
    \includegraphics[width=0.45\textwidth]{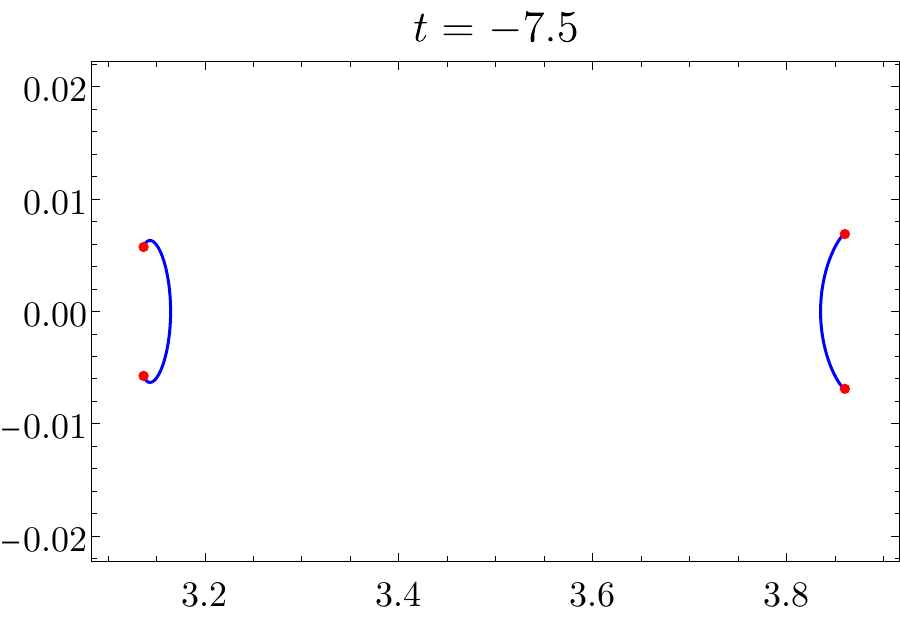}
    \includegraphics[width=0.45\textwidth]{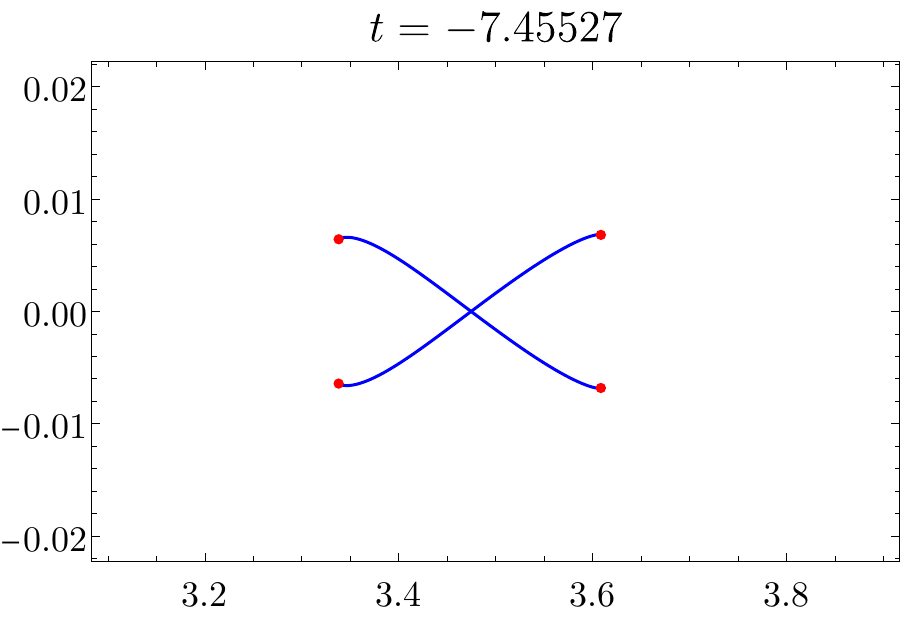}
    \includegraphics[width=0.45\textwidth]{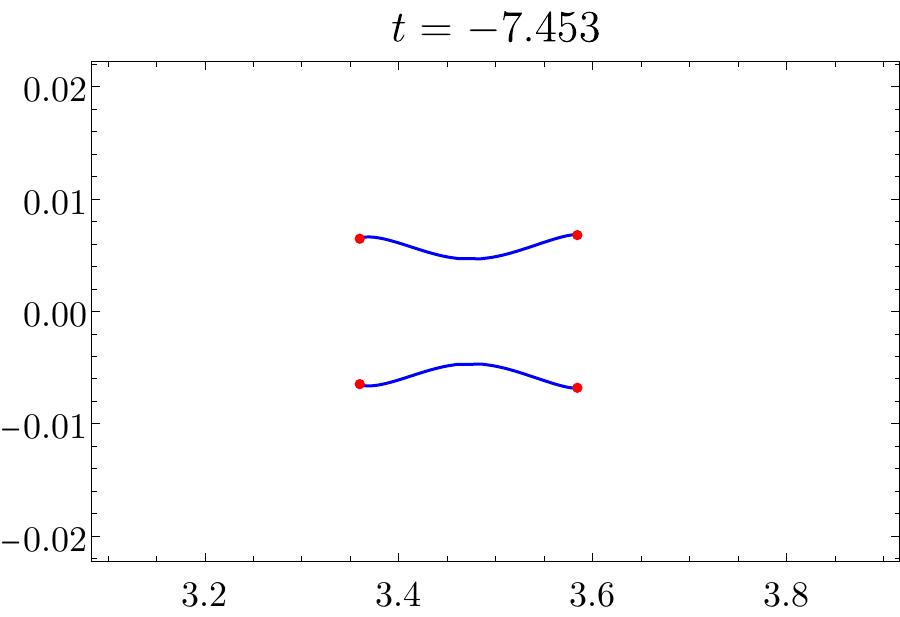}
    \includegraphics[width=0.45\textwidth]{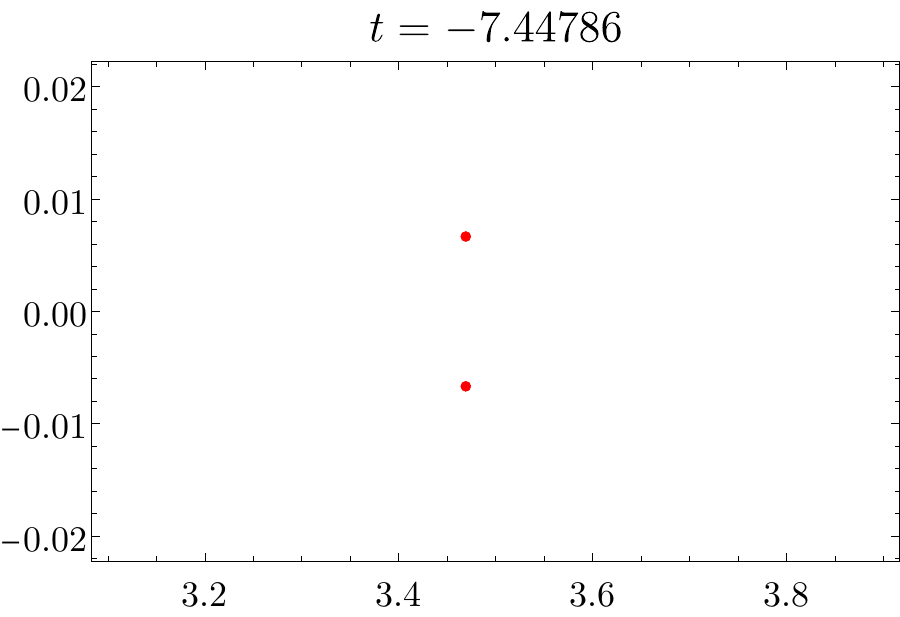}
    \caption{Constant Boyer-Lindquist time-slices of the crease set for an orthogonal merger with a small black hole with $a=1/2$. The blue lines are the creases and the red dots are caustic points. The small black hole is on the left in the first panel.
    The second panel is the crease perestroika (instant of merger). The final panel is the caustic perestroika (annihilation of caustics). These results have been obtained from our perturbative formulae.}
    \label{fig:crease_ts_plots}
\end{figure}

\begin{figure}[h]
    \centering
    \includegraphics[width=0.45\textwidth]{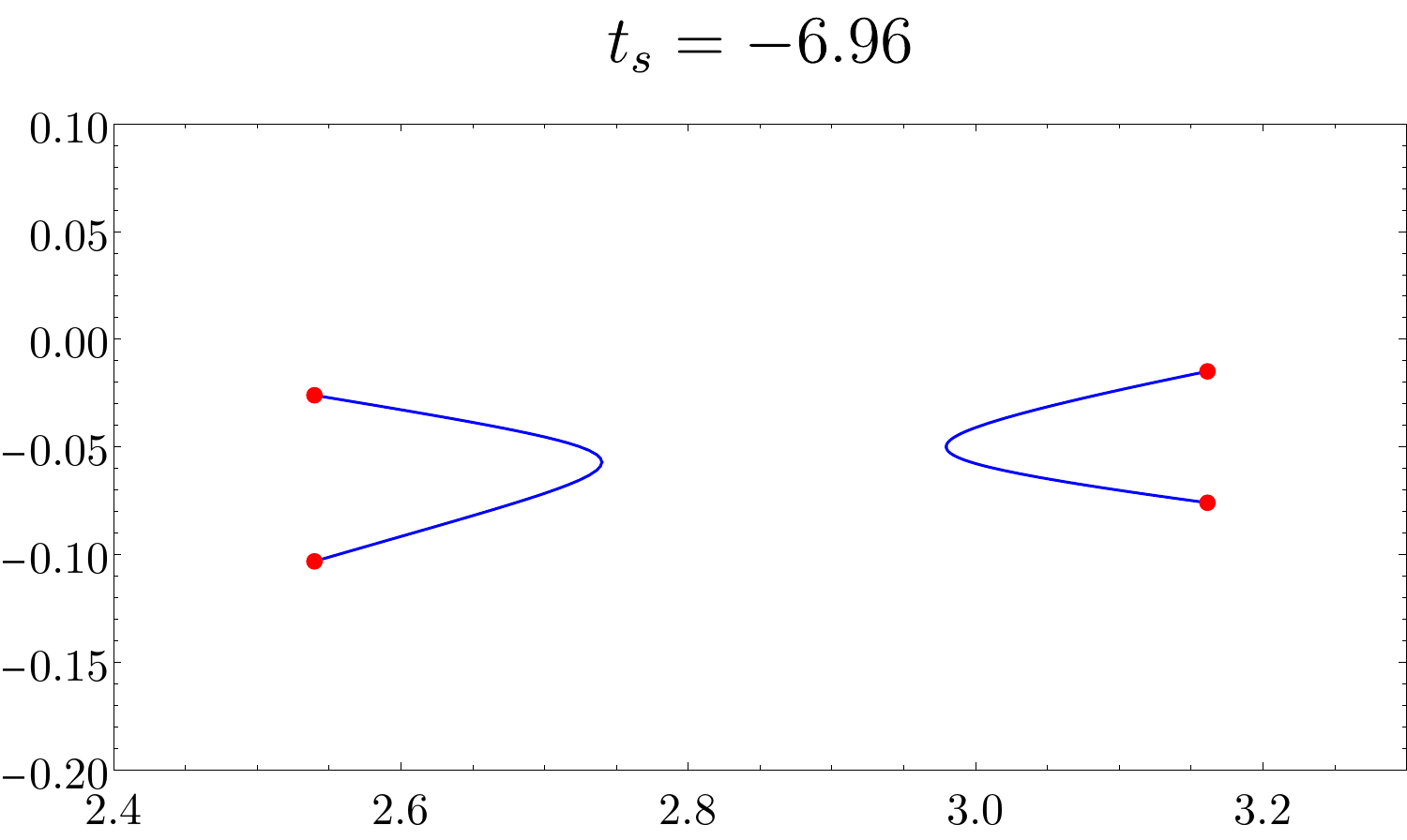}
    \quad
    \includegraphics[width=0.45\textwidth]{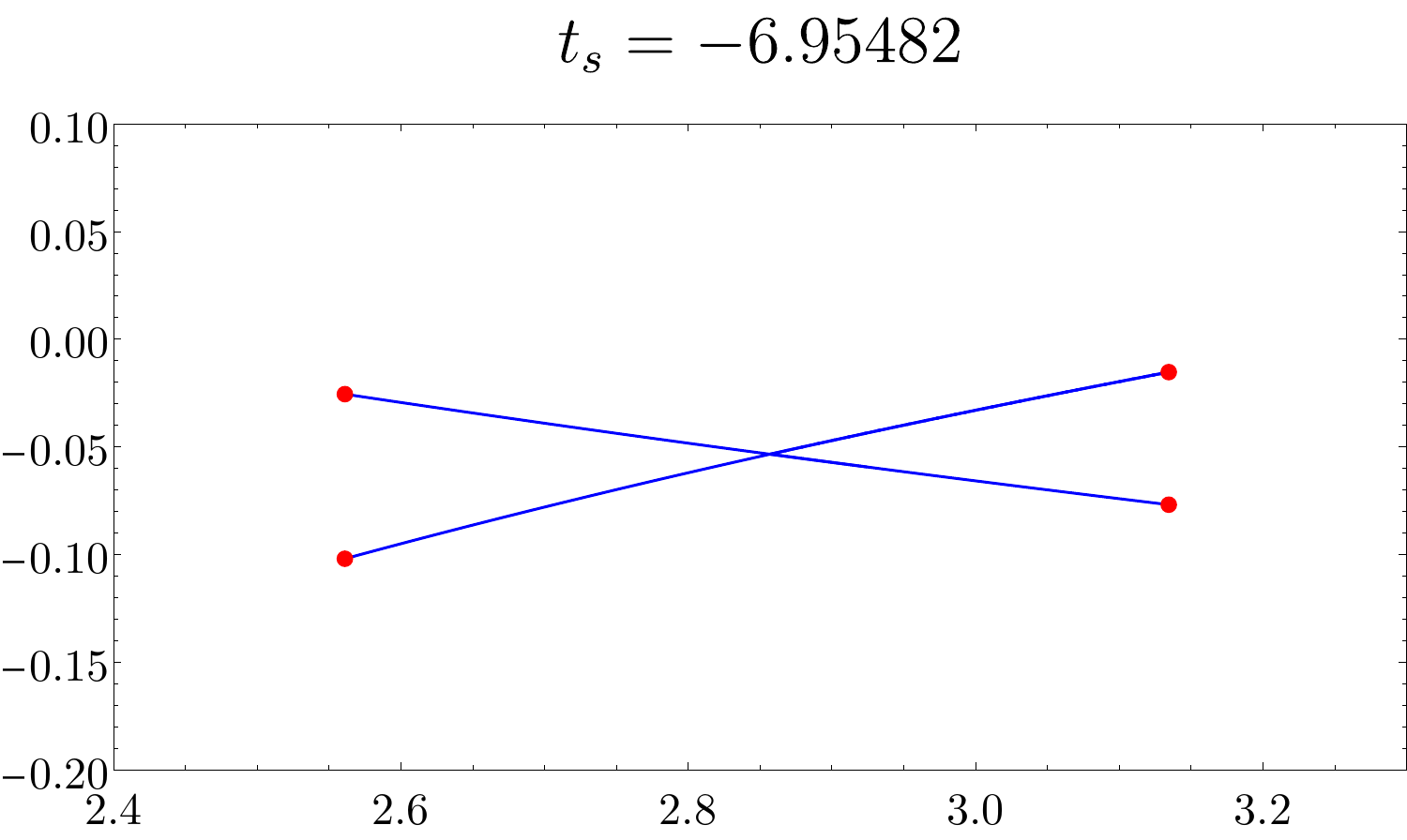}
    \includegraphics[width=0.45\textwidth]{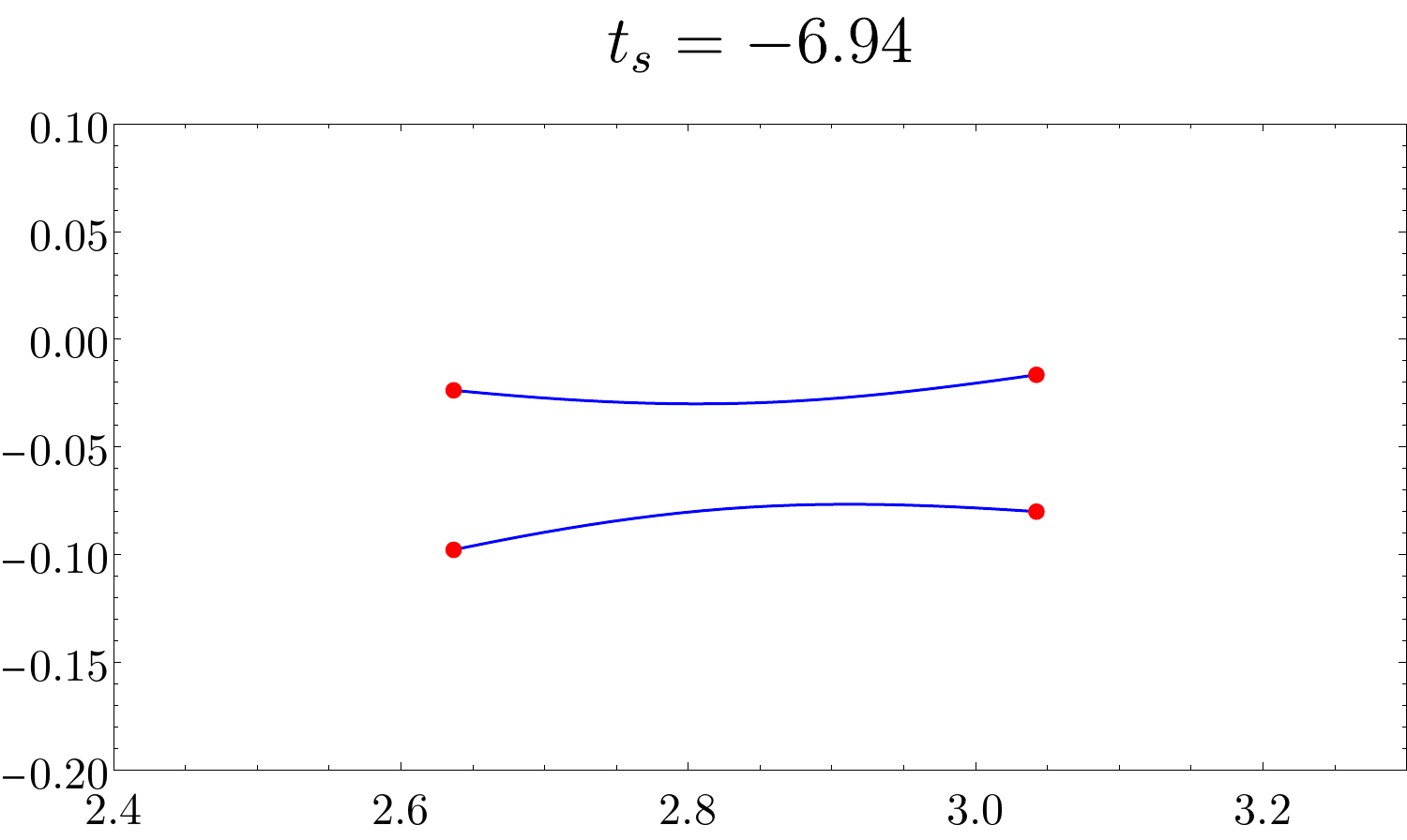}
    \quad 
    \includegraphics[width=0.45\textwidth]{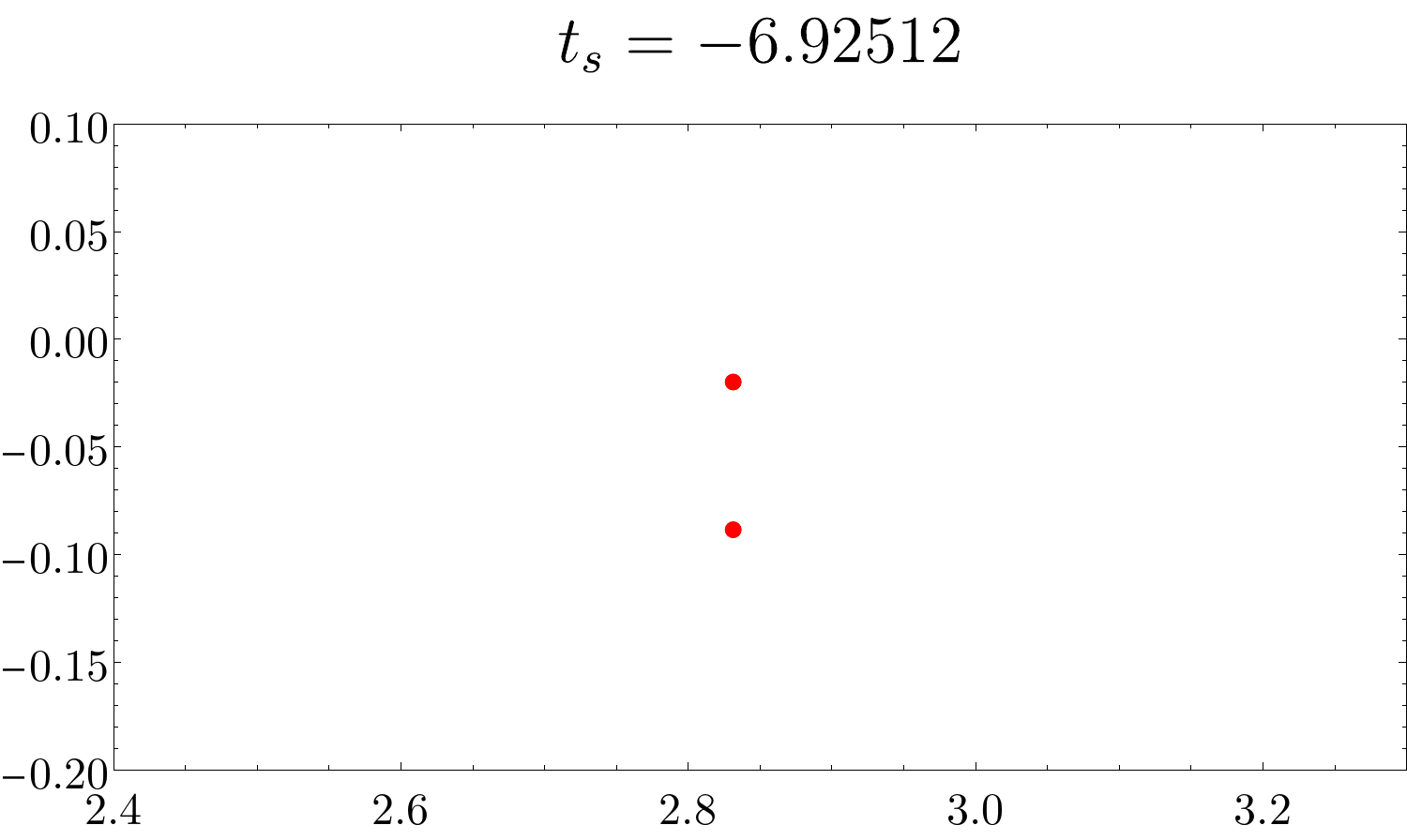}
    \caption{Constant Boyer-Lindquist time-slices of the crease set for an extremal small black hole and merger angle $\cos \theta_o = 0.7$. These plots have been made using our perturbative results, with the Boyer-Lindquist coordinate expansions up to and including $\mathcal{O}(\varepsilon^5)$ corrections. The perturbative results give a correct qualitative description near the moment of merger. Quantitatively, the results are accurate to about one or two significant digits in this regime, depending on the value of $t$. Including higher-order terms improves the quantitative agreement but does not change the qualitative picture in a meaningful way. }
    \label{fig:crease_ts_plots_mu0p7}
\end{figure}

The accuracy of our perturbative results for computing the merger time gives us confidence that these results can be used reliably to study other aspects near the of instant of merger. Taking constant time cross-sections of the crease submanifold near the merger time gives a detailed picture of the behaviour of creases during the merger process. This is shown in Figs \ref{fig:crease_ts_plots}, \ref{fig:crease_ts_plots_mu0p7}. In these plots one sees that the creases have a hyperbolic shape just before and just after the merger, in agreement with the local model of \cite{Gadioux:2023pmw}. We can also see the behaviour of the creases immediately after the merger, where they shrink and disappear in a caustic perestroika \cite{Gadioux:2023pmw}, which corresponds to a plane of constant $t$ that is tangent to a line of caustic points. A non-generic feature of this time slicing is that the caustic perestroikas associated with the two caustic lines occur at the same time. 

At fixed $\varepsilon$ (i.e.~fixed large $r$), our perturbative formulae show that the changes in $\mu_s$ and $\phi_s$ across the crease submanifold are $\mathcal{O}(\varepsilon^4)$ and $\mathcal{O}(\varepsilon^5)$ respectively. The same is true if one converts to taking a cross-section of constant $t$ rather than constant $r$. Hence, in a constant $t$ cross-section, the extent of the crease in the $\theta$ direction (parallel to the small black hole spin) is much greater than its extent in the $\phi$ direction. Our numerical results confirm that the same is true near to the small black hole at early time: at constant $t$, the crease submanifold extends mainly in the $z$-direction with only small variation in the $(x,y)$ directions. 

The crease submanifold extends to (spatial) infinity, corresponding to the large black hole horizon at very early time. Using our perturbative results, we find that the proper lengths of the caustic lines are infinite. So it is not obvious that the crease submanifold has a well-defined area. Nevertheless, using our perturbative results, we find that the part of the crease submanifold with $\varepsilon \le \varepsilon_{\rm max}$ does have a finite area:
\begin{align}
A_{\rm Crease}  &= \frac{15 \pi a^2 (1-\mu_o^2)}{64} \varepsilon_{\rm max} \left[1 + \frac{15 \pi \varepsilon_{\rm max}}{64} 
 + \cdots \right] \, .
\end{align}
It follows that the part of the crease submanifold that lies outside the white hole region must have finite area, although our numerical results are not good enough to compute this area accurately. Ref.~\cite{Gadioux:2023pmw} described how the area of the crease submanifold can be used to improve the area-law bound on the efficiency of energy emission in gravitational waves in a black hole merger. However, our result is not useful for doing this in the infinite mass ratio limit, since a simple computation indicates that the change in area of the large black hole is infinite in this limit \cite{Emparan:2016ylg}. 

How is the area we have computed related to the area of the crease submanifold for a spacetime describing a merger of two black holes with finite masses $M \ll {\cal M}$? It is not clear how the latter area behaves in the infinite mass ratio limit ${\cal M} \rightarrow \infty$ since the description of Emparan {\it et al.}~applies only to the final stage of the merger when the small black hole is an $O(M)$ distance from the large one. In particular, it does not apply to the inspiral phase of the merger, when the small black hole is an $O({\cal M})$ distance from the large hole. So it seems unlikely that the area we have computed will agree with the ${\cal M} \rightarrow \infty$ limit of the area of the crease submanifold for a finite mass merger, or indeed even that the latter limit exists (especially if the black holes start at infinite separation). Our crease submanifold should be a good approximation to the crease submanifold of the finite mass ratio merger in a region of spacetime near to the large black hole, of size much smaller than ${\cal M}$. 

The area of the crease submanifold is a geometric invariant of a dynamical black hole spacetime so it would be very interesting to find a physical interpretation for this area, either for a merger or for a spacetime describing the formation of a black hole in gravitational collapse. (In the latter case, the area law implies that the final (equilibrium) black hole area $A_f$ satisfies $A_f \ge 2A_{\rm crease}$ \cite{Gadioux:2023pmw}.) $A_{\rm crease}$ vanishes for very symmetrical spacetimes such as spherically symmetric gravitational collapse, or an axisymmetric black hole merger. Perhaps this area could be regarded as a measure of the complexity of the process by which the final black hole is formed.

\subsection{Lengths of creases}

\begin{figure}
\centering
    \includegraphics[width=0.7\textwidth]{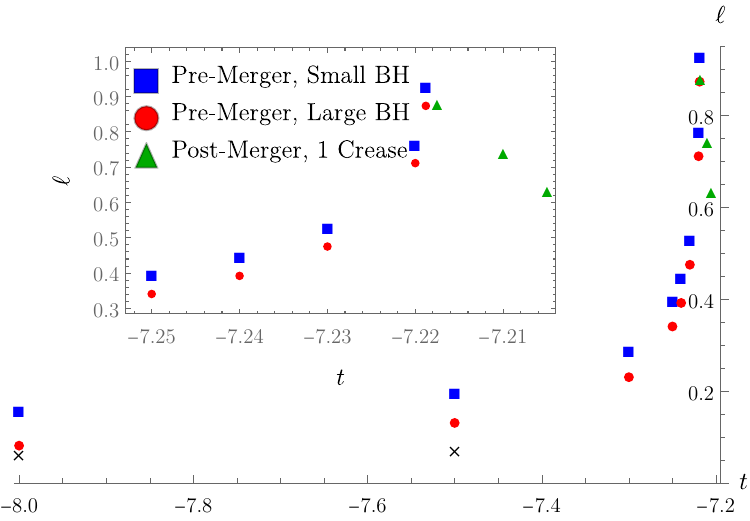}
    \caption{Lengths of the creases in constant $t$ slices, for an orthogonal merger with $M=1$, $a=1$. Blue squares correspond to the small black hole's crease before merger, red discs correspond to the large black hole's crease before merger, and green triangles correspond to one of the creases after merger. The crosses at $t=-8.0$ and $t=-7.5$ are the lengths predicted perturbatively up to order $(-M/t)^{9/2}$. The inset zooms in nearer to the time of merger.} 
    \label{fig:crease_lengths}
\end{figure}

We can quantify the size of each crease on a horizon cross-section using its proper length. Fig.~\ref{fig:crease_lengths} shows how the proper lengths of the creases evolve as a function of time. This figure is generated using our numerical results. In the far past, the length of the crease on the large black hole horizon tends to zero. Our perturbative calculation shows that the length of this crease is given by 
\be 
\ell_{\rm Crease} = -\frac{15 \pi a^2 (1-\mu_o^2)}{128 \, t} + \mathcal{O}(t^{-2}) \, .
\ee

If the crease submanifold extends across the white hole horizon, as we expect, then the length of the crease on the small black hole horizon should remain non-zero as $t\rightarrow -\infty$, which is consistent with the results shown in the figure. The lengths of both creases increase as $t \rightarrow t_\star-$ (with $t_\star$ the time of merger). The maximum length of the creases is roughly $M$, comparable to the size of the small black hole, which comes from the fact that they extend significantly in the $x$-direction (the direction of merger), as seen in Fig.~\ref{fig:shiny_black_extremal_zoom}. Near the instant of merger, the lengths of the two creases appear to be nearly equal. One might try to explain this using the local description of the instant of merger provided by \cite{Gadioux:2023pmw}, which predicts that the creases are related by a reflection isometry\footnote{This is $X^A \rightarrow -X^A$ in the notation of section \ref{sec:local_model} below.}. The equality suggests that most of the crease length arises from the region where this local model is accurate. 
Post merger, the horizon is connected via a ``bridge'', with a pair of creases along its edges, as in the third image of Fig \ref{fig:shiny_black_extremal_zoom}. These creases rapidly shrink and disappear via annihilation of their ($A_3$ caustic) endpoints, as described in \cite{Gadioux:2023pmw}. 

\subsection{Crease angle}

\begin{figure}
     \centering
    \includegraphics[width=\textwidth]{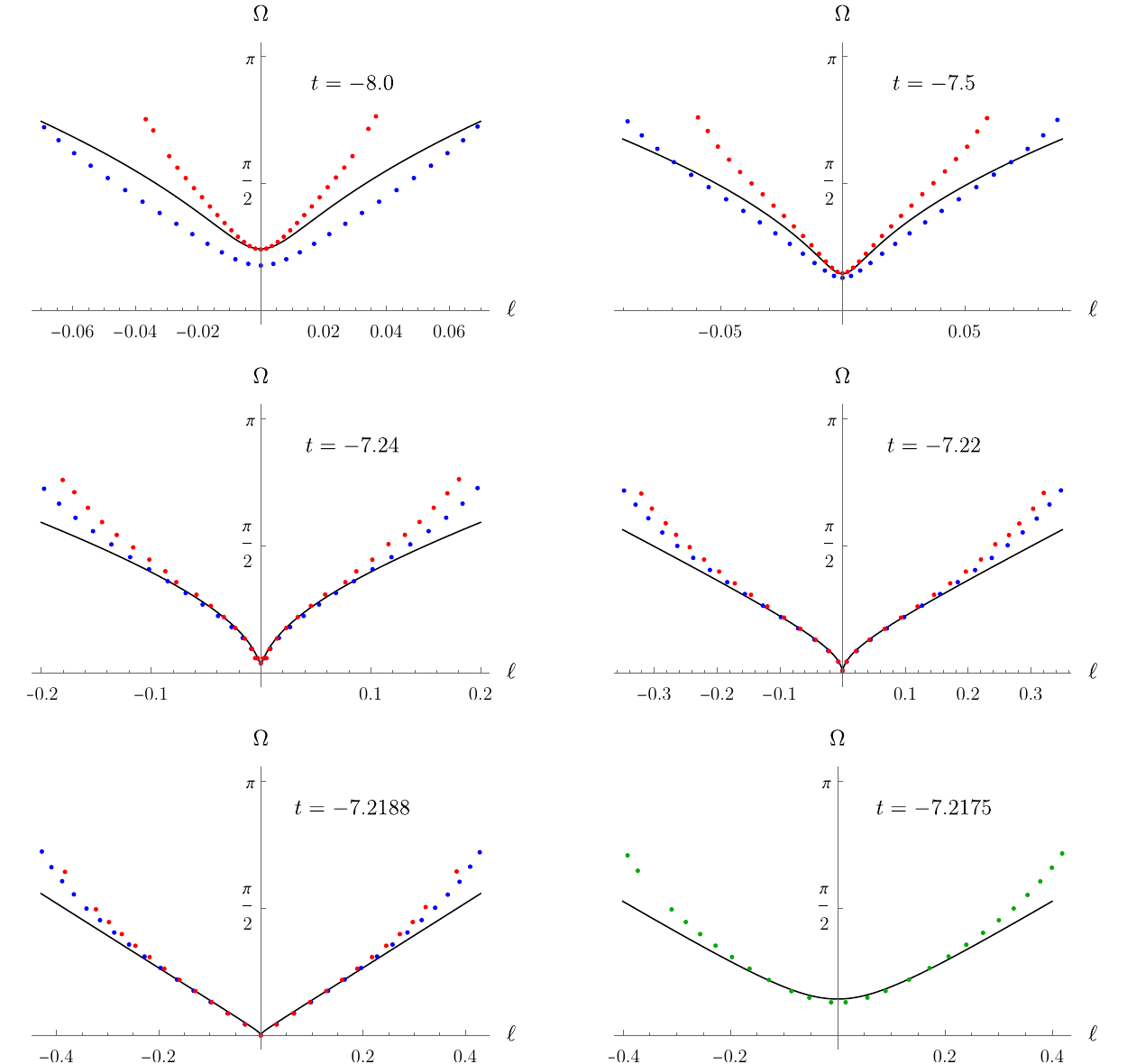}
    \caption{Opening angle $\Omega$ along the creases for various constant $t$ slices before the merger, with parameters $M=1$, $\mu_o=0$ and $a=1$. Blue points correspond to the small black hole's crease; red points correspond to the large black hole. The green dots in the final plot correspond to one crease after the merger. The black line is the best-fit prediction of the local model of \cite{Gadioux:2023pmw}, which is expected to hold in a spacetime region near the point of merger at $t=-7.2188$, $\ell=0$. The horizontal axis is the signed proper distance along the crease from its centre, i.e.~from the point of minimum opening angle. We have omitted the data near the caustic endpoints of each crease (where $\Omega \rightarrow \pi$) because the computation of the angle is less reliable there.}
    \label{fig:crease_angles}
\end{figure}

Another interesting geometrical quantity associated with the creases is the angle $\Omega$ between the two smooth sections of horizon that meet at a crease. In Fig.~\ref{fig:crease_angles} we use our numerical results to show how this angle varies along the creases at different times before the merger. In the far past, we expect the induced metric on the small black hole horizon to be smooth, as explained above (Section \ref{sec:gen_horizon}). Hence we expect $\Omega \rightarrow \pi$ everywhere along the small black hole crease as $t \rightarrow -\infty$. This is consistent with our plots although our numerical data does not extend to very large negative $t$. For the large black hole, our perturbative calculations show that for large negative $t$, $\Omega$ behaves as
\be 
\Omega = \pi - 16 \sqrt{\chi} \left(1-\mu_o^2\right)^{1/4} \sqrt{|\cos \varphi|} \, \left(-\frac{M}{4 t} \right)^{5/4} + \cdots \, ,
\ee
where $\varphi \in (\pi/2,3\pi/2)$ is a parameter along the crease. As $t \rightarrow t_\star$, Fig.~\ref{fig:crease_angles} shows that the two creases behave symmetrically. Near the instant of merger this is explained by the local description of \cite{Gadioux:2023pmw}, but our numerical results show that this agreement is fairly good even away from the instant of merger.

We shall show in Section \ref{sec:local_model} that the local description of \cite{Gadioux:2023pmw} predicts that, near the instant of merger, the crease angle is determined as a function of proper length $\ell$ along the crease by the following parametric equations:
\be
\label{local_Omega_ell}
 \Omega = \sqrt{8k  \lambda_\pm |\tau|} f_\pm(\xi) +\ldots \qquad \ell = \sqrt{\frac{2k|\tau|}{\lambda_\mp}} \int_0^\xi f_\pm(x) dx+\ldots
\ee 
where $\tau = t-t_\star$, $\pm = {\rm sign}(\tau)$, $k,\lambda_+,\lambda_-$ are positive constants ($k$ is dimensionless, $\lambda_\pm$ have dimensions of inverse length) and
\be
\label{fpm_def}
 f_\pm(x) = \left(\cosh^2 x+\frac{\lambda_\mp }{\lambda_\pm} \sinh^2 x\right)^{1/2} \, .
\ee 
The ellipses in equation \eqref{local_Omega_ell} denote terms that are subleading when $|\tau|$ and $\ell$ are small compared to the length scales defined by the spacetime Riemann tensor at the point of merger and the extrinsic curvature of the smooth parts of the horizon near the point of merger. (The latter length scales are closely related to $\lambda_\pm^{-1}$.) These results imply that, for small $\ell$ and $|\tau|$, $\Omega/\sqrt{|\tau|} = g_\pm(\ell/\sqrt{|\tau|})$ for functions $g_\pm$ defined implicitly by the above equations. Hence if we plot $\Omega/\sqrt{|\tau|}$ against $\ell/\sqrt{|\tau|}$ for different fixed values of $\tau$ (of the same sign) then the resulting curves should all lie on top of each other and agree with the functions $f_\pm$ for a particular choice of the parameters $k,\lambda_\pm$. These functions are minimized at $x=0$ and increase linearly at large $|x|$.

\begin{figure}
    \centering
    \includegraphics[width=0.7\textwidth]{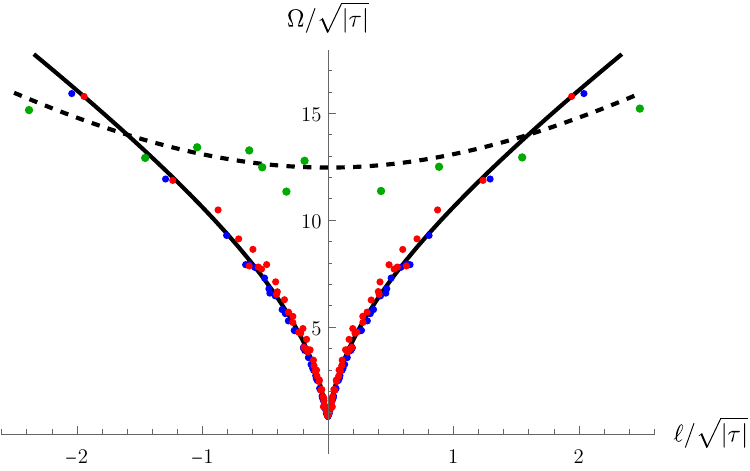}
    \caption{$\Omega/\sqrt{|\tau|}$ plotted against $\ell/\sqrt{|\tau|}$ for the extremal orthogonal merger ($a=1$, $\mu_o=0$) for the small black hole's crease (blue), the large black hole's crease (red), and one crease after merger (green). The black solid line is a best fit of~\eqref{local_Omega_ell} for pre-merger data with $\tau^2+\ell^2\leq0.1^2$. The dashed line is the local model prediction after merger with the same parameters.}
    \label{fig:local_model_angle}
\end{figure}

In Fig.~\ref{fig:local_model_angle} we use our numerical data to confirm this prediction. In this case, the best fit values of the parameters are $k=0.67$, $\lambda_-=0.14$, $\lambda_+=28.7$ for the extremal orthogonal merger, where we included only data with $\tau^2+\ell^2\leq 0.1^2$. The large value of $\lambda_+$ can probably be understood as follows. $\lambda_\pm$ are eigenvalues of a matrix closely related to the $ 2\times 2$ matrices describing the expansion and shear of the horizon generators of the two smooth sections of horizon that intersect at the crease. At an $A_3$ caustic, these matrices have a divergent eigenvalue (Appendix C of \cite{Gadioux:2023pmw}). Since the crease is close to the caustic points, this eigenvalue is large on the crease, which causes the large hierarchy between the values of $\lambda_\pm$. 

\subsection{Crease entropy}

Ref.~\cite{Gadioux:2023pmw} used properties of entanglement entropy across a surface with a crease to motivate the suggestion that a crease makes a contribution to black hole entropy of the form
\be
 S_{\rm crease} = \frac{1}{\sqrt{G\hbar}}\int_{\rm crease} F(\Omega) d\ell
\ee
where the integrand $F(\Omega)$ depends on the types of quantum fields that are present in the spacetime, but has the universal property of having a pole at $\Omega=0$. We can use \eqref{local_Omega_ell} to test this idea. Near the point of merger, $\Omega$ is small so $F$ is dominated by the pole at $\Omega=0$, hence
\be
\sqrt{G\hbar} S_{\rm crease} \approx R \int_{-\ell_0}^{\ell_0} \frac{d\ell}{\Omega}
\ee
where $R$ is the residue of $F$ at $\Omega=0$ and we have restricted to the section of crease with $|\ell|<\ell_0$ where $\ell_0$ is some fixed length much smaller than any length scale associated with the curvature of spacetime or the horizon. Changing variable of integration to $\xi$ using \eqref{local_Omega_ell} now gives simply
\be
\sqrt{G\hbar} S_{\rm crease} \approx R \int_{-\xi_\pm}^{\xi_\pm} \frac{d\xi}{2\sqrt{\lambda_+ \lambda_-}} = \frac{R\xi_\pm}{\sqrt{\lambda_+ \lambda_-}}
\ee
where $\xi_\pm$ is defined by $\ell=\ell_0$ at $\xi = \xi_\pm$, which implies $\exp(\xi_\pm) \propto 1/\sqrt{|\tau|}$ as $\tau \rightarrow 0$. Hence $S_{\rm crease}$ diverges logarithmically in $\tau$ as $\tau \rightarrow 0$. So the suggestion of \cite{Gadioux:2023pmw} gives a contribution to the entropy that diverges at the merger. Hence it appears that creases cannot contribute to black hole entropy in the manner suggested in \cite{Gadioux:2023pmw}.

\subsection{Structure of endpoints in the $(\alpha,\beta)$ plane}

\begin{figure}
    \centering
    \includegraphics[width=\textwidth]{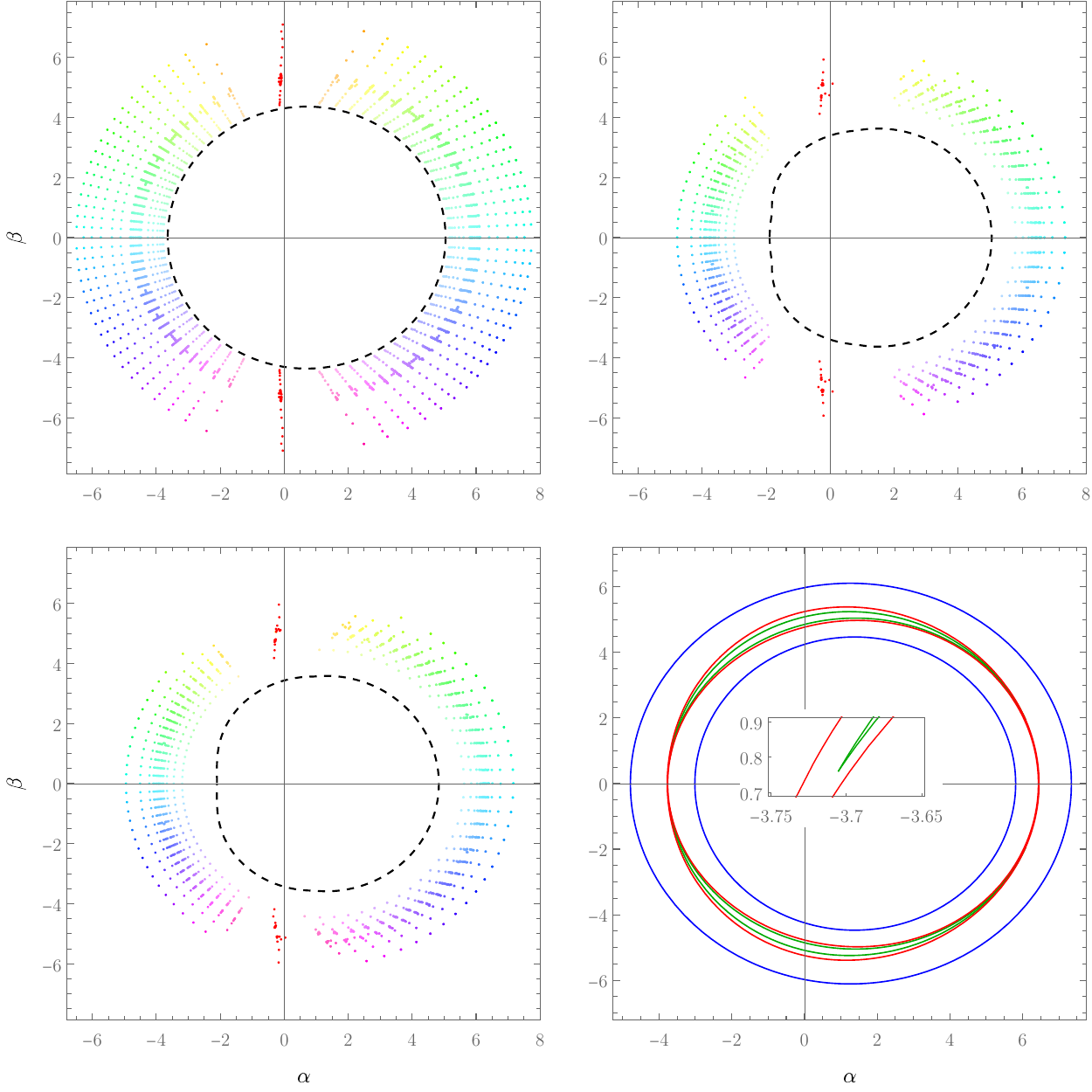}
    \caption{Structure of the endpoint set in the $(\alpha,\beta)$ plane. Each point of this plane corresponds to a horizon generator. The first three quadrants show numerical results for $a=1/2,\mu_o=0$ (top left), $a=1,\mu_o=0$ (top right) and $a=1,\mu_o=1/2$ (bottom left). Each coloured point corresponds to a geodesic whose endpoint we have determined numerically. Generators with caustic endpoints are in red; those with crease endpoints are colour-coded: points with the same colour and lying on the same approximate circle correspond to pairs of intersecting geodesics. The black dashed curve is the false creaseless disc; the true creaseless disc is very slightly inside this disc. The final quadrant gives best fits of the data on constant Boyer-Lindquist time slices for the extremal orthogonal merger: $t=-8$ (blue), $t=-7.2188$ (merger) (red), $t=-7.205$ (green). The inset shows a zoom which confirms that the green curves are disconnected.}
    \label{fig:parameter_space}
\end{figure}

As discussed in Section \ref{sec:gen_horizon}, $(\alpha,\beta)$ can be regarded as coordinates on a late-time cross section of $\cH^+$, with each generator labelled by a unique pair $(\alpha,\beta)$. Each point in the $(\alpha,\beta)$ plane corresponds to one of three types of generator: those with a crease endpoint outside the white hole region, those with a caustic endpoint outside the white hole region and those that emerge from the white hole region. The latter define the creaseless disc in the $(\alpha,\beta)$ plane. Marking each point by type on the $(\alpha,\beta)$ plane reveals an interesting structure shown in Fig.~\ref{fig:parameter_space} for several different mergers.\footnote{
Emparan {\it et al.}~present similar plots but, because of the mistake described above, their plots show the structure of the false crease set rather than the set of endpoints of horizon generators. Their plots have lines of caustic points at $\beta=0$ (exactly) with the pairs of generators meeting at false crease points related by $\beta \rightarrow -\beta$.} 

\begin{figure}
    \centering
    \includegraphics[width=\textwidth,trim={0 0.5cm 0 0cm},clip]{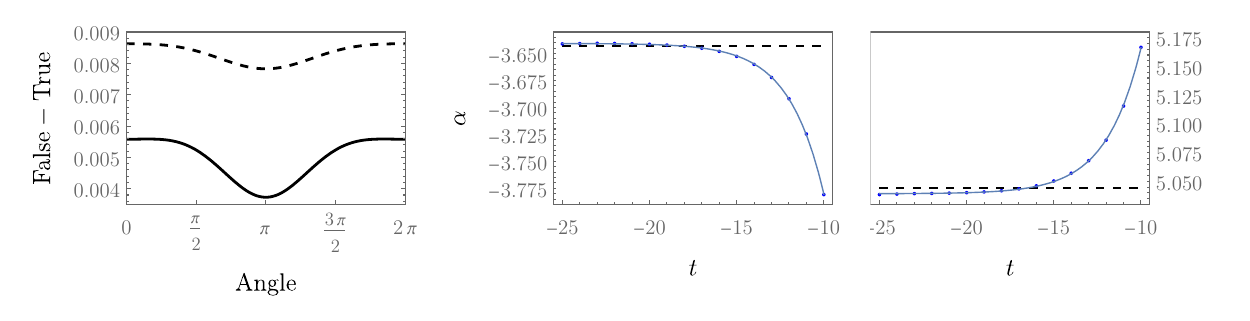}
    \caption{Evidence that the creaseless disc is only very slightly smaller than the false creaseless disc. These results are for an orthogonal merger with $a=1/2$. \textit{Left:} Difference in the radius $\sqrt{\alpha^2+\beta^2}$ of the best fits of constant $t$ cross-sections of the false and true crease submanifolds, as a function of angle about the origin of the $(\alpha,\beta)$ plane.
        The dashed curve corresponds to $t=-8$ and the solid curve to $t=-10$. The extrema lie at or near to the $\beta=0$ line. \textit{Right:} In an orthogonal merger, the map $\theta \rightarrow \pi - \theta$ is a symmetry of $\cH^+$, which can be used to argue that a pair of generators that intersect at the midpoint ($\theta = \pi/2$) of a constant $t$ slice of either pre-merger crease must both have $\beta=0$. For the small black hole's crease, we plot the values $\alpha$ of these pairs of generators as a function of $t$. The best fit is plotted in blue. The limit $t \rightarrow -\infty$ gives the pair of points with $\beta=0$ at the boundary of the creaseless disc. The dashed line gives the values of $\alpha$ of the points with $\beta=0$ on the boundary of the false creaseless disc. The difference in $\alpha$ values between the $\beta=0$ points at the boundaries of the two discs is just $0.002$ on the left and $0.005$ on the right sides.}
    \label{fig:disc_separation}
\end{figure}

In these plots, the dotted line is the boundary of the false creaseless disc, which is easy to determine for the reasons explained in Section \ref{sec:gen_horizon}. This disc has a similar shape to the black hole shadow, but, as discussed in Section \ref{sec:lensing}, the shadow is significantly larger. By definition, the (true) creaseless disc lies inside the false creaseless disc. However, in Fig.~\ref{fig:disc_separation} we show that the difference between the two discs is very small - too small to be visible at the scale of Fig.~\ref{fig:parameter_space}.

Outside the creaseless disc, all points correspond to generators which enter the horizon at the crease submanifold, except for two lines of points whose generators enter $\cH^+$ at caustic points. Generators entering at the crease submanifold do so in pairs. We find that the generators in each pair lie (mostly) on opposite sides of the $\beta$ axis with roughly equal $\beta$. (By comparison, generators that map to the same false crease point are related by a reflection in the $\alpha$ axis \cite{Emparan:2017vyp}.) 

A single generator enters $\cH^+$ at each caustic point \cite{Gadioux:2023pmw}. The lines of caustic points originate from the top and bottom of the creaseless disc. Our numerical results suggest that for positive spin they have slightly negative $\alpha$ where they meet the creaseless disc, and asymptote towards the $\beta$ axis further away. For $\sqrt{\alpha^2 + \beta^2} \gg M$ we can determine these caustic lines perturbatively, with the result:
\be 
\label{A3_pert}
\frac{\alpha}{M} = -\chi \left(\chi^{2} \mu_{o}^{2} + \frac{75\pi^{2}}{32} + \frac{1}{6} \chi^{2}-\frac{51}{4}\right) \frac{M^2}{\beta^2} + \mathcal{O}(\beta^{-3}) \qquad |\beta| \gg M
\ee
where we include only the leading order part --- higher order corrections are presented in Section~\ref{sec:perturbative} and extended to even higher order in Appendix~\ref{app:orthog_pert} for the case of an orthogonal merger.

The final quadrant of Fig.~\ref{fig:parameter_space} depicts the time evolution of the creases in the $(\alpha,\beta)$ plane. This plot shows the generators with endpoints occurring at the same value of $t$, for several different choices of $t$. For $t<t_\star$ this gives $2$ curves, which are nearly circular and concentric (they become slightly ``egg-shaped'' for larger spins). The inner curve corresponds to the small black hole's crease, while the outer curve maps to the large black hole's crease. (Their intersection with the caustic lines correspond to the caustic endpoints of these creases.) The generators that have already entered the horizon at this time are those with $(\alpha,\beta)$ parameters inside the inner curve or outside the outer one. As $t\rightarrow t_\star$, the inner curve grows and the outer curve shrinks. At $t=t_\star$ the curves touch each other at two points on the $\alpha$ axis. These points are the pair of intersecting generators that enter the horizon at the ``point of merger''. For $t>t_\star$, the curves separate into two disconnected crescent-shaped pieces. These correspond to the pair of creases that exist after the merger. As $t$ increases further, these curves shrink and disappear at the caustic lines (the caustic perestroika).

\subsection{Caustic tube}

\label{sec:results_caustic}

As described in Section \ref{sec:lensing}, the caustic tube is important in studies of gravitational lensing by a Kerr black hole. Previous work has determined the caustic tube numerically \cite{rauch, Bozza:2008mi} and perturbatively \cite{Sereno:2006ss,Sereno:2007gd}. Our calculations extend the perturbative results to much higher order than considered previously and provide the first perturbative computation of the Boyer-Lindquist time function along the caustic tube. 

Using the same parameterization in $(\varepsilon, \varphi)$ as we used for the crease submanifold, we find that the caustic tube is determined by the following equations (with $r_s$ given by \eqref{eps_def}):
\begin{align}
\label{ts_caustic}
\frac{t_s}{M} =& - \frac{1}{4 \varepsilon^2} - \frac{15 \pi \varepsilon}{4} - 8 \varepsilon^2 +  \frac{225 \pi^{2} \varepsilon^{2}}{256} -\frac{15 \varepsilon^{3} \pi  \left(375 \pi^{2}+1664\right)}{8192} 
\nonumber\\
&+ \frac{5  \pi  \left(9 (\mu_o^2 - 1)  \cos^{2} \varphi -6 \mu_{o}^{2}+13\right) \chi^{2} \varepsilon^{3}}{32}  + \frac{3 \varepsilon^{4} \left(16875 \pi^{4}-60000 \pi^{2}-1048576\right)}{65536}  
\nonumber
\\
&+  \frac{\left(675  \left(\mu_{o}^{2}-1\right) \pi^{2} \cos^{2} \varphi +16384+\left(-450 \mu_{o}^{2}-1125\right) \pi^{2}\right) \varepsilon^{4} \chi^{2}}{1024} + \mathcal{O}(\varepsilon^5)\, ,
\\
\label{mus_caustic}
\mu_s =& - \mu_o - \frac{15 \pi \chi^2 (1-\mu_o^2)^{3/2}  \varepsilon^4 \sin^3 \varphi}{16} + \mathcal{O}(\varepsilon^6)
\\
\label{phis_caustic}
\phi_s =& \pi \, {\rm sign} (\cos \varphi) -4 \chi  \,\varepsilon^{2}-\frac{5}{4} \pi  \chi  \,\varepsilon^{3} + \frac{\chi  \varepsilon^{4} \left(225 \pi^{2}-2048\right)}{128} + \frac{15 \pi \chi^{2} \varepsilon^{4}   \sqrt{1-\mu_{o}^{2}}\, \cos^{3} \varphi }{16} 
\nonumber
\\
&-\frac{9 \pi \chi \left(2625 \pi^{2}-11392\right) \varepsilon^{5}}{8192} - \frac{3 \chi^{3} \pi  \left(15 \left(1-\mu_{o}^2\right) \cos^{2} \varphi + 10 \mu_{o}^{2}- 3\right) \varepsilon^{5}}{32} + \mathcal{O}(\varepsilon^6).
\end{align}
Here we have written the result to one order beyond~\cite{Sereno:2006ss,Sereno:2007gd}. We have extended the general computations an additional three orders, with results presented in Section~\ref{sec:perturbative}, and an additional eight orders for the special case of an orthogonal merger, as we  discuss in Appendix \ref{app:orthog_pert}.

As before, $\varphi \in [0,2\pi)$ and $0< \varepsilon \ll 1$. Constant $\varepsilon$ cross-sections of the tube are astroids, whose vertices sweep out $4$ spacelike curves in spacetime. These are curves of $A_3$ caustic points, given (at this order) by $\varphi=0,\pi/2,\pi,3\pi/2$. The 2d surfaces of the caustic tube (connecting the $A_3$ lines) are $A_2$ caustic points.\footnote{
The $A_2$ surfaces are null \cite{Gadioux:2023pmw}. Since they are not codimension$-1$ they do not have null geodesic generators.} The lines $\varphi=0,\pi$
correspond to the lines of $A_3$ points that lie at the edges of the false crease set: these points have $\beta=0$ exactly. The lines $\varphi=\pi/2,3\pi/2$ are the $A_3$ lines belonging to $\cH^+$. At higher-order in the $\varepsilon$ expansion the values of $\varphi$ at which these caustics occur are shifted.  In the $(\alpha,\beta)$ plane the shifted values are given to leading order in \eqref{A3_pert}.

From the above equations, we find that all of the $A_3$ caustic lines have infinite proper length as $\varepsilon \rightarrow 0$. 






\section{Predictions of the local model}

\label{sec:local_model}

Ref.~\cite{Gadioux:2023pmw} introduced the notion of a {\it crease perestroika}, which provides an exact {\it local} description of the event horizon near the ``point of merger'' of a generic black hole merger. In this section, we shall use this description to determine how the crease angle $\Omega$ varies along a crease on a horizon cross-section, near to the point of merger. 

\subsection{Summary of the local model}

In a generic merger, the endpoint of a generic generator belongs to the crease submanifold \cite{Gadioux:2023pmw}. Let $\tau$ be a time function. The point of merger w.r.t.~the time foliation defined by $\tau$ is a ``pinch point'': a point at which a hypersurface of constant $\tau$ is tangent to the crease submanifold. The crease submanifold corresponds to a self-intersection of the event horizon. One can describe this exactly in a neighbourhood of a pinch point by introducing Riemannian normal coordinates, and studying the equations describing an arbitrary pair of smooth intersecting null hypersurfaces in these coordinates. The local Lorentz frame at the pinch point is fixed up to rotations by demanding that its time axis is normal to the hypersurface of constant time through this point. A rotation can then be used to simplify the equations for the two null surfaces to the following form near the pinch point\footnote{
Our $(T,X^A,Z,k)$ are written $(t,x^A,z,a)$ in \cite{Gadioux:2023pmw}.}
\cite{Gadioux:2023pmw}:
\be
\label{null_hyps}
 k \tau = Z + \left (b_{AB} - \frac{1}{2} K_{AB} \right) X^A X^B + \ldots \qquad  k \tau = -Z + \left (\hat{b}_{AB} - \frac{1}{2} K_{AB} \right) X^A X^B + \ldots \,.
\ee
Here $\tau$ is the time function, shifted so that the pinch point has $\tau=0$. To apply this to our situation, we shall take $\tau = t-t_\star$ where $t$ is the Boyer-Lindquist time coordinate. The Riemannian normal coordinates are $(T,X^i)$ so that the pinch point is at $T=X^i=0$. The spatial coordinates are $X^i = (X^A,Z)$, $A=1,2$. The positive constant $k$ is defined by the relation between $\tau$ and the normal coordinates:
\be
 T = k\tau + c\tau^2 + d_i \tau X^i + \frac{1}{2}K_{ij} X^i X^j + \mathcal{O}(\tau,X^i)^3
\ee
where $K_{ij}$ is the extrinsic curvature tensor of the surface $\tau=0$ at the pinch point. $c$ and $d_i$ are further constants that will not be important. In \eqref{null_hyps}, $K_{AB}$ are the $X^A$ components of $K_{ij}$, $b_{AB}$ and $\hat{b}_{AB}$ are symmetric matrices describing the expansion and shear of the pair of null hypersurfaces. The area theorem implies that, generically, both of these surfaces have positive expansion so $b_{AA}, \hat{b}_{AA} >0$. The ellipses in \eqref{null_hyps} are $\mathcal{O}(\tau^2)+\mathcal{O}(\tau |X^A|) + \mathcal{O}(|X^A|^3)$. The metric is $g = -dT^2 + dX^A dX^A + dZ^2 + \mathcal{O}(T,X^A,Z)^2$. 

In this description, the event horizon corresponds (locally) to the parts of these null hypersurfaces that lie to the future of their intersection. The intersection provides a local description of the crease submanifold given by the equations
\be
\label{crease_local}
 2k \tau = \mu_{AB} X^A X^B + \mathcal{O}(|X^A|^3) \qquad  2Z = (\hat{b}_{AB} - b_{AB})X^A X^B  + \mathcal{O}(|X^A|^3) 
\ee
where
\be
\label{mu_def}
\mu_{AB} = b_{AB} + \hat{b}_{AB}  - K_{AB}.
\ee
Generically, $\mu_{AB}$ is non-degenerate. A black hole merger corresponds to this matrix having $+-$ signature.\footnote{Crease perestroikas for which it is positive or negative definite describe event horizon nucleation, or collapse of a hole in the horizon \cite{Gadioux:2023pmw}.} Within a surface of constant $\tau$, to leading order, \eqref{crease_local} describes a hyperbola. For $\tau<0$ the two branches of the hyperbola are the creases on the two black holes pre-merger. For $\tau>0$ the two branches of the hyperbola  are the creases lying along the edges of the ``bridge'' that forms in the merger, see Fig.~\ref{fig:shiny_black_extremal_zoom}. 

\subsection{The crease angle in the local model}

Within a surface of constant $\tau$, equations \eqref{null_hyps} describe a pair of smooth intersecting surfaces. At the crease, these have normals (using \eqref{crease_local} to eliminate $\tau$ from the correction terms): 
\bea
 n_1 &=& dZ + [(2b_{AB} - K_{AB} ) X^B +\mathcal{O}(|X^B|^2)] dX^A \nonumber \\ n_2 &=& -dZ + [(2\hat{b}_{AB} - K_{AB} ) X^B +\mathcal{O}(|X^B|^2)] dX^A.
\eea
The crease angle $\Omega$ is defined by
\be
\label{Omega_def}
\cos (\pi -\Omega) = \frac{n_1 \cdot n_2}{||n_1|| ||n_2||} 
\ee
where the inner product and norms are defined w.r.t.~the induced metric $h_{ij}$ on the surface of constant $\tau$. On such a surface we have $dT = \mathcal{O}(\tau,X^i) dX^j$ which implies $h_{ij} = \delta_{ij} + \mathcal{O}(\tau,X^i)^2= \delta_{ij} + \mathcal{O}(|X^A|^2)$ where $(\tau,Z)$ are eliminated using \eqref{crease_local}. This implies
\be
 n_1 \cdot n_2 = -h^{ZZ} \left[ 1- (2b_{AB} - K_{AB} ) X^B (2\hat{b}_{AC} - K_{AC} ) X^C + \mathcal{O}(|X^A|)^3 \right]  \, ,
\ee
\be
 ||n_1||^2 = h^{ZZ} \left[ 1 +  (2b_{AB} - K_{AB} ) X^B (2b_{AC} - K_{AC} ) X^C + \mathcal{O}(|X^A|)^3 \right] \, ,
\ee
and similarly for $||n_2||^2$. Now $h^{ZZ} = 1+\mathcal{O}(|X^A|^2)$, where the quadratic terms depend on the Riemann curvature tensor. So we do not know the quadratic terms in the above expressions. However, when we evaluate \eqref{Omega_def}, $h^{ZZ}$ cancels out and after simplification we obtain
\be
 \Omega^2 = 4\mu_{AB}X^B \mu_{AC} X^C + \mathcal{O}(|X^A|^3) 
\ee
where $\mu_{AB}$ is defined in \eqref{mu_def}. As explained above, $\mu_{AB}$ has signature $+-$ so by a rotation of $X^A$ we can arrange that
\be
 \mu_{AB} = {\rm diag}(-\lambda_-,\lambda_+)
\ee
where $\lambda_\pm>0$. We now have
\be
 \Omega^2 = 4 \left[ \lambda_-^2 (X^1)^2 + \lambda_+^2 (X^2)^2 \right]+ \mathcal{O}(|X^A|^3)
\ee
and the proper length $\ell$ along the crease is defined by
\be
\label{ellX}
 d\ell^2 = h_{ij} dX^i dX^j = (\delta_{AB} + \mathcal{O}(|X^C|^2) dX^A dX^B
\ee
where we used the second equation of \eqref{crease_local} to eliminate $dZ$. The first equation of \eqref{crease_local} is 
\be
\label{crease_local1}
2k\tau = -\lambda_- (X^1)^2 + \lambda_+ (X^2)^2 + \mathcal{O}(|X^A|^3) \, .
\ee
These equations hold for $X^A$ small compared to any length scales defined by the local spacetime curvature (which appears in the higher order terms in Riemannian normal coordinates) and the extrinsic curvature of the null hypersurfaces (e.g. $\lambda_-^{-1},\lambda_+^{-1}$). 

For $\tau<0$ we can solve \eqref{crease_local1} to leading order with
\be
\label{theta_def}
  X^1 =  \sqrt{\frac{2k|\tau|}{\lambda_-}} \cosh \xi \qquad X^2 = \sqrt{\frac{2k|\tau|}{\lambda_+}} \sinh \xi \qquad (\tau<0)
\ee
which is valid for small $X^A$, i.e., small $\sqrt{|\tau|} e^\xi$. For $\tau>0$ we swap the 
the $\cosh$ and $\sinh$ in \eqref{theta_def}.
We then have, with $\pm = {\rm sign}(\tau)$,
\be
\label{Omega_theta}
 \Omega = \sqrt{8 k |\tau|\lambda_\pm } f_\pm(\xi) + \mathcal{O}(|\tau| e^{2\theta})
\ee
where $f_\pm$ is defined in \eqref{fpm_def}. Since $f_\pm(\xi)$ is minized at $\xi=0$, this shows that the minimum value of $\Omega$ along the crease is $\sqrt{8k|\tau| \lambda_\pm}$, i.e.~$\mathcal{O}(\sqrt{|\lambda|})$. 
From \eqref{ellX}, choosing $\ell=0$ at $\xi=0$ we have
\be
\label{ell_theta}
 \ell = \sqrt{\frac{2k|\tau|}{\lambda_\mp}} \int_0^\xi f_\pm (x) dx + \mathcal{O}(|\tau| e^{2\xi}) \, .
\ee
The implications of equations \eqref{Omega_theta} and \eqref{ell_theta} are discussed in Section \ref{sec:results} following \eqref{local_Omega_ell}.

\section{Numerical method}

\label{sec:numerics}

We solve for the horizon generators in the same way as explained in~\cite{Emparan:2017vyp}. In particular, we impose late-time conditions on the null geodesic equations such that the horizon asymptotes to a null plane at finite retarded time. These equations (written in Boyer-Lindquist coordinates) can be solved analytically~\cite{Gralla:2019ceu}, but the solutions are very cumbersome and are difficult to work with. Furthermore, due to the presence of multiple elliptic integrals and Jacobi elliptic functions, they are rather slow to evaluate. Thus, it is less computationally expensive to integrate the equations numerically. An issue with numerical integration is that one cannot evolve from infinity, as one would like. Instead, as is done in~\cite{Emparan:2017vyp}, we expand the geodesic equations in a large affine parameter $\lambda$ and integrate them order by order to obtain our late-time conditions. Often, we will want to determine the spatial coordinates of a generator in a particular time-slice. We find that there is a trade-off between setting a large initial $\lambda_i$ to minimise errors in the late-time conditions and the accuracy at which the value of the affine parameter in the corresponding time-slice can be measured. Taking $\lambda_i=10000$ yields approximately six significant figures of accuracy, so the numerical errors are several orders of magnitude smaller than the dimensions of the crease (in a time-slice).

We evolve each null geodesic backwards in time to arbitrarily early times (i.e.~arbitrarily large negative $t$). To obtain the horizon, the integration must be stopped at the first caustic or crease point encountered along each geodesic. Identifying crease points is particularly difficult, as, in theory, intersections can occur between any two geodesics on the entire $2$-dimensional parameter space $(\alpha,\beta)$. In practice, we find that in a given time-slice, the geodesics that constitute the creases lie on two near-circular curves in this space; one curve for each of the small and large black holes' creases (Fig.~\ref{fig:parameter_space}, final panel). The small black hole's curve lies inside the large black hole's one, but outside the creaseless disc. 

Our algorithm for finding crease points exploits the structure in parameter space described above, but is also partly inspired from the methodology of~\cite{Bohn:2016afc}. Consider a Cauchy surface $\Sigma$ corresponding to a surface of constant Boyer-Lindquist time $t$. All generators which encounter a crease or caustic point on $\Sigma$ lie in a relatively narrow annulus $\mc{A}$ of parameter space, its centre offset along the $\alpha$-axis due to the black hole spin. Henceforth we will use polar coordinates $(\delta,\varsigma)$ in this space, with the centre of $\mc{A}$ as the origin. Now we divide $\mc{A}$ into two parts, say $\mc{A}_L$ and $\mc{A}_R$. The boundary should lie roughly at the angles $\varsigma$ at which the endpoints of the crease occur (these will be caustic points). The centrepiece of the algorithm works as follows: we take one of the two parts, say $\mc{A}_L$,\footnote{In practice, we take the part that maps to a surface of lower curvature in our coordinate system.} and choose an appropriately-dense set of geodesics with parameters $(\delta,\varsigma)\in\mc{A}_L$. We determine their spatial positions within $\Sigma$. Here we find it useful to employ the $(x',y',z')$ coordinate system defined in~\eqref{Emp_Cart_Primed}, since then we no longer need to worry about the periodicity of $\phi$. To remedy the finite density of points, we connect the points by triangles to form a surface. This can be done by projecting the points onto a plane (e.g.~the $y'z'$-plane), performing a Delaunay triangulation~\cite{musin1997properties} and reinstating the third dimension. Now, we fix $\varsigma$ to lie in $\mc{A}_R$. Varying $\delta$, this forms a curve in $\Sigma$. One can then check which triangle this curve intersects. If there are no intersections, then there are no creases at the corresponding values of $\varsigma$ and $t$ (except perhaps if the width of $\mc{A}$ or the boundary separating $\mc{A}_L$ and $\mc{A}_R$ were poorly chosen). If there is an intersection, we take the minimum and maximum values of the coordinates $(\delta,\varsigma)$ of the vertices of the intersected triangle, and repeat the above procedure with a denser sample of geodesics within that subset of parameter space. The denser sampling leads to smaller triangles on $\Sigma$, implying better reproduction of the curvature of the surface, and hence more accurate results. We iterate until the desired accuracy is reached.

We have implemented the above algorithm in Mathematica, which has enabled us to determine crease points arbitrarily close to the point of merger. The algorithm is not perfectly robust; indeed at times the programme does not converge to a crease point when one would expect that it does. Nevertheless, it is sufficient for our demands, and is not too computationally costly. We estimate that the error in the spacetime coordinates of the crease points is on the order of $10^{-4}M$; this is between two and four orders of magnitude smaller than the proper length of the crease in constant $t$ slices.

We have determined crease points for three different mergers: $\{a=1/2,\mu_o=0\}$, $\{a=1,\mu_o=0\}$ and $\{a=1,\mu_o=1/2\}$. In each case we include data before and after the time of merger. We have made this data available alongside the paper. The data is structured as a nested list, with three elements in the first level, corresponding to the three cases above, in that order. Each of these elements contains a list of pairs of crease points, with a pair grouping the two points on the $(\alpha,\beta)$ plane that map to the same spacetime point on the horizon. Each crease point is given as $\{\{\alpha,\beta,\lambda_i-\lambda_c\},\{t,r,\theta,\phi\}\}$ where $\lambda_c$ is the affine parameter at the crease point, assuming a setup as described at the start of this section. ($\lambda$ increases along the future-directed evolution of the geodesic.)

We find that the non-smooth structure is qualitatively similar for all three data sets. The main quantitative differences lie in the size of the crease submanifold (it is largest for the extremal orthogonal merger), and the inclination of the horizon.

It is useful to also compute the caustic endpoints of the crease lines. In theory, it is reasonably straightforward to determine caustic points, as one can check for a local condition, in particular, the vanishing of $\partial x^\mu/\partial \delta \wedge \partial x^\nu/\partial \varsigma \wedge dx^\rho/d\lambda$, where $\lambda$ is the affine parameter along the geodesic. However, we are looking for a specific set of caustic points: those that form the two edges of the caustic tube that are off the $\theta=\pi+\theta_o$ plane. Unfortunately, the map from parameter space to spacetime is very dense near caustic points, and a large variation in $\varsigma$ can lead to only very small displacements in spacetime. Thus, while the spacetime coordinates of the caustic points can be determined accurately with the above method, the coordinates in parameter space are somewhat less reliable. To remedy this, we write for each crease point one of its geodesic's $(\delta,\varsigma)$ coordinates as a function of the second geodesic's coordinates. We are able to extract the fixed point of this function with much better precision than with the former method. Nevertheless, the uncertainties in the data for the caustic points remain higher than for the crease points.

\section{Perturbative method and results}
\label{sec:perturbative}

\subsection{Introduction}

The equations governing null geodesics in the Kerr spacetime are summarized in Appendix \ref{app:kerr_geos}. In this section we shall solve them perturbatively in the regime of large impact parameter, i.e. $\sqrt{\alpha^2 + \beta^2} \gg M$. We shall then use our results to determine the crease submanifold and the caustic tube perturbatively. To do this, we shall have to extend the perturbative calculations to higher order than has been considered previously \cite{Bray:1985ew,Sereno:2006ss,Sereno:2007gd}. 

We wish to track a null geodesic as it propagates from a ``source'' point $\{t_s,r_s,\theta_s, \phi_s\}$ to an ``observer'' $\{t_o, r_o, \theta_o, \phi_o\}$. The trajectory of the null geodesic is given by the following integrals
\begin{align}
\label{geo_rad_polar}
\int_{\rm P} \frac{dr}{\pm \sqrt{R}} &= \int_{\rm P} \frac{d\theta}{\pm \sqrt{\Theta}} \, ,
\\
\label{geo_azimuth}
\phi_o-\phi_s &= \int_{\rm P} \frac{a\left(2Mr-a\lambda\right) dr}{\pm \Delta \sqrt{R}}  + \int_{\rm P} \frac{\lambda d\theta}{\pm \sin^2\theta \sqrt{\Theta}} \, ,
\\
\label{geo_temporal}
t_o - t_s &=  \int_{\rm P} \frac{\left[r^2\left(r^2+a^2\right) + 2 a M r \left(a-\lambda\right)\right] dr}{\pm \Delta \sqrt{R}} + \int_{\rm P} \frac{a^2 \cos^2\theta}{\pm \sqrt{\Theta}} d\theta \, ,
\end{align}
using notation that is summarized in Appendix \ref{app:kerr_geos}. The subscript ``${\rm P}$'' indicates that the integrals are to be evaluated along the path of the photon.  The signs of $\sqrt{R}$ and $\sqrt{\Theta}$ are chosen to match the signs of $dr$ and $d\theta$ along the path. Accordingly, the signs of the integrals will change at turning points of the $r$ and $\theta$ motion. 

We are interested in the case where the ``observer'' is at infinity, $r_o = \infty$. To study the temporal integral in this case it is useful to work with a retarded time coordinate 
\be 
\label{udef}
u = t - r - 2 M \log \left(r/M\right) \, .
\ee
Without loss of generality, we will restrict to an observer with $\phi_o = 0$ and $u_o = 0$. We will evaluate the integrals in the limit of large impact parameter. Instead of working directly with $(\alpha,\beta)$ it is more convenient to use $(\lambda,\eta)$ and define
\be 
b = \sqrt{\eta + \lambda^2} \, ,
\ee
so that a geodesic is uniquely specified by the parameters $(b, \lambda)$.\footnote{It is of course possible to solve the geodesic equations perturbatively in terms of $(\alpha, \beta)$. However in practice we find that the additional square roots introduced result in a slower evaluation than when using $(b, \lambda)$. This becomes problematic at high orders.} 

To consistently track the terms appearing in the perturbative expansions we introduce a parameter $\epsilon$, taking $b \sim \epsilon^{-1}$, $\lambda \sim \epsilon^{-1}$ and $r_s \sim \epsilon^{-2}$. The evaluation of the integrals in the limit of large impact parameter is then achieved by solving them in the limit $\epsilon \to 0$ and at the end of the computation we can simply take $\epsilon = 1$. Note the difference between $\epsilon$ and the parameter $\varepsilon$ defined in \eqref{eps_def}, following \cite{Sereno:2006ss,Sereno:2007gd}. Here $\epsilon$ is simply a counting parameter that allows us to easily collect all terms of a given order in the large impact parameter expansion. But defining $\varepsilon$ (which is ${\cal O}(\epsilon)$) lets us simplify the form of our results. 

The evaluation of the integrals is straightforward but the intermediate and final results are complicated. We present the details of that analysis in Appendix~\ref{app:perturbative_eval}, commenting here on the main features before proceeding to the study of caustics and creases. To evaluate the integrals, we need to know how many turning points in $r$ and $\theta$ occur along a geodesic. These large impact parameter geodesics describe scattering trajectories, which have a single turning point in $r$. The results of Appendix~\ref{app:false_crease} imply that, when evolved backwards from $\scri^+$, geodesics have 
one turning point in $\theta$ before reaching the false crease set, so they have at most one turning point in $\theta$ before reaching the crease set. We find that exactly one turning point is required. 

In the case of generic inclination angle $\theta_o$ the intermediate expressions are sufficiently complicated that our evaluation of the integrals and subsequent analysis of the crease set and caustic tube encounters slowdowns. In practice, our evaluation of the geodesic integrals in this case has terms up to and including corrections at $\mathcal{O}(\epsilon^9)$, which is five orders higher than considered in previous work~\cite{Sereno:2006ss,Sereno:2007gd}. It is possible to evaluate the geodesic integrals to higher order, but we find that aspects of the analysis of the caustic tube and crease set encounters time and memory limitations in our implementation. In the case of an orthogonal merger, i.e.~$\theta_o = \pi/2$, our perturbative results can be extended to much higher order. We have extended the evaluation of the geodesic integrals to include $\epsilon^{18}$ corrections in this case, and we see no fundamental obstacle to working to even higher order. We do not present these higher-order results due to their complexity, though in Appendix~\ref{app:orthog_pert} we include analysis of the caustic tube and crease set to higher order. An ancillary Mathematica notebook contains various results too complicated to be presented here.

In the next section we shall discuss how these perturbative results are used to determine the caustic tube. The following section will then use them to determine the crease submanifold.

\subsection{Caustic tube}

Here we construct the caustic tube perturbatively. We follow the approach of Sereno and de Luca~\cite{Sereno:2006ss,Sereno:2007gd} (with slight differences to be explained below), extending their results to higher perturbative order. This is independent from how the crease submanifold will be determined in the following section, and therefore will provide a non-trivial consistency check of our computations. 

To parameterize the results it is convenient to introduce polar coordinates in the $(\alpha, \beta)$ plane. We define new polar impact parameters $(\rho, \varphi)$ as
\be\label{eqn:polar_impact} 
\alpha = M \rho \cos \varphi  \, , \quad \beta = M \rho \sin \varphi \, ,
\ee
where the factor of $M$ is introduced so that $\rho$ is dimensionless.  Note that polar-like impact parameters were also introduced by Sereno and de Luca~\cite{Sereno:2006ss,Sereno:2007gd}. However, while their parameterization produces the correct results at lowest order in the large impact parameter expansion, the parameter $b_2$ introduced in Eq.~(37) of \cite{Sereno:2006ss} becomes complex whenever $\beta^2 < a^2 \mu_o^2$. No such problem arises with the polar impact parameters \eqref{eqn:polar_impact}. The parameters $(\rho, \varphi)$ can be regarded as coordinates on the observer's sky, just like $(\alpha,\beta)$. Geodesics are labelled by $(\rho, \varphi)$ and we shall take $r_s$ as the parameter along each geodesic. 

Along each geodesic we can regard $(\mu_s,\phi_s)$ as functions of $(r_s,\rho, \varphi)$. In lensing, the magnification of a source at $(r_s,\mu_s,\phi_s)$ is controlled by the Jacobian that relates differential area elements in the $(\mu_s, \phi_s)$ parameter space to the $(\rho, \varphi)$ space. The caustics are those points for which the magnification is infinite. For such points, the Jacobian will vanish:
\be \label{jacobian}
J \equiv \left[\frac{\partial \mu_s \partial \phi_s}{\partial \rho \partial \varphi} \right]_{r_s} = 0 \, .
\ee
We evaluate $J$ using our perturbative solution for the geodesic equations. The result is complicated so we shall not write it explicitly (the first three non-trivial terms can be found in \cite{Sereno:2007gd} --- though recall the difference in the polar impact parameters above). For given $r_s$, Eq.~\eqref{jacobian} determines a curve in the $(\alpha,\beta)$ plane. This curve is a constant $r_s$ cross-section through the caustic tube.

At this stage it is convenient to switch from using $\epsilon$ to using $\varepsilon$ introduced in Eq.~\eqref{eps_def} since this quantity is defined explicitly in terms of $r_s$. We can then solve the condition $J = 0$ to obtain $\rho$ as a function of $\varepsilon$ and $\varphi$ over the caustic tube. Explicitly, we write
\be 
\rho = \frac{{}^{(-1)} \rho_{\rm Cau}}{\varepsilon} + {}^{(0)} \rho_{\rm Cau} + {}^{(1)} \rho_{\rm Cau} \varepsilon + \cdots \, .
\ee
In the expansion of $J$, the first non-trivial contribution arises at $\mathcal{O}(\epsilon^2)$, which allows for the determination of ${}^{(-1)} \rho_{\rm Cau}$. The higher-order terms follow in the obvious way, with the $\mathcal{O}(\epsilon^N)$ term in the expansion of $J$ allowing for the determination of ${}^{(N-3)} \rho_{\rm Cau}$.  Since we have evaluated the geodesic integrals up to and including corrections at $\mathcal{O}(\epsilon^9)$ we can reliably determine the coefficients up to ${}^{(7)} \rho_{\rm Cau}$. Our analysis determines these coefficients to be:
\begingroup
\allowdisplaybreaks
\begin{align}
{}^{(-1)} \rho_{\rm Cau} =& \, 1 \, ,
\\
{}^{(0)} \rho_{\rm Cau} =& \, \frac{15 \pi}{32} +  \chi  \sqrt{1-\mu_{o}^{2}}\, \cos \varphi \, ,
\\
{}^{(1)} \rho_{\rm Cau} =& \, 4 -\frac{675 \pi^{2}}{2048} +\frac{15 \pi \chi \sqrt{1-\mu_{o}^{2}}\, \cos \varphi }{32} - \frac{\chi^{2}  \left(1-\mu_o^2 \right) 
\sin^2 \varphi }{2} \, ,
\\
{}^{(2)} \rho_{\rm Cau} =& \, \frac{135 \pi  \left(25 \pi^{2}-272\right)}{8192} -\frac{\chi \left(225 \pi^{2}-2048\right) \sqrt{1-\mu_{o}^{2}} \cos  \varphi }{256} 
    \nonumber 
    \\
    &-\frac{15 \pi \chi^2 \left[ 3\left(1-\mu_{o}^{2}\right) \cos^{2}\varphi +6 \mu_{o}^{2}+1 \right]}{256} \, ,
\\
{}^{(3)} \rho_{\rm Cau} =& \, 16 + \frac{78525 \pi^{2}}{16384} -\frac{5315625 \pi^{4}}{8388608} + \frac{315 \pi  \chi  \left(375 \pi^{2}-3968\right) \sqrt{1-\mu_{o}^{2}}\, \cos  \varphi}{65536} 
    \nonumber 
    \\
    &- \frac{3 \chi^{2}  \left(1725 \pi^{2}-16384\right) \left(1-\mu_{o}^2\right)  \cos^{2} \varphi }{8192}+\frac{\left(2700 \pi^{2} \mu_{o}^{2}+5175 \pi^{2}-16384 \mu_{o}^{2}-49152\right) \chi^{2}}{8192}
    \nonumber 
    \\
    &+\frac{15  \chi^{3} \left[2 \mu_{o}^{2}+5  -\left(1-\mu_{o}^{2}\right)\cos^{2}\varphi\right] \pi  \sqrt{1-\mu_{o}^{2}}\, \cos \varphi }{256} -\frac{\chi^{4} (1-\mu_o^2)^2 \sin^4 \varphi }{8} \, ,
\\
{}^{(4)} \rho_{\rm Cau} =& \,  -\frac{50025 \pi}{2048} -\frac{541125 \pi^{3}}{65536} + \frac{2278125 \pi^{5}}{2097152} -\frac{\chi \left(253125 \pi^{4}-2383200 \pi^{2}-2097152\right) \sqrt{1-\mu_{o}^{2}}\cos \varphi}{65536 } 
	\nonumber
	\\
	&+\frac{3375 \pi  \chi^2 (1-\mu_o^2) \left(50\pi^{2}-451\right) \cos^2 \varphi}{51200}-\frac{15 \pi  \chi^2 \left(225 \pi^{2} \mu_{o}^{2}+4500 \pi^{2}-8192 \mu_{o}^{2}-37968\right)  }{32768} 
	\nonumber
	\\
	&- \frac{15 \pi \chi^4 \left( \left(25 \mu_{o}^{4}-50 \mu_{o}^{2}+25\right) \cos^{4}\varphi+\left(-32 \mu_{o}^{4}+22 \mu_{o}^{2}+10\right) \cos^{2}\varphi-16 \mu_{o}^{4}+32 \mu_{o}^{2}+5 \right)}{2048} \, ,
	\\
{}^{(5)} \rho_{\rm Cau} =& \, 64 + \frac{15951825 \pi^{2}}{524288} + 	\frac{532389375 \pi^{4}}{33554432} -\frac{34206046875 \pi^{6}}{17179869184} 
	\nonumber
	\\
	&+ \frac{75 \pi \chi \sqrt{1-\mu_{o}^{2}} \left(30405375 \pi^{4}-291014400 \pi^{2}-116260864\right)  \cos \varphi }{268435456 } 
	\nonumber
	\\
	&- \frac{\chi^{2} \left(1-\mu_o^2 \right) \left(47536875 \pi^{4}-451591200 \pi^{2}-134217728\right) \cos^{2}\varphi}{4194304}
	\nonumber
	\\
	&-\frac{\chi^{2}\left(14731875 \pi^{4} \mu_{o}^{2}-96592500 \pi^{4}-90835200 \pi^{2} \mu_{o}^{2}+848822400 \pi^{2}+536870912\right) }{16777216} 
	\nonumber
	\\
	&+ \frac{15 \pi \chi^3  \left(1-\mu_{o}^2\right)^{3/2} \left(87975 \pi^{2}-746752\right) \cos^{3}\varphi}{524288}
	\nonumber
	\\
	&+\frac{15 \pi \chi^3 \left(7650 \pi^{2} \mu_{o}^{2}-106875 \pi^{2}+3584 \mu_{o}^{2}+923392\right) \sqrt{1-\mu_{o}^{2}}\, \cos \varphi}{524288} 
	\nonumber
	\\
	&+ \frac{5 \chi^{4} \left(1-\mu_{o}^2\right)^{2} \left(16515 \pi^{2}-65536\right) \cos^{4}\varphi}{131072}
	\nonumber
	\\
	&-\frac{\left(22950 \pi^{2} \mu_{o}^{4}-82125 \pi^{2} \mu_{o}^{2}-65536 \mu_{o}^{4}+59175 \pi^{2}+393216 \mu_{o}^{2}-327680\right) \chi^{4} \cos^{2}\varphi}{65536}
	\nonumber
	\\
	&-\frac{\left(47700 \pi^{2} \mu_{o}^{4}-900 \pi^{2} \mu_{o}^{2}-196608 \mu_{o}^{4}-35775 \pi^{2}-131072 \mu_{o}^{2}+327680\right) \chi^{4}}{131072}
	\nonumber
	\\
	&+ \frac{15\pi \chi^5 \sqrt{1-\mu_o^2} \cos \varphi}{2048} \bigg[ 87 \left(1-\mu_{o}^2\right)^{2} \cos^{4}\varphi+\left(-156 \mu_{o}^{4}+214 \mu_{o}^{2}-58\right) \cos^{2}\varphi
	\nonumber
	\\
	&+48 \mu_{o}^{4}-12 \mu_{o}^{2}+27 \bigg] - \frac{\left(1-\mu_o^2\right)^3 \chi^6 \sin^6 \varphi }{16} \, .
\end{align}
\endgroup
Higher order terms can be worked out, but they are sufficiently complicated that it is not useful to present them here.\footnote{They can be found in the supplementary Mathematica notebook.} In~\cite{Sereno:2007gd}, the analog of ${}^{(-1)}\rho_{\rm Cau}$ to ${}^{(2)}\rho_{\rm Cau}$ were determined and presented in Eqs.~(10)-(12) of that work. While the above expressions are very similar to the corresponding expressions in~\cite{Sereno:2007gd} there are a few small differences in the coefficients owing to the fact that the parameterization of the impact parameters is different. 

We can then substitute the above results back into our perturbative solutions to the geodesic equations to get the Boyer-Lindquist coordinates of the caustic tube parameterized directly in terms of $\varepsilon$ and $\varphi$. We have presented those results including the leading and next-to-leading corrections in~\eqref{ts_caustic},~\eqref{mus_caustic},~\eqref{phis_caustic} in Section \ref{sec:results_caustic}. The subsequent higher-order corrections read:
\begingroup
\allowdisplaybreaks
\begin{align}
\frac{t_s^{(5)}}{M} =& \, - \frac{465 \pi \chi^4 \left(1-\mu_o^2\right)^2 \cos^4 \varphi}{256} + \frac{15 \pi \chi^2 \left(1-\mu_o^2 \right) \, \cos^2 \varphi}{65536} \bigg[-17408 \chi^{2} \mu_{o}^{2}+23625 \pi^{2}+6656 \chi^{2}
	\nonumber
	\\
	&-142080\bigg] - \frac{15 \pi \chi^4 \left(160 \mu_{o}^{4}-180 \mu_{o}^{2}-1\right)}{1280}  + \frac{118125 \pi \chi^2}{32768} \bigg[\left(\mu_{o}^{2}-\frac{3}{10}\right) \pi^{2}-\frac{9472 \mu_{o}^{2}}{1575}
	\nonumber
	\\
	&+\frac{50048}{13125}\bigg] -\frac{35083125 \pi^{5}}{33554432} +\frac{518625\pi^{3}}{131072} +\frac{4389\pi}{256}  \, ,
\\
\frac{t_s^{(6)}}{M} =& \, \frac{45 \pi \chi^5 \left(1-\mu_o^2\right)^{5/2} \, \cos^5 \varphi}{16} + \frac{12375 \pi^2 \chi^4 \left(1-\mu_o^2\right)^2 \cos^4 \varphi}{16384} 
	\nonumber
	\\
	&+ \frac{5 \pi \chi^3 \left(1-\mu_o^2\right)^{3/2} \cos^3 \varphi}{128} \bigg[156 \chi^{2} \mu_{o}^{2}+225 \pi^{2}-44 \chi^{2}-1224 \bigg] 
	\nonumber
	\\
	&-\frac{225 \pi^2 \chi^2 \left(1-\mu_o^2\right) \, \cos^2 \varphi}{131072} \bigg[-160 \chi^{2} \mu_{o}^{2}+10125 \pi^{2}+496 \chi^{2}-89256 \bigg] 
	\nonumber
	\\
	&- \frac{15 \pi \chi^3 \left(1-\mu_o^2\right)^{3/2}}{512} \bigg[96 \chi^{2} \mu_{o}^{2}+225 \pi^{2}+16 \chi^{2}-1224\bigg] -\frac{1575 \left(4 \mu_{o}^{4}-4 \mu_{o}^{2}-1\right) \pi^{2} \chi^{4}}{16384}
	\nonumber
	\\
	&+\left(-\frac{759375}{65536} \pi^{4} \mu_{o}^{2}+\frac{928125}{131072} \pi^{4}+\frac{836775}{8192} \pi^{2} \mu_{o}^{2}-\frac{1228725}{16384} \pi^{2}+\frac{512}{3}\right) \chi^{2}+\frac{26578125 \pi^{6}}{16777216}
	\nonumber
	\\
	&-\frac{1839375 \pi^{4}}{262144}-\frac{896}{3}-\frac{3615975 \pi^{2}}{65536} \, ,
\\
\mu_s^{(6)} =& \, \frac{15 \pi \chi^2 \left(1-\mu_o^2 \right)^{3/2} \, \sin^3 \varphi }{1024} \bigg[-80 \left(1-\mu_{o}^2\right) \chi^{2} \cos^{2}\varphi -16 \left(6 \mu_{o}^{2}+1\right) \chi^{2}+225 \pi^{2}-1224 \bigg] \, ,
\\
\mu_s^{(7)} =& \, \frac{15 \pi \chi^2 \left(1-\mu_o^2\right)^{3/2} \, \sin^3 \varphi}{131072} \bigg[4096 \sqrt{1-\mu_o^2} \, \cos \varphi \left((1-\mu_o^2)\cos^2 \varphi + 1 + 6 \mu_o^2 \right) 
	\nonumber
	\\
	&+11520   \pi \chi^{2} \left(1-\mu_{o}^{2}\right) \cos^{2} \varphi   +\left(57600 \pi^{2}-313344\right) \chi \sqrt{1-\mu_{o}^{2}}\,\cos \varphi   -111375 \pi^{3}
	\nonumber
	\\
	&+929280 \pi \bigg]
\\
\phi_s^{(6)} =& \frac{75 \chi^{4} \pi  \left(1-\mu_{o}^{2}\right)^{\frac{3}{2}} \cos^{5}\varphi}{64}-\frac{5 \pi \chi^{2} \sqrt{1-\mu_{o}^{2}} \left(-256 \chi^{2} \mu_{o}^{2}+675 \pi^{2}-80 \chi^{2}-3672\right)  \, \cos^{3}\varphi}{1024}
	\nonumber
	\\
	&+\frac{64 \chi^{3}}{3}-\frac{\left(-253125 \pi^{4}+1922400 \pi^{2}+4194304\right) \chi}{49152}
\\
\phi_s^{(7)} =& \, \frac{75 \pi \chi^5 \left(1-\mu_o^2 \right)^2 \, \cos^6 \varphi}{32} + \frac{675 \pi^2 \chi^4 \left(1-\mu_o^2 \right)^{3/2}\,  \cos^5 \varphi}{512} 
	\nonumber
	\\
	&+ \frac{15 \pi \chi^3 \left(1-\mu_o^2 \right) \cos^4 \varphi}{512} \bigg[94 \chi^{2} \mu_{o}^{2}+225 \pi^{2}+18 \chi^{2}-1224\bigg] 
	\nonumber
	\\
	&-\frac{225  \pi^{2} \chi^{2} \sqrt{1-\mu_{o}^{2}} \left(-768 \chi^{2} \mu_{o}^{2}+7425 \pi^{2}+768 \chi^{2}-61952\right) \, \cos^{3}\varphi}{131072} 
	\nonumber
	\\
	&- \frac{15 \pi \chi^3 \left(1-\mu_o^2\right)\, \cos^2 \varphi}{65536} \bigg[-5120 \chi^{2} \mu_{o}^{2}+39375 \pi^{2}+1536 \chi^{2}-220416 \bigg] 
	\nonumber
	\\
	&+ \frac{15 \pi \chi^5 \left(16 \mu_{o}^{4}-4 \mu_{o}^{2}-3\right)}{256}  + \bigg[ -\frac{124875}{32768} \pi^{3} \mu_{o}^{2}+\frac{37125}{65536} \pi^{3}+\frac{2775}{128} \pi  \mu_{o}^{2}-\frac{945}{256} \pi
\bigg]\chi^3 
	\nonumber
	\\
	&+ \frac{\left(325771875 \pi^{5}-2723040000 \pi^{3}-2369126400 \pi \right) \chi}{33554432}\, .
\end{align}
\endgroup
Here $t_s^{(n)}$ represents the coefficient of $\varepsilon^n$ in the perturbative expansion, and similarly for $\mu_s^{(n)}$ and $\phi_s^{(n)}$. We have determined the general results for further orders in $\varepsilon$, but the resulting expressions are too complicated to be usefully presented. 

Remarkably, the results for $(\mu_s, \phi_s)$ up to and including $\mathcal{O}(\varepsilon^4)$ corrections agree precisely with the results of Sereno and de Luca ---c.f.~Eqs.~(13) and (14) of~\cite{Sereno:2007gd}--- despite the fact that the impact parameters used in that work are not well-defined for $\beta = 0$. Important differences arise at higher order, and if the parameterization introduced in~\cite{Sereno:2007gd} is used, the results become inconsistent at~$\mathcal{O}(\varepsilon^6)$.\footnote{We note that, if a comparison is made between the polar angle $\varphi$ we have used and the polar angle defined in~\cite{Sereno:2007gd}, then in the limit of large impact parameters, the two angles agree up to $\mathcal{O}(\epsilon^2)$ corrections. Since terms containing $\varphi$ first appear at $\mathcal{O}(\epsilon^4)$ in $(\mu_s, \phi_s)$, this explains why the results agree at $\mathcal{O}(\epsilon^4)$ and $\mathcal{O}(\epsilon^5)$, but disagree at $\mathcal{O}(\epsilon^6)$ and beyond. }

These equations show that the derivative of the map $(\varepsilon,\varphi) \rightarrow (t_s,r_s,\mu_s,\phi_s)$ has rank $2$ except at $\varphi = 0,\pi/2,\pi,3\pi/2$ where it degenerates to rank $1$. These values of $\varphi$ correspond to the lines of $A_3$ points, i.e.~the edges of the caustic tube. The $A_3$ lines lying at the edges of the false crease set have $\varphi=0,\pi$ {\it exactly}. The equations for the other two pairs of $A_3$ lines (which lie on $\cH^+$) are $\varphi=\pi/2,3\pi/2$ at leading order. At higher order in the $\varepsilon$-expansion, these $A_3$ lines are shifted giving non-trivial curves $\varphi = \varphi(\varepsilon)$ that satisfy the following equation
\begin{align}
\cos \varphi &= \,\chi \sqrt{1-\mu_o^2}  \, \varepsilon^3 \bigg[-\chi^{2} \mu_{o}^{2}-\frac{75}{32} \pi^{2}-\frac{1}{6} \chi^{2}+\frac{51}{4} + \left(\frac{15 \pi  \chi^{2}\mu_{o}^{2}}{32}  +\frac{5 \pi  \chi^{2}}{64} +\frac{120375 \pi^{3}}{8192} -\frac{15285\pi}{128} \right) \varepsilon 
\nonumber\\
&-\frac{\left(96 \chi^{2} \mu_{o}^{2}+225 \pi^{2}+16 \chi^{2}-1224\right) \left(-448 \chi^{2} \mu_{o}^{2}+675 \pi^{2}+112 \chi^{2}-3672\right)}{18432} \varepsilon^2\bigg] + \mathcal{O}(\varepsilon^6) \, ,
\end{align}
which coincides with the extrema of $\mu_s$ as a function of $\varphi$. This result, to leading order, was presented in Eq.~\eqref{A3_pert} in $(\alpha, \beta)$ notation.

It is also straightforward to convert the Boyer-Lindquist coordinates of the caustic tube to the quasi-Cartesian coordinates introduced in~\eqref{Emp_Cart}. We do not reproduce the result to all orders here, instead giving just the leading and subleading terms in those expressions:
\begin{align}
\frac{x}{M} &= - \frac{1}{4 \varepsilon^2} + \frac{5 \pi \chi^2 \left(1-\mu_o^2\right) \varepsilon^3}{4} + \mathcal{O}(\varepsilon^4) \, ,
\\
\frac{y}{M} &= \sqrt{1-\mu_o^2} \bigg[ \chi  + \frac{5 \pi \chi \varepsilon}{16} -  \frac{\left(225 \pi^{2}-2048\right) \varepsilon^{2} \chi}{512} -\frac{15 \pi  \chi^{2} \varepsilon^{2}}{64}  \sqrt{1-\mu_{o}^{2}}\, \cos^3 \varphi 
\nonumber
\\
&+ \frac{9 \pi  \left(2625 \pi^{2}-11392\right) \varepsilon^{3} \chi}{32768} + \frac{3 \pi  \left(10 \mu_{o}^{2}-3\right) \varepsilon^{3} \chi^{3}}{128} -\frac{45 \pi  \left(\mu_{o}^{2}-1\right)  \cos^{2} \varphi \varepsilon^{3} \chi^{3}}{128} + \mathcal{O}(\varepsilon^4) \bigg] 
\\
\frac{z}{M} &=  - \frac{15 \pi \chi^2 \varepsilon^2}{64}  (1-\mu_o^2) \sin^3 \varphi  - \frac{5 \pi \chi^2 \mu_o \sqrt{1-\mu_o^2} \, \varepsilon^3}{4} 
+ \mathcal{O}(\varepsilon^4) \, .
\end{align}
These equations were used to produce Fig.~\ref{fig:caustic_tube}. It is also useful to record 
\be
   \frac{\sqrt{x^2 + y^2}}{M} = \frac{1}{4 \varepsilon^2}  + 2 \chi^2 \left(1-\mu_o^2 \right) \varepsilon^2 + \mathcal{O}(\varepsilon^4) 
\ee
which is used in Fig.~\ref{fig:creasencaustic}.

The faces of the caustic tube are expected to correspond to caustic points of type $A_2$. Ref.~\cite{Gadioux:2023pmw} showed that a surface of $A_2$ caustics should be null (c.f.~the discussion following Eq.~(26) therein). We can use this as a consistency check on our results. To do this, we shall determine the pull-back of the metric to the caustic tube. We substitute the perturbative expansions into the Kerr metric and regard $(\varepsilon, \varphi)$ as the coordinates on the tube. The induced metric has the form
\be\label{induced_metric}
ds^2_{\rm Caustic} = g_{\varepsilon\varepsilon} d \varepsilon^2 + 2 g_{\varepsilon\varphi} d\varepsilon d\varphi + g_{\varphi\varphi} d\varphi^2 \, .
\ee
The determinant is 
\be 
{\rm det} \, g_{\rm Caustic} = g_{\varepsilon\varepsilon} g_{\varphi\varphi} - g_{\varepsilon\varphi}^2 \, .
\ee
When expanding this determinant, a number of non-trivial cancellations occur. Potentially divergent terms are all identically vanishing as a direct consequence of the fact that $(t_s, \mu_s, \phi_s)$ do not depend on $\varphi$ at their lowest orders. The first term that is potentially non-zero arises at $\mathcal{O}(\varepsilon^0)$ and reads,
\be \label{det_e0}
{\rm det} \, g_{\rm Caustic} = \frac{M^2}{4} \left[-\left(\frac{d t_s^{(3)}}{d \varphi} \right)^2 + \frac{M^2}{1-\mu_o^2} \left(\frac{d \mu_s^{(4)}}{d  \varphi} \right)^2 + M^2 (1-\mu_o^2) \left(\frac{d \phi_s^{(4)}}{d \varphi} \right)^2 \right] + \mathcal{O}(\varepsilon)  \, .
\ee
Using the perturbative expansions of $(t_s, \mu_s, \phi_s)$ it can be easily verified that this contribution vanishes. Our results also let us compute several higher-order contributions, up to and including the corrections at $\mathcal{O}(\varepsilon^4)$, which we also find to vanish after a lengthy computation. So our perturbative results are consistent with the expectation that the faces of the caustic tube are null surfaces.

We also study the length of the lines of $A_3$ caustics. At leading order, the $A_3$ lines are curves of constant $\varphi$ equal to integer multiples of $\pi/2$. The length is then
\be 
d \ell = \sqrt{g_{\varepsilon \varepsilon}} d\varepsilon
\ee
where we find
\begin{align}
g_{\varepsilon\varepsilon} &= \frac{4 M^2}{\varepsilon^4} + \frac{15 \pi M^2}{4 \varepsilon^3} + \frac{32 M^2}{\varepsilon^2} - \frac{225 \pi^2 M^2}{128 \varepsilon^2} + \frac{15 \pi  \,M^{2} \left(1125 \pi^{2}-11392\right)}{8192 \varepsilon}  
\nonumber
\\
&+\frac{135 \pi a^{2} \left(1- \mu_{o}^{2}\right)  \cos^{2} \varphi }{32 \varepsilon}-\frac{15 \pi a^{2} \left(2 \mu_{o}^{2}+5\right) }{32 \varepsilon} -\frac{3 M^{2} \left(16875 \pi^{4}-60000 \pi^{2}-1048576\right)}{16384}
\nonumber 
\\
&-\frac{15 \pi a^3 \left(1-\mu_o^2\right)^{3/2} \, \cos^3 \varphi}{4 M} + \frac{675 \pi^2 a^2 \left(1-\mu_o^2\right) \cos^2\varphi}{256} + \frac{225}{256} \pi ^2 a^2 \left(9 \mu_o^2-2\right)-128 a^2 \mu_o^2 + \mathcal{O}(\varepsilon) \, .
\end{align}
To the order of the above series expansions, the lines of $A_3$ caustics occur for $\varphi$ equal to integer multiples of $\pi/2$. Within the domain of validity of the series expansion, i.e.~where $\varepsilon$ is small, the $A_3$ lines are space-like as expected. From the above expression we see that these lines have infinite proper length. At further subleading order, the location of the $A_3$ caustics on $\mathcal{H}^+$ no longer corresponds to constant $\varphi$ and so the computation above requires suitable modifications to capture those corrections. However, since these details arise at subleading order, they do not alter the conclusion that the $A_3$ caustics are lines of infinite proper length.

\subsection{Crease submanifold}

We find crease points as follows.  We fix one null geodesic by specifying $(\lambda, b)$ and $\sgn$ (the sign of $\dot{\theta}$ at the observer). We then seek a second geodesic labelled by $(\lambda', b')$ with the same value of $\sgn$ such that the two geodesics meet at some spacetime point.\footnote{Taking the opposite sign for $\sgn$ ultimately leads to the false crease --- see Appendices \ref{app:false_crease} and \ref{app:false_crease_pert}.} Taking $r_s$ to parameterize the geodesics we have a system of three equations that must be solved (recall $\mu \equiv \cos \theta$):
\be\label{eq:crease_cond} 
t_s(\lambda', b', r_s) = t_s(\lambda, b, r_s) \, , \quad \mu_s(\lambda', b', r_s) = \mu_s(\lambda, b, r_s) \, , \quad \phi_s(\lambda', b', r_s) = \phi_s(\lambda, b, r_s)    \, ,
\ee 
where $\phi \sim \phi + 2\pi$ must be taken into account in the last equation. These equations are solved to determine $b', \lambda'$ and $r_s$ as functions of $(\lambda, b)$. To obtain a solution, we write $b', \lambda'$ and $r_s$ as series expansions in $\epsilon$ (starting at order $\epsilon^{-1}$, $\epsilon^{-1}$, $\epsilon^{-2}$ respectively), substitute those expansions into Eqs.~\eqref{eq:crease_cond} and solve order-by-order in $\epsilon$ to fix the coefficients. 

The first non-trivial equation arises at order $\mathcal{O}(1)$. At this order, the only non-trivial equation is the one involving $t_s$. Solving this determines the leading term in the expansion for $b'$. At order $\mathcal{O}(\epsilon)$ all three equations are non-trivial, but only two of the three equations are independent. These two equations can be solved to determine the next term in the expansion of $b'$ and the first term in the expansion of $r_s$. At orders $\mathcal{O}(\epsilon^2)$ and $\mathcal{O}(\epsilon^3)$ the situation is analogous --- only two of the three equations are independent and can be solved for higher order terms in the expansions of $b'$ and $r_s$. The expansion coefficients of $b'$ determined in this way depend on $(b,\lambda)$ along with the so-far undetermined coefficients in the expansion of $\lambda'$. On the other hand, the expansion coefficients of $r_s$ are fully determined, depending only on $(b,\lambda)$. As such, the relation $r_s = r_s(b, \lambda)$ can be inverted and one is free to use any two of the three parameters $\{r_s, b, \lambda\}$. At order $\mathcal{O}(\epsilon^4)$ the three equations become independent. It is at this order when the first term in the expansion of $\lambda'$ is determined, which in turn fixes the lower order terms in the expansions of $b'$ as well. This general trend then continues to all higher orders. When the equations~\eqref{eq:crease_cond} are solved up to and including $\mathcal{O}(\epsilon^N)$ corrections (with $N \ge 4$) then $N-2$ terms in the expansion of $b'$ are fully determined, $N-3$ terms in the expansion of $\lambda'$ are fully determined, and $N-1$ terms in the expansion of $r_s$ are fully determined. 

To illustrate the above, we present the results obtained when one solves for the crease points when the integrated geodesics are evaluated up to and including $\mathcal{O}(\epsilon^6)$ terms: 
\begin{align}
b' =& \, b + \frac{2 M \chi \lambda}{b} + \frac{15 \pi M^2 \chi \lambda}{16 b^2} -\frac{15 M^{3} \lambda^{2} \pi  \,\chi^{2}}{16 b^{4}}-\frac{M^{3} \lambda  \left(675 \pi^{2}-8192\right) \chi}{512 b^{3}} +  \mathcal{O}\!\left(\epsilon^3 \right) \, ,
\\
\lambda' =& - \lambda - \frac{2 M \chi \lambda^2}{b^2} -\frac{15 \pi  \chi  \,\lambda^{2} M^{2}}{16 b^{3}} +  \mathcal{O}\!\left(\epsilon^2 \right) \, ,
\end{align}
along with the radial location of the crease point
\begin{align}\label{rsCrease}
r_s =& \, \frac{b^2}{4 M} + \frac{\chi \lambda}{2} - \frac{15 \pi b}{64} + \frac{225 \pi^2 M}{1024} - 2 M + \frac{M \chi^2}{4} - \frac{135 \pi M^2 \left(25 \pi^2 - 272 \right)}{16384 b} 
\nonumber\\
&- \frac{\left(225\pi^2 - 2048 \right) M^2 \chi \lambda}{512 b^2} + \frac{15 \pi \left(b^2 +\lambda^2 + 2 \mu_o^2 b^2 \right) M^2 \chi^2}{512 b^3} 
\nonumber\\
&-\frac{M^{3} \left(1575 \pi^{2} b^{2}-3375 \pi^{2} \lambda^{2}-16384 b^{2}+32768 \lambda^{2}\right) \chi^{2}}{4096 b^{4}}+\frac{135 \left(25 \pi^{2}-272\right) \lambda  \pi M^{3} \chi}{4096 b^{3}}
\nonumber \\
&+\frac{3 M^{3} \left(16875 \pi^{4}-60000 \pi^{2}-1048576\right)}{262144 b^{2}}
+\mathcal{O}\!\left(\epsilon^3 \right) \, .
\end{align}

After the equations have been solved in the manner described, it is convenient to convert the results to the polar impact parameters introduced in the previous section.\footnote{One can use the polar impact parameters or $(\alpha,\beta)$ from the outset, but we find this to be less computationally efficient at higher orders.} This facilitates comparison with the caustic tube and is a convenient parameterization for plotting. In terms of $(\rho, \varphi)$ we find that the intersecting geodesic has polar impact parameters:
\begin{align}
\rho' =& \, \rho - 2 \chi \sqrt{1-\mu_o^2} \, \cos \varphi  - \frac{15 \pi \chi}{16} \frac{\sqrt{1-\mu_o^2} \, \cos \varphi}{\rho} + \frac{\left( 675 \pi^2 - 8192 \right) \chi \sqrt{1-\mu_o^2} \, \cos \varphi}{512 \rho^2} 
	\nonumber
	\\
	&- \frac{15 \pi \chi^2 \left(1-\mu_o^2 \right) \cos^2 \varphi}{16 \rho^2} + \frac{\chi^4 \left(1-\mu_o^2\right) \sin^2 \varphi}{3 \rho^3} \bigg[3(1-\mu_o^2)\cos^2\varphi + 1 + 6\mu_o^2 \bigg] 
	\nonumber
	\\
	&- \frac{15 \pi \chi^3 \sqrt{1-\mu_o^2} \, \cos \varphi}{128 \rho^3} \bigg[ 9(1-\mu_o^2) \cos^2 \varphi + 1 + 6 \mu_o^2 \bigg] 
	\nonumber
	\\
	&+ \frac{\chi^2 \left(1-\mu_o^2\right)}{512 \rho^3} \bigg[ \left(3525 \pi^{2}-29440\right) \cos^{2}\varphi-2400 \pi^{2}+13056\bigg] 
	\nonumber
	\\
	&- \frac{15 \pi \chi \left(2025 \pi^2 - 21184 \right)  \sqrt{1-\mu_o^2} \, \cos \varphi}{16384 \rho^3} + \mathcal{O}(\epsilon^4) \, ,
	\\
\cos \varphi' =& \, 	- \cos \varphi + \frac{\chi \sqrt{1-\mu_o^2} \, \sin^2 \varphi}{\rho^3} \bigg[\frac{51}{2} - \frac{75 \pi^2}{16} - \frac{\chi^2}{3} - 2 \chi^2 \mu_o^2 - \left(1-\mu_o^2\right)\chi^2 \cos^2\varphi\bigg] + \mathcal{O}(\epsilon^4) \, .
\end{align}
We have calculated $\rho'$ up to and including $\mathcal{O}(\epsilon^6)$ and $\varphi'$ up to and including $\mathcal{O}(\epsilon^5)$ but the results are too complicated to display here. Non-trivial cancellations for the $\mathcal{O}(\epsilon)$ and  $\mathcal{O}(\epsilon^2)$ terms in $\cos \varphi'$ result in an interesting leading-order relationship $\cos \varphi' = -\cos \varphi + \mathcal{O}(\epsilon^3)$. However this relationship fails at $\mathcal{O}(\epsilon^3)$, as seen above. 

In addition to the above, we have a relationship between $\rho$ and $\varepsilon$ that takes the form of an expansion
\be 
\rho = \sum_{i=-1}^\infty {}^{(i)} \rho_{\rm Cr} \, \varepsilon^i \, 
\ee
with the first few coefficients in the expansion reading
\begingroup
\allowdisplaybreaks
\begin{align}
{}^{(-1)} \rho_{\rm Cr} =& \, 1 \, , 
\\
{}^{(0)} \rho_{\rm Cr} =& \, \frac{15 \pi}{32} +  \chi  \sqrt{1-\mu_{o}^{2}}\, \cos \varphi \, ,
\\
{}^{(1)} \rho_{\rm Cr} =& \, 4 -\frac{675 \pi^{2}}{2048} +\frac{15 \pi \chi \sqrt{1-\mu_{o}^{2}}\, \cos \varphi }{32} - \frac{\chi^{2}  \left(1-\mu_o^2 \right) 
\sin^2 \varphi }{2} \, ,
\\
{}^{(2)} \rho_{\rm Cr} =& \, \frac{135 \pi  \left(25 \pi^{2}-272\right)}{8192} -\frac{\chi \left(225 \pi^{2}-2048\right) \sqrt{1-\mu_{o}^{2}} \cos  \varphi }{256} 
    \nonumber 
    \\
    &-\frac{15 \pi \chi^2 \left[\left(1-\mu_{o}^{2}\right) \cos^{2}\varphi +6 \mu_{o}^{2}+1 \right]}{256} \, ,
\\
{}^{(3)} \rho_{\rm Cr} =& \,  16 + \frac{78525 \pi^{2}}{16384} - \frac{5315625 \pi^{4}}{8388608} + \frac{315 \pi  \chi \sqrt{1-\mu_{o}^{2}} \left(375 \pi^{2}-3968\right) \cos \varphi }{65536} 
	\nonumber 
    \\
    &- \frac{3 \chi^{2} \left(1-\mu_{o}^2\right) \left(1875 \pi^{2}-16384\right) \cos^{2}\varphi}{8192} + \frac{\chi^{2} \left(2700 \pi^{2} \mu_{o}^{2}+5175 \pi^{2}-16384 \mu_{o}^{2}-49152\right)}{8192} 
    \nonumber 
    \\
    &+ \frac{15 \pi \chi^3 \sqrt{1-\mu_o^2} \left(\left(-5625 \pi^{2}+49152\right) \cos^{2}\varphi+5175 \pi^{2}-49152 \right) \cos\varphi}{256} 
    \nonumber 
    \\
    &- \frac{\chi^4 \left(1-\mu_o^2\right)^2 \sin^4 \varphi}{8}  \, ,
\\
{}^{(4)} \rho_{\rm Cr} =& \, -\frac{50025\pi}{2048}  -\frac{541125 \pi^{3}}{65536}  + \frac{2278125\pi^{5}}{2097152}  -\frac{\chi  \sqrt{1-\mu_{o}^{2}} \left(253125 \pi^{4}-2383200 \pi^{2}-2097152\right) \cos \varphi }{65536} 
	\nonumber 
    \\
    &+ \frac{15 \pi \chi^2 (1-\mu_o^2) \left(6525 \pi^2 - 64144 \right) \cos^2\varphi}{32768} -\frac{15 \pi \chi^2   \left(225 \pi^{2} \mu_{o}^{2}+4500 \pi^{2}-8192 \mu_{o}^{2}-37968\right)}{32768} 
    \nonumber 
    \\
    &-\frac{75 \pi \chi^{4} \left(1-\mu_{o}^2\right)^{2}   \cos^{4}\varphi}{2048}-\frac{5 \pi \chi^4 \left(1-\mu_{o}^2\right) \left(16 \mu_{o}^{2}+5\right) \cos^{2}\varphi}{1024}+\frac{15 \pi \chi^{4} \left(16 \mu_{o}^{4}-32 \mu_{o}^{2}-5\right) }{2048} \, ,
\\
{}^{(5)} \rho_{\rm Cr} =& \,   64+\frac{15951825}{524288} \pi^{2}+\frac{532389375}{33554432} \pi^{4} -\frac{34206046875}{17179869184} \pi^{6} + 75 \pi \chi \sqrt{1-\mu_o^2} \cos \varphi \bigg[\frac{30405375}{268435456} \pi^{4}
	\nonumber 
    \\
    &-\frac{1136775}{1048576} \pi^{2}-\frac{887}{2048} \bigg] - 32 \chi^2 \left(1-\left(1-\mu_o^2\right)\cos^2\varphi\right) - \frac{12467475 \pi^2 \chi^2}{131072} \bigg[\left(\mu_{o}^{2}-1\right) \cos^{2}\varphi
    \nonumber 
    \\
    &+\frac{29473}{55411}-\frac{3154 \mu_{o}^{2}}{55411} \bigg] + \frac{81860625 \pi^4 \chi^2}{8388608} \bigg[\left(\mu_{o}^{2}-1\right) \cos^{2}\varphi+\frac{318}{539}-\frac{97 \mu_{o}^{2}}{1078}\bigg] 
    \nonumber 
    \\
    &+ \frac{\chi^3 \left(1-\mu_o^2\right)^{3/2} \cos^3 \varphi}{524288} \bigg[1204875 \pi^{3}-11255040 \pi\bigg] 
    \nonumber 
    \\
    &- \frac{15 \pi \chi^3 \sqrt{1-\mu_o^2} \, \cos \varphi}{524288} \bigg[11550 \pi^{2} \mu_{o}^{2}+87675 \pi^{2}-108032 \mu_{o}^{2}-818944\bigg] 
    \nonumber 
    \\
    &+ \frac{5 \chi^4 \left(1-\mu_o^2\right)^2 \cos^4 \varphi}{131072} \bigg[12915 \pi^{2}-65536 \bigg] + \chi^4 \cos^2 \varphi \bigg[-\frac{8625}{32768} \pi^{2} \mu_{o}^{4}+\mu_{o}^{4}
    \nonumber 
    \\
    &+\frac{67575}{65536} \pi^{2} \mu_{o}^{2}-6 \mu_{o}^{2}-\frac{50325}{65536} \pi^{2}+5 \bigg]  + \chi^4 \bigg[-\frac{11925}{32768} \pi^{2} \mu_{o}^{4}+\frac{3}{2} \mu_{o}^{4}+\frac{225}{32768} \pi^{2} \mu_{o}^{2}+\mu_{o}^{2}
    \nonumber 
    \\
    &+\frac{35775}{131072} \pi^{2}-\frac{5}{2}\bigg] + \frac{5 \pi \chi^5 \sqrt{1-\mu_o^2} \, \cos \varphi}{2048} \bigg[ 105 \left(1-\mu_{o}^2\right)^{2} \cos^{4}\varphi
    \nonumber 
    \\
    &+2 \left(1-\mu_{o}^2\right) \left(94 \mu_{o}^{2}-73\right) \cos^{2}\varphi+48 \mu_{o}^{4}+44 \mu_{o}^{2}+97 \bigg] - \frac{\chi^6 \left(1-\mu_o^2\right)^3 \sin^6 \varphi}{16} \, .
\end{align}
\endgroup 
We have determined ${}^{(N)}\rho_{\rm Cr}$ for two additional orders ---up to ${}^{(7)}\rho_{\rm Cr}$--- but the resulting expressions become too complicated to present.

The expansion of the polar impact parameter can be used to determine the Boyer-Lindquist coordinates of the crease. Using just the terms presented above it is possible to work out the Boyer-Lindquist coordinates of the crease up to and including $\mathcal{O}(\varepsilon^7)$ for $\mu_s$ and $\phi_s$ and up to and including $\mathcal{O}(\varepsilon^6)$ for $t_s$. The leading order terms were presented in Eqs.~\eqref{eqn:t_crease},~\eqref{eqn:mu_crease}, and~\eqref{eqn:phi_crease}.  Here we present several subsequent orders:
\begingroup
\allowdisplaybreaks
\begin{align}
\frac{t_s^{(5)}}{M} =& -\frac{45 \pi \chi^4 \left(1-\mu_o^2\right)^2 \, \cos^4 \varphi}{256} + \frac{5 \pi \chi^4 \left(34 \mu_{o}^{4}-47 \mu_{o}^{2}+13  \right) \cos^2\varphi}{128} - \frac{3 \pi \chi^4 \left( 160 \mu_{o}^{4}-180 \mu_{o}^{2}-1 \right)}{256}   \,  ,
\\
\frac{t_s^{(6)}}{M} =&  \, -\frac{5 \pi \chi^5 \left(1-\mu_o^2\right)^{3/2} \sin^2 \varphi \, \cos\varphi}{32} \bigg[3 \left(1-\mu_o^2\right) \cos^2\varphi + 6 \mu_o^2 + 1 \bigg] + \frac{5175 \pi^2 \chi^4 \left(1-\mu_o^2 \right)^2 \, \cos^4 \varphi}{16384} -
	\nonumber 
    \\
    & \frac{75 \pi^2 \chi^4 \left(10\mu_o^4 - 41 \mu_o^2 + 31\right) \cos^2 \varphi}{8192} - \frac{1575 \pi^2 \chi^4 \left(4 \mu_o^4 - 4 \mu_o^2 - 1 \right)}{16384}  
    \nonumber 
    \\
    &- \frac{45 \pi \chi^3 \left(1-\mu_o^2\right)^{3/2} \left(25 \pi^2 - 136 \right) \, \sin^2\varphi \,  \cos \varphi}{512} - \frac{225 \pi^2 \chi^2 \left(1-\mu_o^2\right)\, \left(3375 \pi^2 - 29752\right) \cos^2 \varphi}{131072} 
    \nonumber 
    \\
    &+ \frac{512 \chi^2}{3}+\frac{ \pi^{4} \chi^2 \left(-4556250 \mu_{o}^{2}+2784375\right)}{393216}+\frac{\pi^{2} \chi^2 \left(40165200 \mu_{o}^{2}-29489400\right) }{393216} 
    \nonumber 
    \\
    &+ \frac{26578125 \pi^{6}}{16777216} -\frac{1839375  \pi^{4}}{262144}-\frac{3615975 \pi^{2}}{65536} -\frac{896}{3} \, ,
\\
\mu_s^{(5)} =& \,  0 \, ,
\\
\mu_s^{(6)} =& \, \frac{25 \pi \chi^2 \left(1-\mu_o^2\right)^{3/2} \sin \varphi}{5120} \bigg[80 \left(\mu_{o}^{2}-1\right) \chi^{2} \cos^{2}\varphi-288 \chi^{2} \mu_{o}^{2}-48 \chi^{2}+675 \pi^{2}-3672\bigg] \, ,
\\
\mu_s^{(7)} =& \frac{15 \pi \chi^2 \left(1-\mu_o^2\right)^{3/2} \, \sin \varphi}{32} \bigg[\left(1-\mu_{o}^{2}\right)^{\frac{3}{2}} \chi^{3} \cos^{3}\varphi+\frac{15 \pi  \left(1-\mu_{o}^{2}\right) \chi^{2} \cos^{2}\varphi}{16}
	\nonumber 
    \\
    &+\frac{\left(96 \chi^{2} \mu_{o}^{2}+225 \pi^{2}+16 \chi^{2}-1224\right) \chi  \sqrt{1-\mu_{o}^{2}}\, \cos\varphi}{48}-\frac{111375 \pi^{3}}{4096}+\frac{1815 \pi}{8}
 \bigg] \, ,
\\
\phi_s^{(5)} =& \frac{15 \pi \chi}{32} \bigg[\left(\mu_{o}^{2}-1\right) \chi^{2} \cos^{2}\varphi-2 \chi^{2} \mu_{o}^{2}+\frac{3 \chi^{2}}{5}-\frac{1575 \pi^{2}}{256}+\frac{267}{10} \bigg] \, ,
\\
\phi_s^{(6)} =& \frac{\chi}{49152} \bigg[253125 \pi^{4}-1922400 \pi^{2}+1048576 \chi^{2}-4194304\bigg] \, ,
\\
\phi_s^{(7)} =& \frac{75 \pi \chi}{256} \bigg[ \bigg( \left(-\mu_{o}^{4}+2 \mu_{o}^{2}-1\right) \cos^{4}\varphi+\left(\frac{68}{15} \mu_{o}^{4}-\frac{22}{5} \mu_{o}^{2}-\frac{2}{15}\right) \cos^{2}\varphi-\frac{16 \mu_{o}^{4}}{5}+\frac{4 \mu_{o}^{2}}{5}+\frac{3}{5}\bigg)\chi^4 
	\nonumber 
    \\
    &+ \left(\left(\frac{\left(-3525 \mu_{o}^{2}+3525\right) \pi^{2}}{1280}+\frac{83 \mu_{o}^{2}}{5}-\frac{83}{5}\right) \cos^{2}\varphi+\frac{\left(16650 \mu_{o}^{2}-2475\right) \pi^{2}}{1280}-74 \mu_{o}^{2}+\frac{63}{5} \right)\chi^2 
    \nonumber 
    \\
    &+ 241-\frac{4343625\pi^{4}}{131072} +\frac{141825\pi^{2}}{512} \bigg] \, .
\end{align}
\endgroup
These Boyer-Lindquist coordinates can also be converted to the quasi-Cartesian coordinates~\eqref{Emp_Cart} we use in plotting. The terms presented above are sufficient to evaluate the quasi-Cartesian coordinates up to and including corrections at $\mathcal{O}(\varepsilon^5)$. Including all the corrections we have computed, this can be further extended to $\mathcal{O}(\varepsilon^7)$. Since the conversion is straightforward, we present here only the results up to and including  $\mathcal{O}(\varepsilon^3)$ corrections. The crease is resolved at $\mathcal{O}(\varepsilon^2)$, so this includes one subleading correction. The results are:
\begin{align}
    \frac{x}{M} =& \, - \frac{1}{4 \varepsilon^2} + \frac{5 \pi  \,\chi^{2}\left(1-\mu_{o}^{2}\right)    \,\varepsilon^{3}}{4} + \mathcal{O}(\varepsilon^4)\, ,
    \\
    \frac{y}{M} =& \sqrt{1-\mu_o^2} \, \chi \left[1 +\frac{5 \pi  \varepsilon}{16} + \left(4-\frac{225 \pi^{2}}{512}\right)  \,\varepsilon^{2} + \frac{9 \pi \left(2625 \pi^{2}-11392\right)     \,\varepsilon^{3}}{32768}  \right.
    \nonumber\\
    &\left. -\frac{3 \pi  \left(5\left(\mu_{o}^{2}-1\right) \cos^{2} \varphi -10 \mu_{o}^{2}+3\right) \chi^{2} \,\varepsilon^{3}}{128} \right] + \mathcal{O}(\varepsilon^4) \, ,
    \\
    \frac{z}{M} =& \,  - \frac{15  \pi \chi^{2}  \left(1-\mu_{o}^{2}\right) \sin \! \left(\varphi \right)  \varepsilon^{2}}{64}-\frac{5 \pi \mu_{o} \sqrt{1-\mu_{o}^{2}} \,\chi^{2} \,\varepsilon^{3} }{4} + \mathcal{O}(\varepsilon^4) \, ,
\end{align}
and we also introduce the polar-like coordinate 
\begin{align}
 \frac{\sqrt{x^2 + y^2}}{M} = \frac{1}{4 \varepsilon^2} + 2 \left(1-\mu_o^2\right) \chi^2 \varepsilon^2  + \mathcal{O}(\varepsilon^4) \, .
\end{align}

Finally, let us comment on the location of the $A_3$ caustics, which form the boundary of the crease set. By restricting to the crease set and computing the Jacobian as in Eq.~\eqref{jacobian}, we can show that in the perturbative regime the caustics are reached when 
\begin{align} 
\cos \varphi_c =&\, \chi \sqrt{1-\mu_o^2} \varepsilon^3 \bigg[ \frac{51}{4} - \frac{75 \pi^2}{32} - \left(\mu_o^2 + \frac{1}{6} \right)\chi^2  + \frac{15 \pi}{128} \left(\frac{15 \left(6\mu_{o}^{2}+1\right) \pi  \,\chi^{2}}{192}+\frac{120375 \pi^{3}}{8192}-\frac{15285 \pi}{128}\right) \varepsilon 
	\nonumber 
    \\
    &-\frac{\left(-448 \chi^{2} \mu_{o}^{2}+675 \pi^{2}+112 \chi^{2}-3672\right) \left(96 \chi^{2} \mu_{o}^{2}+225 \pi^{2}+16 \chi^{2}-1224\right)}{18432} \varepsilon^2 \bigg]  + \mathcal{O}(\varepsilon^6) \, .
\end{align}
This result is exactly the same as we found for the lines of $A_3$ caustics in the previous section, and therefore this provides a non-trivial consistency check of our computations. At leading order we have the caustics located at $\varphi = \pi/2$ and $\varphi = 3 \pi/2$, with subleading corrections arising at $\mathcal{O}(\varepsilon^3)$ shifting these values. For the leading order results this shift is not important, but it must be accounted for when working to high order in the perturbative expansions, as in Appendix~\ref{app:orthog_pert}.

The expressions presented above are sufficient to produce the results of Section~\ref{sec:results}. These results give qualitatively correct depictions of the situation even in the vicinity of the moment of merger. Higher-order perturbative results give a much better quantitative agreement near the moment of merger. We discuss this agreement in more detail for an orthogonal merger in Appendix~\ref{app:orthog_pert}.

It is also of interest to understand the constant time slices of the crease set. For this, we must invert the Boyer-Lindquist time function at the crease to obtain $\varepsilon$ as a function of $t_s$. This involves solving a polynomial equation of degree depending on the expansion order of the series. In practice, we find that this function has always (at least) two roots, one corresponding to the large black hole crease at the specified moment of time and the other the crease of the small black hole. When the equation $t_s = t_s(\varepsilon, \varphi)$ is solved numerically, it is possible to obtain both of these creases. However, when the inversion is carried out perturbatively for small $\varepsilon$, only the crease of the large black hole can be captured. 
The result is simplified if we introduce a quantity $\tau$, which we take to be positive and satisfying
\be 
t \equiv - \frac{M}{4 \tau^2} \, .
\ee
Then we have
\begin{align}\label{eqn:vareps_of_tau}
\varepsilon =&\, \tau + \frac{15 \pi }{2} \tau^4 + \left(16 - \frac{225 \pi^2}{128} \right) \tau^5 + \bigg[\frac{195 \pi}{32} + \frac{5625 \pi^3}{4096}  - \frac{5 \pi \chi^2}{16} \left( 13 - 6 \mu_o^2 - 3(1-\mu_o^2) \cos^2 \varphi \right) \bigg] \tau^6 
\nonumber
\\
&+ \bigg[96 + \frac{149625 \pi^{2}}{1024} -\frac{50625 \pi^{4}}{32768} -32 \chi^2 + \frac{\left(450 \mu_{o}^{2}+1125\right) \pi^{2} \chi^2}{512} 
\nonumber
\\
&+ \frac{\left(-225 \mu_{o}^{2}+225\right) \pi^{2} \chi^2 \cos^{2} \varphi}{512} \bigg]\tau^7 + \mathcal{O}(\tau^8) \, .
\end{align}
In our ancillary Mathematica notebook, additional terms are computed up to $\mathcal{O}(\tau^{11})$. We can use the above to replace $\varepsilon$ in the Cartesian-like coordinates to obtain them as functions of time. 

\begin{figure}
    \centering
    \includegraphics[width=0.45\textwidth]{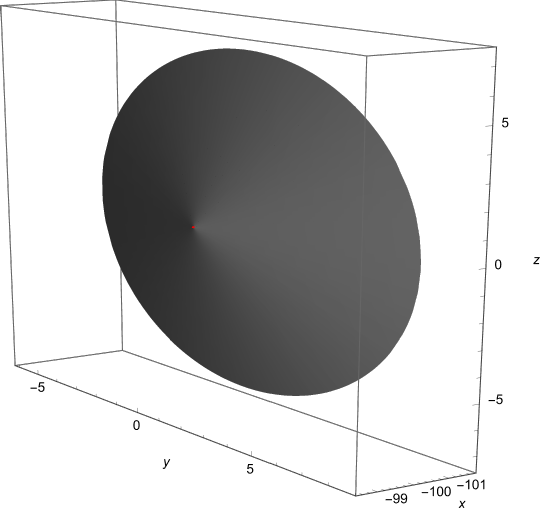}
    \quad\quad
    \includegraphics[width=0.45\textwidth]{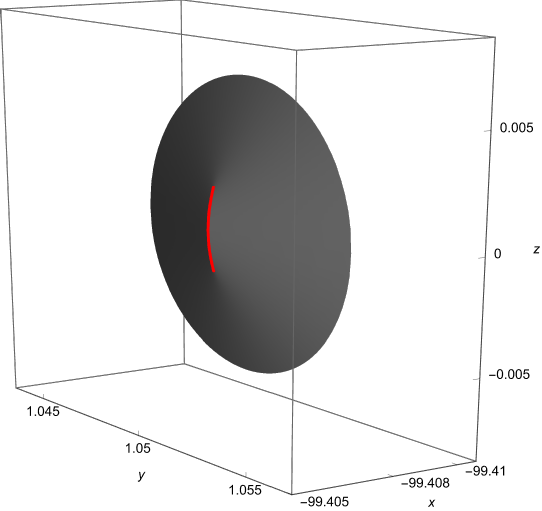}
    \caption{Cross section of the large black hole event horizon constructed from the perturbative expressions. The right figure is a zoomed-in version of the left, focusing on the crease which is shown in red. In this case we have $a = M$ for an orthogonal merger ($\mu_o = 0$) and the time-slice corresponds to $\tau = 1/20$ ($t = -100$). The units are such that $M = 1$. }
    \label{fig:pert_horizon}
\end{figure}

In Fig.~\ref{fig:pert_horizon} we show a horizon cross-section of the large black hole constructed from the perturbative analysis. The figure shows an orthogonal merger with $\mu_o = 0$ and $a = M$. For these parameters, the relative size of the crease is greatest. In this case, we have focused on a cross-section with $\tau = 1/20$, which corresponds to $t = -100$, for which convergence of the perturbative expansions is robust. The left plot shows a portion of the large black hole horizon, while the right plot shows a zoom-in on the crease --- shown here in red.

\subsection{Geometrical properties of the crease set}

We can compute the area of the crease set from our perturbative results. The induced metric on the crease set takes the same form as in~\eqref{induced_metric}, with different metric components owing to the fact that the expansions of $(t_s, \mu_s, \phi_s)$ are different. As in the case of the caustic, the first term that could be non-zero is the $\mathcal{O}(\varepsilon^0)$ one, which reads the same as~\eqref{det_e0}. However, unlike for the caustic, this term evaluates to something nonzero for the crease. Our perturbative results allow us to compute the leading and several subleading corrections to the determinant of the induced metric on the crease set, with the first few terms reading
\be 
{\rm \det} \, g_{\rm Crease} = \frac{225 \pi^2  a^4 \left(1 - \mu_o^2\right)^2 \cos^4 \varphi}{1024}  \left[1 + \frac{15 \pi \varepsilon}{16} + \mathcal{O}(\varepsilon^2)\right] \, .
\ee
Note that the determinant is manifestly positive over the full range of parameters, as it should be for a space-like submanifold of the full metric. To this order in the expansion, one can safely assume that the $A_3$ caustics are located at $\varphi = \pi/2$ and $\varphi = 3\pi/2$. Near these points the determinant approaches zero.

Computing the area involves an integration over $\varepsilon$ and $\varphi$. For $\varphi$ the appropriate range is $(\pi/2, 3\pi/2)$. For $\varepsilon$ we take the integration range over an interval $\varepsilon \in [0, \varepsilon_{\rm max}]$. Here $0$ corresponds to $r_s \to \infty$, while $\varepsilon_{\rm max}$ is ultimately determined by the accuracy of the series or the question of interest. To capture the area of large black hole crease, $r_s$ should be approximately at the scale of the small black hole horizon which corresponds to $\varepsilon_{\rm max}$ between $\sqrt{2}/4$ for a non-rotating black hole and $1/2$ in the extremal limit. In any case, we will keep this parameter arbitrary. The result for the area is then 
\begin{align}
A_{\rm Crease}  =&\, \frac{15 \pi a^2 (1-\mu_o^2)}{64} \varepsilon_{\rm max} \bigg[1 + \frac{15 \pi \varepsilon_{\rm max}}{64} 
 +  \left(\frac{185}{24} - \frac{2625 \pi^2}{2048} + \frac{\left(23 + 61 \mu_o^2 \right) \chi^2}{144} \right)\varepsilon_{\rm max}^2 
 	\nonumber
 	\\
 	&+ \frac{5 \pi}{16384} \left(3375 \pi^2 - 29752 + 4 \chi^2 \left(13 + 71 \mu_o^2 \right) \right)  \varepsilon_{\rm max}^3  + \cdots \bigg] \, .
\end{align}
The area is largest for an orthogonal merger involving an extremal black hole. In Appendix~\ref{app:orthog_pert} we extend this computation several additional orders for the case of an orthogonal merger.

To get a sense for the magnitude of the crease area, we can compare it to the area of the small black hole. Taking only the leading order term, the ratio is
\be 
\frac{A_{\rm Crease}}{ A_{\rm Kerr}} = \frac{15 \pi}{512} \left( \frac{\chi^2}{1+ \sqrt{1-\chi^2} } \right) \varepsilon_{\rm max} \left(1-\mu_o^2 \right)\, .
\ee
The ratio is maximal for orthogonal mergers at extremality, where it is approximately equal to $0.09204 \,  \varepsilon_{\rm max}$.

We can also quite easily work out the length of the crease on the large black hole in a given time-slice. To do this, we first convert the induced metric from the $(\varepsilon, \varphi)$ parameter space to the $(\tau, \varphi)$ or $(t, \varphi)$ parameter space using~\eqref{eqn:vareps_of_tau}. (Note, because the time function depends on both $\varepsilon$ and $\varphi$ the metric component $g_{\varphi \varphi}$ changes when this change of coordinates is made). Then, the length is given by integrating
\begin{align} 
\ell_{\rm Crease} = \int \sqrt{g_{\varphi\varphi}} \, d\varphi  =& \,  -\frac{15 \pi \chi^2 \left(1-\mu_o^2\right)}{128 t} \bigg[1  -\frac{\mu_o^2 M \chi ^2}{3 t}-\frac{5 M \chi ^2}{48 t}-\frac{523M}{96 t}+\frac{225 \pi ^2 M}{256 t} 
	\nonumber
	\\
	&+ \frac{5 \pi  \left(22275 \pi ^2-157184\right)}{65536}  \left(-\frac{M}{t}\right)^{3/2} + \cdots \bigg] \, .
\end{align}
Note that the term proportional to $t^{-3/2}$ vanishes identically, with the sub-leading correction  entering at $\mathcal{O}(t^{-2})$.

Finally, we compute the angle of the crease. We give an overview of the basic procedure, omitting complicated and ultimately unnecessary intermediate results. We work in a slice of constant Boyer-Lindquist time, inverting the general expression $t(r_s, b, \lambda)$ to obtain a relation $r_s(t, b_, \lambda)$ which we then substitute into $\mu_s$ and $\phi_s$ to obtain the parameterization of the horizon as a function of $(t,b,\lambda)$. At the crease, two segments of the horizon --- $\mathcal{N}_1$ and $\mathcal{N}_2$ --- intersect. One of these segments will have coordinates $(t, b, \lambda)$ at the crease, while the other will have coordinates $(t, b', \lambda')$. The relationship between $(b', \lambda')$ and $(b, \lambda)$ at the crease was discussed above. We begin by constructing the normal and tangent vectors to each of these null hypersurfaces. At a moment of constant Boyer-Lindquist time, the tangent vectors are
\begin{align} 
T_1 &\equiv \frac{\partial}{\partial b} = \frac{\partial r}{\partial b} \frac{\partial}{\partial r} + \frac{\partial \mu}{\partial b} \frac{\partial}{\partial \mu} + \frac{\partial \phi}{\partial b} \frac{\partial}{\partial \phi}
\\
T_2 &\equiv \frac{\partial}{\partial \lambda} = \frac{\partial r}{\partial \lambda} \frac{\partial}{\partial r} + \frac{\partial \mu}{\partial \lambda} \frac{\partial}{\partial \mu} + \frac{\partial \phi}{\partial \lambda} \frac{\partial}{\partial \phi} \, ,
\end{align}
and similarly for $\mathcal{N}_2$ replacing $(b,\lambda) \to (b', \lambda')$. The normal vectors $N$ and $N'$ are then easily and uniquely constructed by determining a unit normal vector orthogonal to $(T_1, T_2)$ and $(T_1', T_2')$. The angle of the crease $\Omega$ is then determined through the relationship
\be 
\cos \left(\pi - \Omega \right) = h_{ij} N^i N'^j \, ,
\ee
where $h_{ij}$ is the induced metric on a slice of constant Boyer-Lindquist time. Carrying out this computation we find that the first non-trivial contribution arises at $\mathcal{O}(\varepsilon^5)$ and reads 
\be 
\cos \left(\pi - \Omega \right) = 1 - 128 \chi \sqrt{1-\mu_o^2} \, |\cos \varphi| \, \varepsilon^5 + \cdots \, .
\ee
Solving for the angle gives
\be 
\Omega = \pi - 16 \sqrt{\chi} \left(1-\mu_o^2\right)^{1/4} \sqrt{|\cos \varphi|} \, \varepsilon^{5/2} + \cdots  \, .
\ee
This can equivalently be written in terms of the Boyer-Lindquist time 
\be 
\Omega = \pi - 16 \sqrt{\chi} \left(1-\mu_o^2\right)^{1/4} \sqrt{|\cos \varphi|} \, \left(-\frac{M}{4 t} \right)^{5/4} + \cdots \, .
\ee
To this order in $\varepsilon$, the $A_3$ points are located at $\varphi = \pi/2$ and $\varphi = 3\pi/2$. Thus we see that the angle of the crease goes to $\pi$ as the caustic is approached.

\section*{Acknowledgments}

We are grateful to Roberto Emparan and Harald Pfeiffer for helpful discussions. RAH is grateful to the
Department of Applied Mathematics and Theoretical Physics at the University of Cambridge for hospitality during various stages of this work. MG is supported by an STFC studentship and a Cambridge Trust Vice-Chancellor’s Award. The work of RAH received the support of a fellowship from ``la Caixa” Foundation (ID 100010434) and from the European Union’s Horizon 2020 research and innovation programme under the Marie Skłodowska-Curie grant agreement No 847648 under fellowship code LCF/BQ/PI21/11830027. HSR is supported by STFC grant no.~ST/X000664/1. HSR acknowledges support of the Institut Henri Poincar\'e (UAR 839 CNRS-Sorbonne Universit\'e), and LabEx CARMIN (ANR-10-LABX-59-01).

\newpage

\FloatBarrier
\appendix

\section{Null geodesics in the Kerr spacetime}

\subsection{Geodesic equations}

\label{app:kerr_geos}

In the Kerr spacetime we define
\be 
\Delta = r^2 - 2 M r + a^2 \, , \quad \Sigma = r^2 + a^2 \cos^2\theta \, .
\ee
Let $p^\mu$ be tangent to an affinely parameterized null geodesic. The geodesic equations are (see e.g. \cite{Gralla:2019ceu})
\be
\frac{\Sigma}{E} p^r = \pm_r \sqrt{R(r)}
\ee
\be
\frac{\Sigma}{E} p^\theta = \pm_\theta \sqrt{\Theta(\theta)}
\ee
\be
 \frac{\Sigma}{E} p^\phi =\frac{a}{\Delta} (r^2 + a^2 - a\lambda)+ \frac{\lambda}{\sin^2 \theta} -a
\ee
\be
\label{pt}
 \frac{\Sigma}{E} p^t = \frac{r^2+a^2}{\Delta}(r^2 + a^2 - a\lambda)+a(\lambda - a \sin^2 \theta)
\ee
where $E,\lambda,\eta$ are the conserved quantities along the geodesic and
\begin{align}
\label{Rdef}
R(r) &= r^4 + \left(a^2-\lambda^2-\eta \right)r^2 + 2 M \left[ \left(\lambda - a\right)^2 + \eta \right] r - a^2 \eta \, ,
\\
\label{Thetadef}
\Theta(\theta) &= \left(a^2-\lambda^2 \csc^2\theta \right) \cos^2\theta + \eta \, ,
\end{align}
By rescaling the affine parameter we shall set $E=1$ (null geodesics with $E=0$ cannot escape the ergosphere and therefore cannot correspond to our horizon generators). $\pm_r$ and $\pm_\theta$ change sign when the geodesic moves through a turning point of the $r$ or $\theta$ motion. 

The impact parameters $(\alpha,\beta)$ are related to the conserved quantities $(\lambda,\eta)$ and the angle $\theta_o$ of the line of sight to the distant observer by \cite{Bardeen:1973tla,Gralla:2019drh}
\be
\label{alpha_beta_def}
 \alpha = -\frac{\lambda}{\sin \theta_o} \qquad \qquad \beta = \pm_o \sqrt{\Theta(\theta_o)}
\ee
where $\pm_o$ is the sign of $p^\theta$ when the geodesic reaches the observer.

We shall sometimes make use of the non-affine Mino parameter $\tau$ defined by \cite{Mino:2003yg} $d\tau/d\mu = E/\Sigma$ (where $\mu$ is the affine parameter), for which $dx^\mu/d\tau = (\Sigma/E)p^\mu$, so the factors of $\Sigma/E$ then drop out of the geodesic equations.   

\subsection{False crease}

\label{app:false_crease}

Consider a pair of null geodesics ending at the point $(u_o=0,\theta_o,\phi_o=0)$ of future null infinity, with parameters $(\alpha,\beta)$ and $(\alpha,-\beta)$ where $\beta > 0$. Based on numerical evidence, Ref.~\cite{Emparan:2017vyp} observed that this pair of geodesics intersects at a point with $\theta = \pi - \theta_o$. In this section we shall present an analytical explanation of this result. We'll see that, evolving backwards in time, each geodesic has a single turning point in $\theta$; one above the equatorial plane and the other below the equatorial plane. So, roughly speaking, one geodesic travels mostly over the ``northern hemisphere'' and the other mostly over the ``southern hemisphere''. 

For each of our geodesics we can choose the Mino time parameter $\tau$ to vanish at infinity with $\tau<0$ along the geodesic. The equation for $r(\tau)$ does not depend on the sign of $\beta$ and hence $r(\tau)$ is the same function for both geodesics, with $r(0)=\infty$. 

Geodesics with $\eta<0$ occupy a small region at the centre of the $(\alpha,\beta)$ plane (Fig.~1 of \cite{Gralla:2019drh}) which we expect to lie inside the false creaseless disc so we assume $\eta>0$. The function $\Theta(\theta)$ then has two roots. These can be written as $\theta_\star \in (0,\pi/2)$ and $\pi-\theta_\star$. These roots correspond to turning points in $\theta$ along the geodesic. The function $\theta(\tau)$ oscillates periodically between these turning points. We'll see that, evolving backwards from $\tau=0$, the geodesics with parameters $(\alpha,\pm \beta)$ intersect after half a period of this oscillation. 

Since $\Theta$ must be non-negative along the geodesic, we have $\theta_\star<\theta_o<\pi-\theta_\star$. To be concrete, we'll assume $\theta_o \le \pi/2$, i.e.~the geodesics reach infinity above (or on) the equatorial plane. Straightforward modifications to the following argument can be used to extend it to $\theta_o>\pi/2$. Let $\theta_+(\tau)$ be the solution of the $\theta$ equation with parameters $(\alpha,\beta)$. Since $\dot{\theta}_+(0)>0$, if we evolve the geodesic backwards in time then $\theta_+$ decreases, reaches the turning point at $\theta_\star$, and then increases, reaching $\theta=\pi-\theta_o$ at some value $\tau=\tau_i<0$ (``i'' for initial). Now define $\theta_-(\tau) = \pi - \theta_+(\tau_i-\tau)$. We claim that this is the solution for $\theta$ along the geodesic with parameters $(\alpha,-\beta)$. To justify this, observe that $\dot{\theta}_-(\tau) = \dot{\theta}_+(\tau_i -\tau)$ so $\dot{\theta}_-(\tau)^2 = \Theta(\theta_+(\tau_i-\tau))^2 = \Theta(\pi-\theta_-(\tau))^2=\Theta(\theta_-(\tau))^2$ using the symmetry $\Theta(\theta)=\Theta(\pi-\theta)$. Furthermore ${\theta}_-(0)=\theta_o$ and $\dot{\theta}_-(0)=-\dot{\theta}_+(0)$ so $\theta_-$ satisfies the appropriate final conditions at $\tau=0$. Hence $\theta_-$ is a solution of the $\theta$-equation with parameters $(\alpha,-\beta)$. Viewed forwards in time, $\theta_\pm$ both start at $\pi-\theta_o$ with $\theta_+$ decreasing to a minimum and then increasing to $\theta_o$ and $\theta_-$ increasing to a maximum and then decreasing to $\theta_o$.

Let $\phi_\pm(\tau)$ be the solution for $\phi(\tau)$ along the two geodesics. Integrating the $\phi$-equation from $\tau=\tau_i$ to $\tau=0$ gives
\be
 -\phi_\pm(\tau_i) = \int_{\tau_i}^0 \left[\frac{a}{\Delta}(r^2+a^2-a\lambda) -a \right] d\tau +  \int_{\tau_i}^0 \frac{\lambda}{\sin^2 \theta_\pm(\tau)} d\tau
\ee
The first integral is the same for each geodesic because they have the same $r(\tau)$. The second integral is easily seen to be the same for both geodesics using the relation between $\theta_-$ and $\theta_+$. Hence we have $\phi_+(\tau_i)=\phi_-(\tau_i)$. 

Finally let $u_\pm(\tau)$ be the solution for $u(\tau)$ along the two geodesics. Using the definition of $u$ (Eq.~\eqref{udef}), Eq.~\eqref{pt} gives an equation for $du/d\tau$. Integrating this from $\tau=\tau_i$ to $\tau=0$ gives 
\be
 -u_\pm(\tau_i) =  -a^2 \int_{\tau_i}^0 \sin^2 \theta_\pm(\tau) d\tau + \ldots
\ee
where the ellipsis denotes an integral where the integrand involves only $r(\tau)$. So, just as for $\phi$, we see that the RHS takes the same value for the two geodesics, hence $u_+(\tau_i) = u_-(\tau_i)$. It follows that the two geodesics intersect at $\tau=\tau_i$ and the point of intersection has $\theta = \pi-\theta_o$. 

This proves the observation of Ref.~\cite{Emparan:2017vyp} modulo one subtlety, which is that the parameter $\tau$ is bounded below. For example, a ``scattering'' geodesic, which has $r \rightarrow \infty$ in the far past has $\tau \rightarrow \tau_\infty$ in the far past, where $\tau_\infty<0$ is finite. The above proof works only if $\tau_i>\tau_\infty$, i.e.~the geodesics undergo at least one half-period of $\theta$-oscillation before they reach past null infinity. The results of \cite{Emparan:2017vyp}, or our perturbative and numerical results, confirm that this is indeed the case for $(\alpha,\beta)$ lying outside the false creaseless disc.

\section{Perturbative evaluation of geodesic integrals}
\label{app:perturbative_eval}
The turning points of the geodesic motion will play an important role in our analysis. The turning point of the radial motion is the largest root $R(r_{\rm min}) = 0$. While this can be solved for analytically, in practice we use the result in its series form. The first few orders of this expansion read
\begin{align}
\frac{r_{\rm min}}{b} =& \, 1 - \frac{M}{b} -  \frac{M^2 \left( \chi \lambda - 3 b \right) \left(\chi \lambda - b \right)}{2 b^4}  -  \frac{2 M^3 \left(\chi \lambda - 2 b \right) \left( \chi \lambda -  b \right)}{b^5}  
\nonumber\\
& + \frac{M^4 \left[\left(16 \chi^{2}-105\right) b^{4}-16 \chi  \lambda  \left(\chi^{2}-12\right) b^{3}+\left(4 \chi^{4} \lambda^{2}-118 \chi^{2} \lambda^{2}\right) b^{2}+32 \chi^{3} \lambda^{3} b -5 \chi^{4} \lambda^{4} \right]}{8 b^8}
\nonumber\\
&
+\frac{M^5 \left[12 b^{4} \chi^{2}-48 b^{4}-7 \chi  \lambda  \left(2 \chi^{2}-15\right) b^{3}+4 \chi^{2} \lambda^{2} \left(\chi^{2}-22\right) b^{2}+35 \chi^{3} \lambda^{3} b -6 \chi^{4} \lambda^{4} \right]}{b^9} + \mathcal{O}\left(\epsilon^5 \right).
\end{align}
The polar function $\Theta$ is simpler if we write it in terms of the direction cosine $\mu \equiv \cos \theta$. We can then write
\be 
\Theta = \frac{a^2 \left(\mu_+^2 - \mu^2 \right) \left(\mu^2 + \mu_-^2 \right)}{1-\mu^2} 
\ee
with
\be 
\mu_\pm^2 = \frac{\sqrt{\left(a^{2}+2 a \lambda +b^{2}\right) \left(a^{2}-2 a \lambda +b^{2}\right)} \pm  \left( a^2-b^2 \right)}{2 a^2} \, .
\ee
The turning points of the polar motion are then $\pm \mu_+$, with $\mu_+$ corresponding to $\theta_{\rm min}$.

We evaluate the integrals in the weak deflection limit. The radial motion proceeds from $r_s$ to $r_{\rm min}$, which is the turning point satisfying $R(r_{\rm min}) = 0$, and then from $r_{\rm min}$ to $r = \infty$. The polar motion will proceed from $\theta_s$ to either $\theta_{\rm max}$ or $\theta_{\rm min}$ and then to the observer at $\theta_o$. Which turning point is reached will depend on whether $\theta$ is initially increasing or decreasing.

\subsection{Evaluation of integral~\eqref{geo_rad_polar}}

Let us begin by evaluating the radial integral of~\eqref{geo_rad_polar}. Since the integrand falls off sufficiently fast as $r \to \infty$ we can massage this integral into the following form
\be
\int_{\rm P} \frac{dr}{\pm \sqrt{R}} = 2 \int_{r_{\min}}^\infty \frac{dr}{\sqrt{R}} - \int_{r_s}^\infty \frac{dr}{\sqrt{R}} \, .
\ee
The second term can be evaluated by expanding the integral in a large-$r$ series and performing the integration term-by-term. The first integral requires a bit more care. We begin by introducing a new integration parameter 
\be 
x \equiv \frac{r_{\rm min}}{r}
\ee
and then isolate the leading behaviour of $R$ in the limit $\epsilon \to 0$, which gives
\be 
R_{\rm lead} = \frac{(1-x^2) b^4}{x^4} \, .
\ee
Pulling off the leading behaviour, we expand $\sqrt{R/R_{\rm lead}}$ in a small $\epsilon$ series. The integral is then performed term-by-term,
\be 
\int_{r_{\min}}^\infty \frac{dr}{\sqrt{R}} = \int_{0}^1 \frac{r_{\rm min} dx }{x^2 \sqrt{R_{\rm lead}}} \left[1 + \frac{\left(x^{2}+2 x +2\right) M \epsilon}{b \left(x +1\right)} + \cdots \right] \, .
\ee
We do not present the intermediate result for the integral. 

Next we consider the polar integral of~\eqref{geo_rad_polar}. We introduce the change of integration parameters
\be 
u \equiv \frac{\mu}{\mu_+} \, ,
\ee
after which the integral becomes 
\be 
\int_{\rm P} \frac{d\theta}{\pm \sqrt{\Theta}} = \int_{\rm P} \frac{d u}{\pm a \mu_+ \sqrt{\left(1-u^2\right)\left(u^2 + \frac{\mu_-^2}{\mu_+^2} \right)}} .
\ee
Along any segment of the path, this integral has a relatively simple form written in terms of special functions,
\be 
\int \frac{d u}{\pm a \mu_+ \sqrt{\left(1-u^2\right)\left(u^2 + \frac{\mu_-^2}{\mu_+^2} \right)}} = \frac{1}{a \mu_-} {\rm F} \left(\arcsin u \bigg| - \frac{\mu_+^2}{\mu_-^2} \right) \, ,
\ee
where $F(x | k)$ is the elliptic integral of the first kind. The integral along the path can then be written as
\be 
\int_{\rm P} \frac{d\theta}{\pm \sqrt{\Theta}} = \frac{1}{a \mu_-} \left\{ 2 {\rm F} \left( \frac{\pi}{2} \bigg| - \frac{\mu_+^2}{\mu_-^2} \right) + \pm_o \left[{\rm F} \left( \arcsin \frac{\mu_s}{\mu_+} \bigg| - \frac{\mu_+^2}{\mu_-^2} \right) + {\rm F} \left( \arcsin \frac{\mu_o}{\mu_+} \bigg| - \frac{\mu_+^2}{\mu_-^2} \right)\right] \right\} \, .
\ee
We then have all the information required to evaluate the left-hand and right-hand sides of Eq.~\eqref{geo_rad_polar}. This equation should be regarded as a constraint that determines $\mu_s$ in terms of the given data $\mu_o$ along with $r_s$. By using the Jacobi elliptic functions, the equation can be formally manipulated into a somewhat simpler form where all instances of $\mu_o$ and $\mu_s$ are explicit, i.e.~not inside special functions. This result is
\be 
\mu_s = - \frac{\pm_o \sqrt{1 - \mu_o^2} \sqrt{1- \frac{\mu_o^2}{\mu_-^2}} \, {\rm sn}(x | y) + \mu_o {\rm cn}(x|y) {\rm dn} (x | y)}{1 - \frac{\mu_o^2}{\mu_-^2} {\rm sn}^2(x | y)} 
\ee 
where ${\rm sn}$, ${\rm cn}$ and ${\rm dn}$ are the Jacobi elliptic functions and for brevity we have defined
\begin{align}
x = a \mu_- \int_{\rm P} \frac{dr}{\pm \sqrt{R}} \, , \quad y = - \frac{\mu_+^2}{\mu_-^2} \, .
\end{align}
While it is possible to express $\mu_s$ in this way, it is not useful for the purposes of direct computation. In that case, it is best to solve the equation perturbatively, expanding the above in a small $\epsilon$ series. In this limit the elliptic functions become ordinary trig functions to leading order, and we can collect all terms in the following form
\be\label{mus_perturb} 
\mu_{\sigma, s} = - \mu_{\sigma, o} \cos \delta + \pm_o \sqrt{1- \mu_{\sigma, o}^2} \sin \delta
\ee
where we have introduced for simplicity both here and later the short-hand
\be 
\mu_{\sigma} \equiv \, \frac{\mu \,  b}{\sqrt{b^2- \lambda^2}} \, 
\ee
and the parameter $\delta$ is determined order-by-order with the result to $\mathcal{O}(\epsilon^5)$ given by
\begin{align}
\delta =& \, \frac{4 M}{b} - \frac{b}{r_s} + \frac{M^2 \left(15 \pi b - 32 \chi \lambda \right)}{4 b^3} + \frac{M^3 \left[ 128b + 6 \left(\mu_{\sigma, o}^2 - 2 \right) \chi^2 b - 45 \pi \chi \lambda  \right]}{3 b^4}
\nonumber
\\ 
&-\frac{2 M^{3} \chi^{2} \lambda^{2} \left(\mu_{\sigma ,o}^{4}-8 \mu_{\sigma ,o}^{2}+8\right)}{\left(\mu_{\sigma ,o}^{2}-1\right) b^{5}} -\frac{M^{2} \mu_{\sigma ,o}^{2} \chi^{2} }{2 b r_{s} }-\frac{M^{2} \chi^{2}  \lambda^{2} \mu_{\sigma ,o}^{4}}{2 \left(1-\mu_{\sigma ,o}^{2}\right) b^{3} r_s} - \frac{b^3}{6 r_s^3} 
\nonumber
\\
&-\frac{4 M^4 \lambda  \left(\mu_{\sigma ,o}^{2}-4\right)  \chi^{3}}{b^{5}}+\frac{15 \pi M^4  \left(4 \mu_{\sigma ,o}^{2}-19\right) \chi^{2}}{32 b^{4}}-\frac{256  M^4\lambda \chi}{b^{5}}+\frac{3465 \pi  M^{4}}{64 b^{4}}
\nonumber 
\\
&-\frac{M^{4} \chi^{2} \lambda^{2} \left[ \left(60 \pi  b -128 \chi  \lambda \right) \mu_{\sigma ,o}^{4}+\left(-1425 \pi  b +1024 \chi  \lambda \right) \mu_{\sigma ,o}^{2}+1425 \pi  b -1024 \chi  \lambda \right]}{32 \left(\mu_{\sigma ,o}^{2}-1\right) b^{7}}
\nonumber 
\\ 
&-\frac{\pm_o \, \chi^{2} M^{2} \mu_{\sigma ,o} \left(b^{2}-4 M r_s \right)^{2}   \left[\left(\mu_{\sigma ,o}^{4}-2 \mu_{\sigma ,o}^{2}+1\right) b^{2}-\mu_{\sigma ,o}^{2} \lambda^{2} \left(\mu_{\sigma ,o}^{2}-2\right)\right] }{2 \left(-\mu_{\sigma ,o}^{2}+1\right)^{\frac{3}{2}} b^{6} r_s^{2}} 
\nonumber 
\\ 
&-\frac{\pm_o \, 15 \left(\pi  b -\frac{32 \chi  \lambda}{15}\right) \left(\left(\mu_{\sigma ,o}^{4}-2 \mu_{\sigma ,o}^{2}+1\right) b^{2}-\mu_{\sigma ,o}^{2} \lambda^{2} \left(\mu_{\sigma ,o}^{2}-2\right)\right) \chi^{2} M^{4} \left(M r_{s}-\frac{b^{2}}{4}\right) \mu_{\sigma ,o}}{\left(-\mu_{\sigma ,o}^{2}+1\right)^{\frac{3}{2}} r_{s} b^{8}} 
\nonumber 
\\ 
&-\frac{15 \left(4 \chi^{2} \mu_{\sigma ,o}^{2}-27 \chi^{2}+231\right) \pi  \,M^{5} \chi  \lambda}{8 b^{6}}-\frac{3 b^{5}}{40 r_{s}^{5}}+\frac{M \,b^{3}}{4 r_{s}^{4}}+\frac{\chi^{2} \mu_{\sigma ,o}^{2} M^{2} b}{4 r_{s}^{3}}
\nonumber 
\\ 
&-\frac{2 \chi^{2} M^{3} \left(2 \mu_{\sigma ,o}^{6}-5 \mu_{\sigma ,o}^{4}+4 \mu_{\sigma ,o}^{2}-1\right)}{\left(\mu_{\sigma ,o}-1\right)^{2} \left(\mu_{\sigma ,o}+1\right)^{2} r_{s}^{2} b}+\frac{M^{4} \chi^{2} \left(\chi^{2} \mu_{\sigma ,o}^{4}+128 \mu_{\sigma ,o}^{2}-64\right)}{8 r_{s} b^{3}}
\nonumber 
\\ 
&-\frac{M^{5} \left(-\frac{7168}{5}+\left(\mu_{\sigma ,o}^{4}+4 \mu_{\sigma ,o}^{2}-8\right) \chi^{4}+320 \chi^{2}\right)}{2 b^{5}} 
\nonumber 
\\ 
&-\frac{\left(M \left(\mu_{\sigma ,o}^{8}+16 \mu_{\sigma ,o}^{6}-144 \mu_{\sigma ,o}^{4}+256 \mu_{\sigma ,o}^{2}-128\right) r_{s}-\frac{\mu_{\sigma ,o}^{8} b^{2}}{4}\right) \chi^{4} \lambda^{4} M^{4}}{2 r_{s} b^{9} \left(\mu_{\sigma ,o}-1\right)^{2} \left(\mu_{\sigma ,o}+1\right)^{2}}
\nonumber 
\\ 
&+\frac{15 M^{5} \lambda^{3} \pi  \left(4 \mu_{\sigma ,o}^{4}-63 \mu_{\sigma ,o}^{2}+63\right) \chi^{3}}{8 b^{8} \left(\mu_{\sigma ,o}^{2}-1\right)} + \frac{\lambda^{2} \chi^{2} M^{2}}{r_{s}^{3} b^{7} \left(\mu_{\sigma ,o}-1\right)^{2} \left(\mu_{\sigma ,o}+1\right)^{2}} \bigg[M^{2} \chi^{2} r_{s}^{2} \left(M r_{s}-\frac{b^{2}}{4}\right) \mu_{\sigma ,o}^{8}
\nonumber 
\\ 
&+\left(9 M^{3} \chi^{2} r_{s}^{3}+M^{2} b^{2} \left(\frac{\chi^{2}}{4}-16\right) r_{s}^{2}+4 M \,b^{4} r_{s}-\frac{b^{6}}{4}\right) \mu_{\sigma ,o}^{6}
\nonumber 
\\ 
&+\left(\left(-66 \chi^{2}+992\right) M^{3} r_{s}^{3}+40 M^{2} b^{2} r_{s}^{2}-10 M \,b^{4} r_{s}+\frac{3 b^{6}}{4}\right) \mu_{\sigma ,o}^{4}
\nonumber 
\\ 
&+\left(\left(104 \chi^{2}-1984\right) M^{3} r_{s}^{3}-48 M^{2} b^{2} r_{s}^{2}+12 M \,b^{4} r_{s}-b^{6}\right) \mu_{\sigma ,o}^{2}-48 M^{3} \left(\chi^{2}-\frac{64}{3}\right) r_{s}^{3}\bigg]
\nonumber 
\\ 
&+ \mathcal{O}(\epsilon^6) \, .
\end{align}
This intermediate result is important for the evaluation of all subsequent integrals.

\subsection{Evaluation of integral \eqref{geo_azimuth}}

We begin with the radial integral. The large-$r$ fall-off of the integral is sufficiently fast that the integral can be re-written as
\be 
\int_{\rm P} \frac{a\left(2Mr-a\lambda\right) dr}{\pm \Delta \sqrt{R}} = 2 \int_{r_{\rm min}}^\infty \frac{a\left(2Mr-a\lambda\right) dr}{ \Delta \sqrt{R}} - \int_{r_s}^\infty \frac{a\left(2Mr-a\lambda\right) dr}{\Delta \sqrt{R}} \, .
\ee
The second integral above is performed by working in a large $r$ series and integrating term-by-term. The result of this, up to fifth order in $\epsilon$, is
\begin{align}
\int_{r_s}^\infty \frac{a\left(2Mr-a\lambda\right) dr}{\Delta \sqrt{R}} &= \, \frac{a M}{r_{s}^{2}} \epsilon^{4}-\frac{1}{3} \frac{a^{2} \lambda}{r_{s}^{3}} \epsilon^{5}+\mathcal{O}\! \left(\epsilon^{6}\right) \, .
\end{align}
To evaluate the first integral above, we transform the variable of integration to $x  = r_{\rm min}/r$ and isolate the leading-order behaviour of the integrand for small $\epsilon$. We then evaluate,
\begin{align} 
2 \int_{r_{\rm min}}^\infty &\frac{a\left(2Mr-a\lambda\right) dr}{ \Delta \sqrt{R}} = \int_0^1 \frac{2 r_{\rm min}}{x^2 \sqrt{R_{\rm lead}}} \left[-\frac{a \left(a x \lambda -2 b M \right) x}{b^{2}} + \cdots \right] \, ,
\\
&= \frac{a \left(-\pi  a \lambda +8 b M \right) \epsilon^{2}}{2 b^{3}}+\frac{a \left(5 M \pi  b -8 a \lambda \right) M \epsilon^{3}}{b^{4}}
\nonumber\\
&+\frac{a \left(-285 M^{2} \pi  a b^{2} \lambda +9 \pi  a^{3} b^{2} \lambda -15 \pi  a^{3} \lambda^{3}+1024 b^{3} M^{3}-64 a^{2} b^{3} M +256 a^{2} b \lambda^{2} M \right) \epsilon^{4}}{16 b^{7}}
\nonumber\\
&+\frac{a \left(2079 M^{3} \pi  b^{3}-243 M \pi  a^{2} b^{3}+1215 M \pi  a^{2} b \lambda^{2}-8192 M^{2} a b^{2} \lambda +384 a^{3} b^{2} \lambda -768 a^{3} \lambda^{3}\right) M \epsilon^{5}}{24 b^{8}} 
\nonumber\\
&+ \mathcal{O}(\epsilon^6) \, .
\end{align}
It is easy to work out higher order terms, but they become increasingly complicated.

We now turn to the evaluation of the angular integral. As before, the integrand can be massaged into the much simpler form
\be 
\int_{\rm P} \frac{\lambda d\theta}{\pm \sin^2\theta \sqrt{\Theta}} = \int_{\rm P} \frac{\lambda \, du}{a \mu_+ (1- \mu_+^2 u^2) \sqrt{\left(1-u^2\right)\left(u^2 + \frac{\mu_-^2}{\mu_+^2} \right)}}
\ee
which admits an anti-derivative,
\be 
\int \frac{\lambda \, du}{a \mu_+ (1- \mu_+^2 u^2) \sqrt{\left(1-u^2\right)\left(u^2 + \frac{\mu_-^2}{\mu_+^2} \right)}} =  \frac{\lambda}{a \mu_-} \Pi \left(\mu_+^2; \arcsin u  \bigg | -\frac{\mu_+^2}{\mu_-^2} \right) \, ,
\ee
where $\Pi(n; \phi | m)$ is the incomplete elliptic integral of the third kind.  In terms of this result, the exact angular integral takes the form
\begin{align}
\int_{\rm P} \frac{\lambda d\theta}{\pm \sin^2\theta \sqrt{\Theta}} = \frac{\lambda}{a \mu_-} \bigg\{ &2 \Pi \left(\mu_+^2; \frac{\pi}{2}  \bigg | -\frac{\mu_+^2}{\mu_-^2} \right) 
\nonumber\\
&  + \pm_o \left[\Pi \left(\mu_+^2; \arcsin \frac{\mu_s}{\mu_+}  \bigg | -\frac{\mu_+^2}{\mu_-^2} \right) + \Pi \left(\mu_+^2; \arcsin \frac{\mu_o}{\mu_+} \bigg | -\frac{\mu_+^2}{\mu_-^2} \right) \right] \bigg\} \, .
\end{align}
 These integrals can be expanded in a small $\epsilon$ expansion to give the following result
\begin{align}
\int_{\rm P} \frac{\lambda d\theta}{\pm \sin^2\theta \sqrt{\Theta}}  =& \pi \,  {\rm sign}(\lambda) +\frac{\pi  \,a^{2} \lambda  \,\epsilon^{2}}{2 b^{3}}  -\frac{9 a^{4} \left(3 b^{2}-5 \lambda^{2}\right) \pi  \lambda  \,\epsilon^{4}}{48 b^{7}} + \pm_o \bigg[\arctan \! \left(\frac{\lambda  \mu_{\sigma ,s}}{b \sqrt{1-\mu_{\sigma ,s}^{2}}\, }\right) 
\nonumber\\
&-\frac{\left(-\arcsin \! \left(\mu_{\sigma ,s}\right) \sqrt{1-\mu_{\sigma ,s}^{2}}+\mu_{\sigma ,s}\right) a^{2} \lambda  \,\epsilon^{2}}{2 b^{3} \sqrt{1-\mu_{\sigma ,s}^{2}}} 
\nonumber\\
&-\frac{3  \lambda  \,a^{4} \left(\mu_{\sigma ,s}^{3}+3 \arcsin \! \left(\mu_{\sigma ,s}\right) \sqrt{1-\mu_{\sigma ,s}^{2}}-3 \mu_{\sigma ,s}\right)\epsilon^{4}}{16 \sqrt{1-\mu_{\sigma ,s}^{2}}\, b^{5}}  
\nonumber\\
&-\frac{3 \left(15 \left(\mu_{\sigma ,s}^{2}-1\right) \arcsin \! \left(\mu_{\sigma ,s}\right) \sqrt{1-\mu_{\sigma ,s}^{2}}+3 \mu_{\sigma ,s}^{5}- 20 \mu_{\sigma ,s}^{3}+15 \mu_{\sigma ,s}\right) a^{4} \epsilon^{4} \lambda^{3}}{48 \left(1-\mu_{\sigma ,s}^{2}\right)^{\frac{3}{2}} b^{7}}
\nonumber\\
&+ \left(\mu_{\sigma, s} \rightarrow \mu_{\sigma, o} \right) \bigg] + \mathcal{O}(\epsilon^6) \, ,
\end{align}
where $\left(\mu_{\sigma, s} \rightarrow \mu_{\sigma, o} \right)$ indicates that all terms that appear with instances of $\mu_{\sigma, s}$ should be replicated with $\mu_{\sigma, s}$ replaced by $\mu_{\sigma, o}$. It is then necessary to substitute for $\mu_{\sigma,s}$ from the results Eq.~\eqref{mus_perturb}. Finally, this can be combined with the radial integrals above to give the final answer for the azimuthal integral, 
\begin{align}
    -\phi_s =& \, \pi \,  {\rm sign}(\lambda) - \frac{\lambda \left(b^2 - 4 M r_s \right)}{\left((1-\mu_{\sigma, o}^2) b^2 + \mu_{\sigma, o}^2 \lambda^2 \right) r_s}  -\frac{ \sigma b  \mu_{\sigma ,o} \lambda  \sqrt{-\mu_{\sigma ,o}^{2}+1}\, \left(b -\lambda \right) \left(b +\lambda \right) \left(b^{2}-8 r_{s} M \right)}{\left(\left(\mu_{\sigma ,o}^{2}-1\right) b^{2}-\lambda^{2} \mu_{\sigma ,o}^{2}\right)^{2} r_{s}^{2}} 
    \nonumber 
    \\
    &+\frac{M^2 \left[16\left(\mu_{\sigma ,o}^{2}- 1\right) \chi b^{2}-15\pi  b  \lambda-16 \chi  \,\lambda^{2} \left(\mu_{\sigma ,o}^{2}-2\right)\right]}{4\left(\left(\mu_{\sigma ,o}^{2}-1\right) b^{2}-\lambda^{2} \mu_{\sigma ,o}^{2}\right) b^{2}}  -\frac{16 M^{2} \sigma    \mu_{\sigma ,o} \sqrt{1-\mu_{\sigma ,o}^{2}}\, \lambda \left(b^{2}-\lambda^{2}\right)}{b \left(\left(\mu_{\sigma ,o}^{2}-1\right) b^{2}-\lambda^{2} \mu_{\sigma ,o}^{2}\right)^{2}}
    \nonumber 
    \\
    &-\frac{b^{4} \lambda  \left[3\left(\mu_{\sigma ,o}^{4}-1\right) b^{4}-2 \left(3\mu_{\sigma ,o}^{4}-3\mu_{\sigma ,o}^{2}-1\right) \lambda^{2} b^{2}+3\mu_{\sigma ,o}^{2} \left(\mu_{\sigma ,o}^{2}-2\right) \lambda^{4}\right]}{6 \left(\left(\mu_{\sigma ,o}^{2}-1\right) b^{2}-\lambda^{2} \mu_{\sigma ,o}^{2}\right)^{3} r_{s}^{3}} 
    \nonumber 
    \\
    &+ \frac{4 M \lambda  \left(b^2 -\lambda^2 \right)\left(b^{2}-4 r_{s} M \right) \left[\left(2\mu_{\sigma ,o}^{4}- \mu_{\sigma ,o}^{2}-1\right) b^{2}-\mu_{\sigma ,o}^{2} \lambda^{2} \left(2\mu_{\sigma ,o}^{2}-3\right)\right] }{r_{s}^{2} \left(\left(\mu_{\sigma ,o}^{2}-1\right) b^{2}-\lambda^{2} \mu_{\sigma ,o}^{2}\right)^{3}}
    \nonumber
    \\ 
    &+\frac{\pm_o \lambda \, M^{2}  \mu_{\sigma ,o}  \sqrt{1-\mu_{\sigma ,o}^{2}} \left(b^2-4 M r_{s}\right) \left(15 \pi  b -32 \chi  \lambda \right) \left(b^{2}-\lambda^{2}\right)}{2 b^{3} \left(b^{2} \mu_{\sigma ,o}^{2}-\lambda^{2} \mu_{\sigma ,o}^{2}-b^{2}\right)^{2} r_{s}}
    \nonumber 
    \\ 
    &+\frac{64 \left(3\left(\mu_{\sigma ,o}^{2}-1\right) b^{4}+b^{2} \lambda^{2}-3\mu_{\sigma ,o}^{2} \lambda^{4}\right) \lambda  \,M^{3}}{3b^{2} \left(\left(\mu_{\sigma ,o}^{2}-1\right) b^{2}-\lambda^{2} \mu_{\sigma ,o}^{2}\right)^{3}} +  \frac{5 \pi \chi  \,M^{3} \left(\left(\mu_{\sigma ,o}^{2}-1\right) b^{2}-\lambda^{2} \left(\mu_{\sigma ,o}^{2}-3\right)\right) }{b^{3} \left(\left(\mu_{\sigma ,o}^{2}-1\right) b^{2}-\lambda^{2} \mu_{\sigma ,o}^{2}\right)}
    \nonumber 
    \\ 
    &-\frac{4 M^{3} \chi^{2} \lambda  \left(2 b^{2} \mu_{\sigma ,o}^{2}-2 \lambda^{2} \mu_{\sigma ,o}^{2}-3 b^{2}+4 \lambda^{2}\right)}{b^{4} \left(\left(\mu_{\sigma ,o}^{2}-1\right) b^{2}-\lambda^{2} \mu_{\sigma ,o}^{2}\right)} + \mathcal{O}(\epsilon^4) \, .
\end{align}


\subsection{Evaluation of integral \eqref{geo_temporal}}

In this case, the integral is not convergent at large $r$, but this is easily resolved by changing the integral from $t$ to the retarded time $u$,
\be 
u_o - u_s = r_s - r_o + 2 M \log \frac{r_s}{r_o} + \int_{\rm P} \frac{\left[r^2\left(r^2+a^2\right) + 2 a M r \left(a-\lambda\right)\right] dr}{\pm \Delta \sqrt{R}} + \int_{\rm P} \frac{a^2 \cos^2\theta}{\pm \sqrt{\Theta}} d\theta \, .
\ee
Then we can further manipulate the radial part of the expression to give
\begin{align}
u_o - u_s =&\,  \int_{\rm P} \left\{ \frac{\left[r^2\left(r^2+a^2\right) + 2 a M r \left(a-\lambda\right)\right]}{\pm \Delta \sqrt{R}} - 1 - \frac{2M}{r} \right\} dr + 2 r_s - 2 r_{\rm min} + 2 M \log \frac{r_s^2}{r_{\rm min}^2} 
\nonumber
\\
&+ \int_{\rm P} \frac{a^2 \cos^2\theta}{\pm \sqrt{\Theta}} d\theta \, .
\end{align}
To arrive at this we have added and subtracted the integral of $(1 + 2M/r)$ along the path. This makes the remaining radial integral presented above convergent at large $r$ and allows it to be treated in the same manner we have treated the preceding ones. Namely, we split the integral up as 
\begin{align}
\int_{\rm P} \left\{ \frac{\left[r^2\left(r^2+a^2\right) + 2 a M r \left(a-\lambda\right)\right]}{\pm \Delta \sqrt{R}} - 1 - \frac{2M}{r} \right\} dr &= 2 \int_{r_{\rm min}}^\infty \left\{ \cdots \right\} dr - \int_{r_s}^\infty \left\{ \cdots \right\} dr \, ,
\end{align}
where we have used $\cdots$ to avoid unnecessary clutter in the repetition of the integrand. As before, the second integral is evaluated by expanding the integrand in a large-$r$ series and integrating term-by-term. This produces the result,
\begin{align}
&\int_{r_s}^\infty \left\{\frac{\left[r^2\left(r^2+a^2\right) + 2 a M r \left(a-\lambda\right)\right]}{\pm \Delta \sqrt{R}} - 1 - \frac{2M}{r} \right\} dr = \frac{b^{2}}{2 r_{s}}-\left(\frac{-32 M^{2} r_{s}^{2}+4 a^{2} r_{s}^{2}-b^{4}}{8 r_{s}^{3}} \right)\epsilon^2
\nonumber
\\
&\quad\quad + \left(\frac{b^{6}}{16 r_{s}^{5}}-\frac{3 M \,b^{4}}{16 r_{s}^{4}}-\frac{a^{2} b^{2}}{12 r_{s}^{3}}+\frac{24 M^{3} r_{s}-6 M \,a^{2} r_{s}-a^{2} \lambda^{2}}{6 r_{s}^{3}} \right)\epsilon^4 + \frac{a \,b^{2} M \lambda  \,\epsilon^{5}}{2 r_{s}^{4}} + \mathcal{O}(\epsilon^6) \, .
\end{align}

Let's now turn to the remaining radial integral. We transform the integration variable to $x = r_{\rm min}/r$ and then isolate the leading behaviour of the integrand. We have to evaluate the following integral, which yields
\begin{align} 
2 \int_{r_{\rm min}}^\infty &\left\{\frac{\left[r^2\left(r^2+a^2\right) + 2 a M r \left(a-\lambda\right)\right]}{\pm \Delta \sqrt{R}} - 1 - \frac{2M}{r} \right\}dr = \int_0^1 \frac{2 r_{\rm min}\, dx}{x^2 \sqrt{R_{\rm lead}}} \left[-\frac{\left(-1+\sqrt{-x^{2}+1}\right) b^{2}}{x^{2}} + \cdots \right]
\nonumber\\
&= 2 r_{\rm min}  + 4 M \log \frac{2 r_{\rm min}}{b} + 2 M +\frac{\left(\pi  \left(15 M^{2}-a^{2}\right) b^{2}-16 a b \lambda  M +a^{2} \lambda^{2} \pi \right) \epsilon}{2 b^{3}}
\nonumber
\\
&+\frac{\left(-15 M \pi  a b \lambda +64 M^{2} b^{2}-6 a^{2} b^{2}+12 a^{2} \lambda^{2}\right) M \,\epsilon^{2}}{b^{4}} +  \frac{\pi  \left(1155 M^{4}-190 a^{2} M^{2}+3 a^{4}\right) \epsilon^{3}}{16 b^{3}}
\nonumber
\\
&-\frac{16 M a \lambda  \left(4 M -a \right) \left(4 M +a \right) \epsilon^{3}}{b^{4}}+\frac{3 \lambda^{2} \left(95 M^{2}-3 a^{2}\right) a^{2} \pi  \,\epsilon^{3}}{8 b^{5}}-\frac{64 a^{3} \lambda^{3} M \,\epsilon^{3}}{3 b^{6}}+\frac{15 a^{4} \lambda^{4} \pi  \,\epsilon^{3}}{16 b^{7}} 
\nonumber
\\
&+ \frac{\left(2688 M^{4}-640 a^{2} M^{2}+15 a^{4}\right) M \,\epsilon^{4}}{3 b^{4}}-\frac{45 M^{2} \lambda  a \pi  \left(77 M^{2}-9 a^{2}\right) \epsilon^{4}}{8 b^{5}}
\nonumber
\\
&+\frac{40 \lambda^{2} \left(64 M^{2}-3 a^{2}\right) a^{2} M \,\epsilon^{4}}{3 b^{6}}-\frac{675 a^{3} \lambda^{3} \pi  \,M^{2} \epsilon^{4}}{8 b^{7}}+\frac{40 a^{4} \lambda^{4} M \,\epsilon^{4}}{b^{8}} 
\nonumber
\\
&+ \frac{3 \pi  \left(51051 M^{6}-15939 M^{4} a^{2}+717 M^{2} a^{4}-5 a^{6}\right)\epsilon^{5}}{128 b^{5}}-\frac{8 M \lambda  \left(896 M^{4}-160 a^{2} M^{2}+3 a^{4}\right) a \,\epsilon^{5}}{b^{6}}
\nonumber
\\
&+\frac{45 \left(5313 M^{4}-478 a^{2} M^{2}+5 a^{4}\right) \lambda^{2} a^{2} \pi \,\epsilon^{5}}{128 b^{7}}-\frac{32 M \,\lambda^{3} a^{3} \left(80 M^{2}-3 a^{2}\right)\epsilon^{5}}{b^{8}}
\nonumber
\\
&+\frac{105 \lambda^{4} \left(239 M^{2}-5 a^{2}\right) a^{4} \pi \,\epsilon^{5}}{128 b^{9}}-\frac{384 a^{5} \lambda^{5} M \epsilon^{5}}{5 b^{10}}+\frac{315 \lambda^{6} a^{6} \pi \,\epsilon^{5}}{128 b^{11}} + \mathcal{O}(\epsilon^6)
\end{align}
which is accurate up to $\mathcal{O}(\epsilon^6)$ corrections. It is easy to compute higher-order corrections, but they become increasingly complicated.

Let us evaluate the angular integral next. After performing the standard transformations, the angular integral can be expressed in the following form
\be 
\int_{\rm P} \frac{a^2 \cos^2\theta}{\pm \sqrt{\Theta}} d\theta = a \mu_+ \int_{\rm P} \frac{u^2 du}{\sqrt{\left(1-u^2\right) \left(u^2 + \frac{\mu_-^2}{\mu_+^2} \right)}} \, .
\ee
This integral can be expressed in terms of elliptic functions,
\be 
\int \frac{u^2 du}{\sqrt{\left(1-u^2\right) \left(u^2 + \frac{\mu_-^2}{\mu_+^2} \right)}} = a \mu_- \left[{\rm E} \left(\arcsin u \bigg| - \frac{\mu_+^2}{\mu_-^2} \right) - {\rm F} \left(\arcsin u \bigg| - \frac{\mu_+^2}{\mu_-^2} \right)\right] \, ,
\ee
where $F(\phi|m)$ is the elliptic integral of the first kind and $E(\phi|m)$ is the elliptic integral of the second kind. This allows for full integral along the path to be expressed as
\begin{align}
\int_{\rm P}  \frac{a^2 \cos^2\theta}{\pm \sqrt{\Theta}} d\theta =&\, a \mu_- \bigg\{2 {\rm E} \left( \frac{\pi}{2} \bigg| - \frac{\mu_+^2}{\mu_-^2} \right) - 2{\rm F} \left(\frac{\pi}{2} \bigg| - \frac{\mu_+^2}{\mu_-^2} \right) 
\nonumber
\\
&+ \pm_o \bigg[{\rm E} \left(\arcsin \frac{\mu_s}{\mu_+} \bigg| - \frac{\mu_+^2}{\mu_-^2} \right) - {\rm F} \left(\arcsin \frac{\mu_s}{\mu_+} \bigg| - \frac{\mu_+^2}{\mu_-^2} \right) 
\nonumber\\
&+ {\rm E} \left(\arcsin \frac{\mu_o}{\mu_+} \bigg| - \frac{\mu_+^2}{\mu_-^2} \right) - {\rm F} \left(\arcsin \frac{\mu_o}{\mu_+} \bigg| - \frac{\mu_+^2}{\mu_-^2} \right)\bigg] \bigg\} \, .
\end{align}
 The elliptic functions can be expanded at small $\epsilon$ to give the following result
\begin{align}
\int_{\rm P}  &\frac{a^2 \cos^2\theta}{\pm \sqrt{\Theta}} d\theta =\, \frac{\left(b^{2}-\lambda^{2}\right) a^{2} \pi  \epsilon}{2 b^{3}}-\frac{3 \left(b^{2}-\lambda^{2}\right) a^{4} \pi  \left(b^{2}-5 \lambda^{2}\right) \epsilon^{3}}{16 b^{7}}
\nonumber
\\
&+\frac{15 \left(b^{2}-\lambda^{2}\right) a^{6} \pi  \left(b^{4}-14 b^{2} \lambda^{2}+21 \lambda^{4}\right) \epsilon^{5}}{128 b^{11}} + \pm_o \bigg[-\frac{\left(b^{2}-\lambda^{2}\right) a^{2} \left(\mu_{\sigma ,s} \sqrt{-\mu_{\sigma ,s}^{2}+1}-\arcsin \! \left(\mu_{\sigma ,s}\right)\right) \epsilon}{2 b^{3}} 
\nonumber
\\
&-\frac{\left(b^{2}-\lambda^{2}\right)^{2} a^{4} \epsilon^{3} \mu_{\sigma ,s}^{5}}{8 \sqrt{-\mu_{\sigma ,s}^{2}+1}\, b^{7}}-\frac{\left(b^{2}-5 \lambda^{2}\right) a^{4} \left(b^{2}-\lambda^{2}\right) \epsilon^{3} \mu_{\sigma ,s}^{3}}{16 \sqrt{-\mu_{\sigma ,s}^{2}+1}\, b^{7}}+\frac{3 \left(b^{2}-5 \lambda^{2}\right) a^{4} \left(b^{2}-\lambda^{2}\right) \epsilon^{3} \mu_{\sigma ,s}}{16 \sqrt{-\mu_{\sigma ,s}^{2}+1}\, b^{7}}
\nonumber
\\
&-\frac{3 \arcsin \! \left(\mu_{\sigma ,s}\right) \left(b^{2}-5 \lambda^{2}\right) a^{4} \left(b^{2}-\lambda^{2}\right) \epsilon^{3}}{16 b^{7}} -\frac{\left(b -\lambda \right)^{3} \left(b +\lambda \right)^{3} a^{6} \epsilon^{5} \mu_{\sigma ,s}^{9}}{16 \left(-\mu_{\sigma ,s}^{2}+1\right)^{\frac{3}{2}} b^{11}}
\nonumber
\\
&+\frac{3 \left(b -\lambda \right)^{2} \left(b +\lambda \right)^{2} \left(b^{2}+3 \lambda^{2}\right) a^{6} \epsilon^{5} \mu_{\sigma ,s}^{7}}{64 \left(-\mu_{\sigma ,s}^{2}+1\right)^{\frac{3}{2}} b^{11}}-\frac{3 \left(b^{4}-14 b^{2} \lambda^{2}+21 \lambda^{4}\right) a^{6} \epsilon^{5} \left(b^{2}-\lambda^{2}\right) \mu_{\sigma ,s}^{5}}{128 \left(-\mu_{\sigma ,s}^{2}+1\right)^{\frac{3}{2}} b^{11}}
\nonumber
\\
&+\frac{5 \left(b^{4}-14 b^{2} \lambda^{2}+21 \lambda^{4}\right) a^{6} \epsilon^{5} \left(b^{2}-\lambda^{2}\right) \mu_{\sigma ,s}^{3}}{32 \left(-\mu_{\sigma ,s}^{2}+1\right)^{\frac{3}{2}} b^{11}}-\frac{15 \left(b^{4}-14 b^{2} \lambda^{2}+21 \lambda^{4}\right) a^{6} \epsilon^{5} \left(b^{2}-\lambda^{2}\right) \mu_{\sigma ,s}}{128 \left(-\mu_{\sigma ,s}^{2}+1\right)^{\frac{3}{2}} b^{11}}
\nonumber
\\
&
+\frac{15 \arcsin \! \left(\mu_{\sigma ,s}\right) \left(b^{4}-14 b^{2} \lambda^{2}+21 \lambda^{4}\right) a^{6} \epsilon^{5} \left(b^{2}-\lambda^{2}\right) \mu_{\sigma ,s}^{2}}{128 \left(\mu_{\sigma ,s}^{2}-1\right) b^{11}}
\nonumber
\\
&-\frac{15 \arcsin \! \left(\mu_{\sigma ,s}\right) \left(b^{4}-14 b^{2} \lambda^{2}+21 \lambda^{4}\right) a^{6} \epsilon^{5} \left(b^{2}-\lambda^{2}\right)}{128 \left(\mu_{\sigma ,s}^{2}-1\right) b^{11}}+ \left(\mu_{\sigma, s} \rightarrow \mu_{\sigma, o} \right) \bigg] + \mathcal{O}(\epsilon^6) \, ,
\end{align}
where $\left(\mu_{\sigma, s} \rightarrow \mu_{\sigma, o} \right)$ indicates that all terms that appear with instances of $\mu_{\sigma, s}$ should be replicated with $\mu_{\sigma, s}$ replaced by $\mu_{\sigma, o}$. It is then necessary to substitute for $\mu_{\sigma,s}$ from the results Eq.~\eqref{mus_perturb}. Finally, this can be combined with the radial integrals above to give the final answer for the retarded time integral, 
\begin{align}
    -u_s =&\, 2 r_s + 4 M \log \left( \frac{2 r_s}{b} \right) + \frac{4 r_s M - b^2}{2 r_s} +\frac{15 M^{2} \pi}{2 b} -\frac{8 \chi  M^{2} \lambda}{b^{2}} + \frac{2 \left(2 b^{2} \mu_{o}^{2}-3 b^{2}+6 \lambda^{2}\right) \chi^{2} M^{3}}{b^{4}}
    \nonumber
    \\
    &-\frac{\left(2 \mu_{o}^{2}-1\right) \chi^{2} M^{2}}{2 r_{s}} -\frac{15 \lambda  \pi  \,M^{3} \chi}{b^{3}}+\frac{64 M^{3}}{b^{2}}-\frac{4 M^{2}}{r_{s}}-\frac{b^{4}}{8 r_{s}^{3}} 
    \nonumber
    \\
    &-\frac{\pm_o \, 16 \sqrt{b^2(1- \mu_{o}^{2})-\lambda^{2}}\, M^{2} \chi^{2} \left(r_{s} M -\frac{b^{2}}{4}\right)^{2} \mu_{o}}{b^{4} r_{s}^{2}} + \frac{\left(-24 \lambda  \left(\mu_{o}^{2}-2\right) M^{4} b^{2}-64 \lambda^{3} M^{4}\right) \chi^{3}}{3 b^{6}}
    \nonumber
    \\
    &+\frac{5 M^{4} \pi  \left(6 b^{2} \mu_{o}^{2}-19 b^{2}+57 \lambda^{2}\right) \chi^{2}}{8 b^{5}}-\frac{256 \lambda  \,M^{4} \chi}{b^{4}}+\frac{1155 \pi  \,M^{4}}{16 b^{3}} 
    \nonumber
    \\
    &-\frac{\pm_o \, 30 \chi^{2} \left(r_{s} M -\frac{b^{2}}{4}\right) \mu_{o} M^{4} \sqrt{-b^{2} \mu_{o}^{2}+b^{2}-\lambda^{2}}\, \left(\pi  b -\frac{32 \chi  \lambda}{15}\right)}{b^{6} r_{s}} 
    \nonumber
    \\
    &-\frac{4 M^{5} \left(\left(\mu_{o}^{2}-\frac{5}{4}\right) b^{4}+\left(-4 \mu_{o}^{2}+10\right) \lambda^{2} b^{2}-10 \lambda^{4}\right) \chi^{4}}{b^{8}}-\frac{15 M^{5} \pi  \lambda  \left(8 b^{2} \mu_{o}^{2}-27 b^{2}+45 \lambda^{2}\right) \chi^{3}}{8 b^{7}}
    \nonumber
    \\
    &-\frac{64 M^{5} \chi^{2} \left(3 b^{2}-13 \lambda^{2}\right)}{b^{6}}+\frac{16 M^{4} \chi^{2} \left(2 b^{2} \mu_{o}^{2}-b^{2}+\lambda^{2}\right)}{b^{4} r_{s}}-\frac{M^{3} \chi^{2} \left(8 b^{2} \mu_{o}^{2}-5 b^{2}+4 \lambda^{2}\right)}{b^{2} r_{s}^{2}}
    \nonumber
    \\
    &+\frac{M^{2} \chi^{2} \left(2 b^{2} \mu_{o}^{2}-b^{2}+2 \lambda^{2}\right)}{4 r_{s}^{3}}-\frac{3465 M^{5} \pi  \lambda  \chi}{8 b^{5}}+\frac{896 M^{5}}{b^{4}}-\frac{4 M^{3}}{r_{s}^{2}}+\frac{3 M \,b^{4}}{16 r_{s}^{4}}-\frac{b^{6}}{16 r_{s}^{5}} 
    \nonumber
    \\
    &+ \frac{\pm_o \, M^{2}}{16 r_{s}^{3} b^{8} \sqrt{-b^{2} \mu_{o}^{2}+b^{2}-\lambda^{2}}\, } \bigg[ -225 \chi^{2} \left(b +\lambda \right) M \left(b -\lambda \right) \mu_{o} \bigg( \frac{64 b^{8}}{225}-\frac{512 b^{6} M r_{s}}{225}+\frac{128 M^{2} b^{4} \chi^{2} r_{s}^{2}}{225}
    \nonumber
    \\
    &+\frac{32 M^{2} \pi  \,b^{3} \chi  \lambda  r_{s}^{2}}{15}+M^{2} \left(r_{s} \left(\pi^{2}-\frac{512 \chi^{2}}{225}+\frac{4096}{225}\right) M -\frac{512 \chi^{2} \lambda^{2}}{225}\right) r_{s}^{2} b^{2}-\frac{64 M^{3} \pi  b \chi  \lambda  r_{s}^{3}}{5}
    \nonumber
    \\
    &+\frac{1024 M^{3} \chi^{2} \lambda^{2} r_{s}^{3}}{75}\bigg) + b^{2} \chi^{2} M \mu_{o}^{3} \bigg(  64 b^{8}-8 r_{s} b^{6} \left(\chi^{2}+64\right) M +192 M^{2} b^{4} \chi^{2} r_{s}^{2}+480 M^{2} \pi  \,b^{3} \chi  \lambda  r_{s}^{2}
    \nonumber
    \\
    &+225 M^{2} \left(M \left(\pi^{2}-\frac{128 \chi^{2}}{45}+\frac{4096}{225}\right) r_{s}-\frac{512 \chi^{2} \lambda^{2}}{225}\right) r_{s}^{2} b^{2}-2880 M^{3} \pi  b \chi  \lambda  r_{s}^{3}+3072 M^{3} \chi^{2} \lambda^{2} r_{s}^{3}\bigg)
    \nonumber
    \\
    &+128 \left(r_{s} M -\frac{b^{2}}{4}\right)^{2} r_{s} \chi^{4} M^{2} b^{4} \mu_{o}^{5}\bigg] -\frac{3 M^{6} \left(380 \chi^{4} \mu_{o}^{2}-717 \chi^{4}+2810 \mu_{o}^{2} \chi^{2}+13379 \chi^{2}-51051\right) \pi}{128 b^{5}}
    \nonumber
    \\
    &+\frac{16 \lambda  \,M^{6} \left(\left(\mu_{o}^{2}-\frac{3}{2}\right) \chi^{4}+72 \chi^{2}-448\right) \chi}{b^{6}}+\frac{15 \lambda^{2} M^{6} \pi  \,\chi^{2} \left(380 \mu_{o}^{2} \chi^{2}-1434 \chi^{2}+15427\right)}{128 b^{7}}
    \nonumber
    \\
    &-\frac{32 \lambda^{3} M^{6} \chi^{3} \left(\left(\mu_{o}^{2}-3\right) \chi^{2}+76\right)}{b^{8}}+\frac{25095 \pi  \,\chi^{4} \lambda^{4} M^{6}}{128 b^{9}}-\frac{384 \chi^{5} \lambda^{5} M^{6}}{5 b^{10}}
    \nonumber
    \\
    &+\frac{1}{r_s} \bigg[\frac{60 \chi^{2} M^{5} \pi  \left(\mu_{o}^{2}-\frac{1}{2}\right)}{b^{3}}-\frac{64 \chi^{3} M^{5} \lambda  \left(2 \mu_{o}^{2}-1\right)}{b^{4}}+\frac{30 \chi^{2} M^{5} \pi  \,\lambda^{2}}{b^{5}}-\frac{64 \chi^{3} M^{5} \lambda^{3}}{b^{6}}\bigg]
    \nonumber
    \\
    &+\frac{1}{r_{s}^{2}} \bigg[-\frac{15 \chi^{2} \pi  \left(\mu_{o}^{2}-\frac{1}{2}\right) M^{4}}{2 b}+\frac{8 \chi^{3} \lambda  \left(2 \mu_{o}^{2}-1\right) M^{4}}{b^{2}}-\frac{15 \chi^{2} \pi  \,\lambda^{2} M^{4}}{4 b^{3}}+\frac{8 \chi^{3} \lambda^{3} M^{4}}{b^{4}} \bigg]-\frac{\chi  \,M^{2} \lambda  \,b^{2}}{2 r_{s}^{4}}
    \nonumber
    \\
    &+ \mathcal{O}(\epsilon^6).
\end{align}

\section{Higher-order perturbative results for orthogonal mergers}
\label{app:orthog_pert}

In the case of an orthogonal merger, i.e.~with $\mu_o = 0$, the evaluation of the perturbative integrals dramatically simplifies and it is possible to go to very high order with the analysis. In practice, we have evaluated the Boyer-Lindquist coordinates of the crease up to and including $\mathcal{O}(\varepsilon^{13})$ corrections in the time function, with $(\phi_s, \mu_s)$ known to one higher order. The resulting expressions are rather complicated. Therefore, rather than presenting them here we include them in an ancillary Mathematica notebook and discuss some relevant results that can be efficiently presented.

In the analysis of the main text, the $A_3$ caustics corresponded to $\varphi = 0, \pi/2, \pi$, and $3\pi/2$, with $\varphi = \pi/2$ and  $3\pi/2$ corresponding to the $A_3$ caustics that belong to the horizon. While higher-order corrections do not displace the $A_3$ caustics that occur at $\varphi = 0$ and $\pi$, the remaining two $A_3$ caustics are shifted. Here we present, for the case of the orthogonal merger, higher-order corrections to the equation governing this displacement:
\begin{align}\label{caustic_shift}
    \cos \varphi =&\, \left(-\frac{1}{6} \chi^{3}-\frac{75}{32} \chi  \,\pi^{2}+\frac{51}{4} \chi \right) \varepsilon^3 + \left(\frac{5}{64} \pi  \,\chi^{3}+\frac{120375}{8192} \chi  \,\pi^{3}-\frac{15285}{128} \chi  \pi \right) \varepsilon^4  + \left(-\frac{85 \chi ^5}{288}  
    \right. 
    \nonumber
    \\
    &\left.-\frac{3575 \pi ^2 \chi ^3}{4096}+\frac{59 \chi ^3}{12}-\frac{18343125 \pi ^4 \chi}{262144}+\frac{5501175 \pi ^2 \chi }{8192}-\frac{1015 \chi }{4} \right) \varepsilon^5 + \left(\frac{875 \pi  \chi ^5}{3072}
    \right. 
    \nonumber
    \\
    &\left.-\frac{81375 \pi ^3 \chi ^3}{32768}+\frac{5825 \pi  \chi
   ^3}{256}+\frac{2530490625 \pi ^5 \chi }{8388608}-\frac{224332875 \pi ^3 \chi }{65536}+\frac{10489185 \pi 
   \chi }{2048} \right) \varepsilon^6 
   \nonumber
   \\
   &+ \left(-\frac{415 \chi ^7}{864}-\frac{268775 \pi ^2 \chi ^5}{196608}+\frac{8173 \chi ^5}{1152}+\frac{1171456875 \pi
   ^4 \chi ^3}{16777216}-\frac{2765975 \pi ^2 \chi ^3}{4096}+\frac{494893 \chi ^3}{1536}
   \right. 
    \nonumber
    \\
    &\left.-\frac{666400921875
   \pi ^6 \chi }{536870912}+\frac{550154278125 \pi ^4 \chi }{33554432}-\frac{2798663025 \pi ^2 \chi
   }{65536}+\frac{5947641 \chi }{1024} \right) \varepsilon^7 
   \nonumber 
   \\ 
   &+ \left(\frac{44695 \pi  \chi ^7}{73728}+\frac{782375 \pi ^3 \chi ^5}{1048576}-\frac{190955 \pi  \chi
   ^5}{49152}-\frac{11010853125 \pi ^5 \chi ^3}{16777216}+\frac{7907563875 \pi ^3 \chi
   ^3}{1048576}
   \right. 
    \nonumber
    \\
    &\left.-\frac{207077635 \pi  \chi ^3}{16384}+\frac{42812518359375 \pi ^7 \chi
   }{8589934592}-\frac{2532137540625 \pi ^5 \chi }{33554432}+\frac{143795005875 \pi ^3 \chi
   }{524288}
   \right. 
    \nonumber
    \\
    &\left.-\frac{4166266755 \pi  \chi }{32768}\right) \varepsilon^8 + \mathcal{O}(\varepsilon^9) \, .
\end{align}
We can also translate the above result into values for $(\alpha, \beta)$. The first several orders read
\begin{align}
\frac{\alpha}{M} &= \left(-\frac{1}{6} \chi^{3}-\frac{75}{32} \chi  \,\pi^{2}+\frac{51}{4} \chi\right) \varepsilon^2 + \left(\frac{111375}{8192} \chi  \,\pi^{3}-\frac{1815}{16} \chi  \pi \right) \varepsilon^3 + \mathcal{O}(\varepsilon^4) \, ,
\\ 
\frac{\beta}{M} &= \pm \bigg[ \frac{1}{\varepsilon} + \frac{15 \pi}{32} + \left(-\frac{675 \pi^{2}}{2048}-\frac{\chi^{2}}{2}+4\right)\varepsilon + \left(\frac{3375}{8192} \pi^{3}-\frac{15}{256} \pi  \,\chi^{2}-\frac{2295}{512} \pi\right)\varepsilon^2 
\nonumber
\\
&+ \left(-\frac{5315625}{8388608} \pi^{4}-\frac{14025}{8192} \pi^{2} \chi^{2}+\frac{78525}{16384} \pi^{2}-\frac{7}{24} \chi^{4}+\frac{27}{4} \chi^{2}+16\right) \varepsilon^3 + \mathcal{O}(\varepsilon^4)\bigg] \, .
\end{align}
In particular, note that the value of $\alpha$ is the same for both of the $A_3$ caustics that belong to the horizon, while the $\beta$ values differ only by a sign. 

We have obtained the results above via two independent computations. In the first case, we have worked directly with the caustic tube, as determined by solving for the vanishing of the Jacobian for the coordinates $(\mu_s, \phi_s)$ as functions of the polar impact parameters $(\rho, \varphi)$. To identify the location of the relevant $A_3$ points, we determined the value of $\varphi$ for which the $z$-coordinate along the caustic tube reaches its maximum value. In the second case, we started with the coordinates $(\mu_s, \phi_s)$ restricted to the crease set. Subject to this restriction, we computed the Jacobian and required it to vanish order by order in the $\epsilon$ expansion.  Setting the expansion coefficients of the Jacobian to zero determines $\cos \varphi$ as a series expansion in $\epsilon$. The results of the two methods are in precise agreement.

\begin{figure}[htp]
    \centering
    \includegraphics[width=1\textwidth]{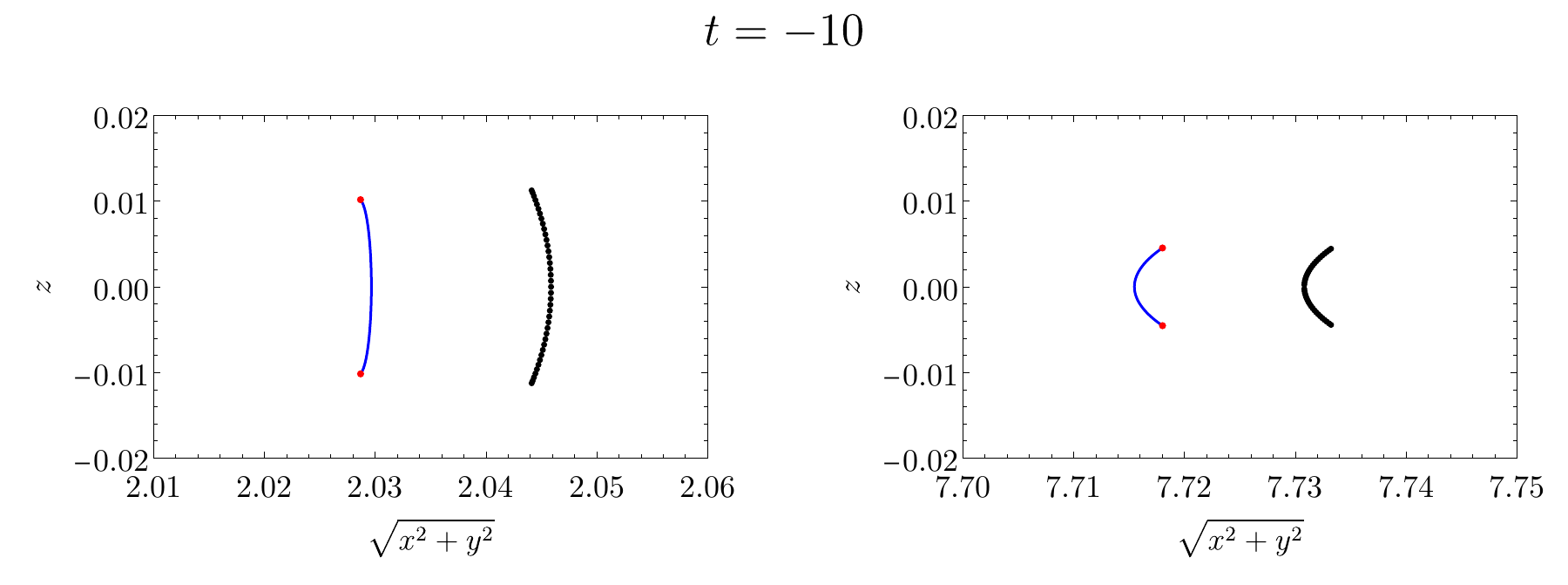}
    \includegraphics[width=0.45\textwidth]{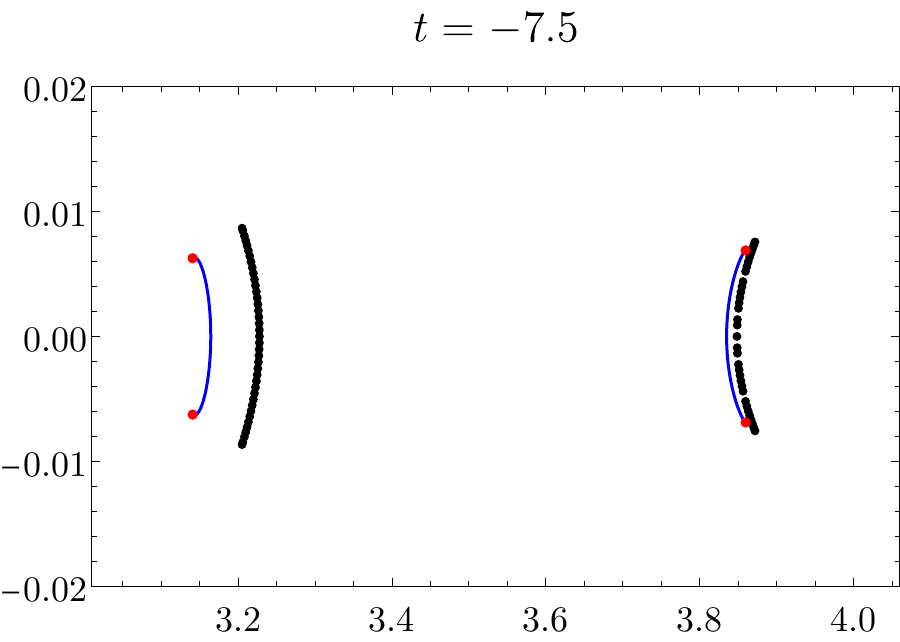}
    \includegraphics[width=0.45\textwidth]{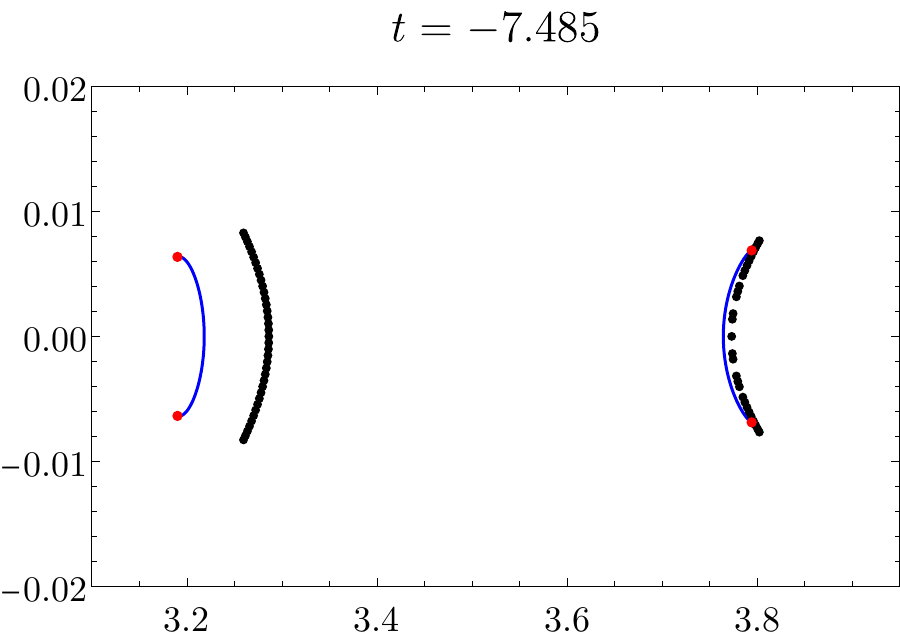}
    \includegraphics[width=0.45\textwidth]{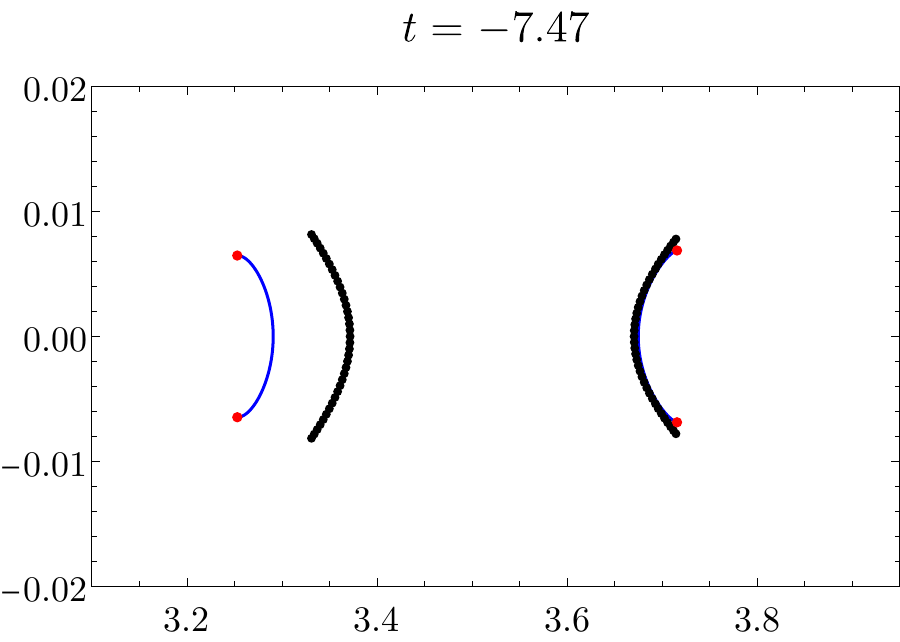}
    \caption{Constant Boyer-Lindquist time-slices of the crease set for an orthogonal merger with a small black hole with $a=1/2$. The blue lines are the creases and the red dots are caustic points obtained from our perturbative results including terms up to $\mathcal{O}(\varepsilon^{13})$. The black dots are the numerical data. Here we show a comparison between the perturbative and numerical results for several values of $t$, illustrating good agreement between the two methods for the small and large black hole. Note the different scale on the $t = -10$ plots relatively to the others.}
    \label{fig:slice_errors}
\end{figure}

A relevant question is how well the perturbative results describe the situation in the vicinity of the small black hole. In the case of the orthogonal merger, we can more easily test this.  One simple measure we use to test the convergence of the series expansions in this vicinity is the calculated time of merger, i.e.~the time at which a time-slice of the horizon transitions from consisting of two components to one component. From the analytical expressions, this time can be computed simply by looking for a maximum of the time function at fixed $\varphi = 0$. We illustrated this comparison in Fig.~\ref{fig:merger_error} as a function of the number of terms included in the series expansion. In this case, we have been able to evaluate the time function to include terms up to $\varepsilon^{13}$ and the computed time of merger agrees with the numerically determined expression to three significant digits. We have also checked that the term-by-term differences in the series expansions are stable at three significant figures. It therefore appears the series approximations \textit{do} converge near the time of merger, albeit very slowly. Of course, convergence of the series is more rapid further away from the merger.

To further illustrate the convergence properties of the perturbative results,  we will compare constant-time cross-sections of the crease obtained from the perturbative results with those obtained from the numerics. We present representative results of this analysis in Fig.~\ref{fig:slice_errors}.

As for the geometrical features, we can compute the area of the crease and the length of the crease within a time-slice to higher order as well. We skip over the details of this computation and present the main results.

To simplify the presentation of the area, we write it as an expansion in $\varepsilon_{\rm Max}$,
\be 
A_{\rm Crease} = \sum_{i = 1}^{\infty} A_{\rm Crease}^{(i)} \varepsilon_{\rm Max}^i \, ,
\ee
with the first several terms reading:
\begingroup
\allowdisplaybreaks
\begin{align}
\frac{A_{\rm Crease}^{(1)}}{M^2} =& \, \frac{15 \pi^2 \chi^2}{64} \, ,
\\
\frac{A_{\rm Crease}^{(2)}}{M^2} =& \, \frac{225 \pi^3 \chi^2}{4096} \, ,
\\
\frac{A_{\rm Crease}^{(3)}}{M^2} =& \, - \frac{5 \pi^2 \chi^2}{393216} \big[23625 \pi ^2-128 \left(23 \chi ^2+1110\right)\bigg] \, ,
\\
\frac{A_{\rm Crease}^{(4)}}{M^2} =& \, \frac{75 \pi^3 \chi^2}{1048576} \bigg[52 \chi ^2+10125 \pi ^2-89256 \bigg] \, ,
\\
\frac{A_{\rm Crease}^{(5)}}{M^2} =& \, \frac{\pi^2 \chi^2}{536870912} \bigg[230400 \pi ^2 \left(86 \chi ^2+34461\right)+8192 \left(1497 \chi ^4-8002 \chi
   ^2+378840\right)
   \nonumber
   \\
   &-847006875 \pi ^4\bigg] \, ,
\\
\frac{A_{\rm Crease}^{(6)}}{M^2} =& \, \frac{5 \pi \chi^2}{1207959552 \left(16 \chi ^2+225 \pi ^2-1224\right)} \bigg[-113246208 \chi ^7-12925440 \pi ^2 \chi ^6
    \nonumber
    \\
    &-65536 \left(24975 \pi ^2-136288\right) \chi ^5-4320 \pi ^2 \left(437025 \pi ^2-4027768\right) \chi ^4
    \nonumber
    \\
    &-3072
   \left(7983104-85752000 \pi ^2+9163125 \pi ^4\right) \chi ^3-2700 \pi ^2 \left(467118976-86251080 \pi ^2
    \right. 
    \nonumber
    \\
    &\left.+4066875 \pi ^4\right) \chi ^2-54
   \left(-5650776064+651943321600 \pi ^2-174732960000 \pi ^4
   \right. 
    \nonumber
    \\
    &\left.+11303296875 \pi ^6\right) \chi +6075 \pi ^2 \left(72626176+1603880000 \pi ^2-461148000 \pi
   ^4
   \right. 
    \nonumber
    \\
    &\left.+30121875 \pi ^6\right) \bigg] \, ,
\\
\frac{A_{\rm Crease}^{(7)}}{M^2} =& \, \frac{\pi^2 \chi^2 }{23089744183296 \left(16 \chi ^2+225 \pi ^2-1224\right)^2} \bigg[133738168057856 \chi ^{10}-13529146982400 \chi ^9
    \nonumber
    \\
    &+6291456 \left(622656075 \pi ^2-3430604096\right) \chi ^8-23488102400 \left(277155 \pi
   ^2-2133056\right) \chi ^7
   \nonumber
    \\
    &+491520 \left(1845893469184-921960313920 \pi ^2+89535162375 \pi ^4\right) \chi ^6
    \nonumber
    \\
    &-165150720 \left(35034558464-6004019200
   \pi ^2+346055625 \pi ^4\right) \chi ^5
   \nonumber
    \\
    &+2304 \left(377882516389888+8073042212505600 \pi ^2-2267962971120000 \pi ^4
    \right. 
    \nonumber
    \\
    &\left.+155778573515625 \pi ^6\right) \chi
   ^4+1290240 \left(75273352511488+7212664934400 \pi ^2
   \right. 
    \nonumber
    \\
    &\left.-8908821360000 \pi ^4+695241984375 \pi ^6\right) \chi ^3+64800
   \left(-4826220810207232
   \right. 
    \nonumber
    \\
    &\left.-7713013802762240 \pi ^2+3554805730464000 \pi ^4-481362922125000 \pi ^6
    \right. 
    \nonumber
    \\
    &\left.+21785968828125 \pi ^8\right) \chi ^2-907200
   \left(-4921044326416384-14906978106736640 \pi ^2
   \right. 
    \nonumber
    \\
    &\left.+5491341504768000 \pi ^4-629528105700000 \pi ^6+23709984609375 \pi ^8\right) \chi 
    \nonumber
    \\
    &-3645 \left(136-25
   \pi ^2\right)^2 \left(-21467714551808+20380628582400 \pi ^2
   \right. 
    \nonumber
    \\
    &\left.-38307060000000 \pi ^4+3682851046875 \pi ^6\right) \bigg] \, .
\end{align}
\endgroup
We refrain from presenting higher-order terms as, for example, the result for $A_{\rm Crease}^{(8)}$ would fill a page on its own. For the area, the fact that the caustics are displaced from $\varphi = \pi/2$ and $\varphi = 3\pi/2$ has consequences at $\mathcal{O}(\varepsilon_{\rm Max}^6)$ in the above. If this important detail is not accounted for, one would find that the area is divergent. 

The second geometric quantity of interest is the length of the crease in a surface of constant Boyer-Lindquist time. In this case, the length admits a simple expansion in the parameter $\tau$ defined above Eq.~\eqref{eqn:vareps_of_tau}, i.e., $t = - M/(4 \tau^2)$. In terms of $\tau$, the length can be expanded as a series
\be 
\ell_{\rm Crease} = \sum_{i = 2}^\infty \ell_{\rm Crease}^{(i)} \tau^i \, ,
\ee
with the first several coefficients in the series reading:
\begingroup
\allowdisplaybreaks
\begin{align}
\frac{\ell_{\rm Crease}^{(2)}}{M} =& \, \frac{15 \pi \chi^2}{32} \, ,
\\
\frac{\ell_{\rm Crease}^{(3)}}{M} =& \, 0 \, ,
\\
\frac{\ell_{\rm Crease}^{(4)}}{M} =& \, \frac{5 \pi \chi^2}{2048} \bigg[80 \chi ^2-675 \pi ^2+4184\bigg] \, ,
\\
\frac{\ell_{\rm Crease}^{(5)}}{M} =& \, \frac{75 \pi^2 \chi^2 \left( 22275 \pi^2 - 157 184 \right)}{262144} \, ,
\\
\frac{\ell_{\rm Crease}^{(6)}}{M} =& \, - \frac{\pi \chi^2}{12582912} \bigg[36000 \pi ^2 \left(344 \chi ^2+60651\right)+1024 \left(2271 \chi
   ^4-53567 \chi ^2+864501\right)
	\nonumber
	\\
	&-239203125 \pi ^4 \bigg] \, ,
\\
\frac{\ell_{\rm Crease}^{(7)}}{M} =& \, \frac{5 \pi^2 \chi^2}{536870912} \bigg[-576000 \pi ^2 \left(1964 \chi ^2+105879\right)-32768 \left(184 \chi
   ^4-292678 \chi ^2-1469343\right)
   \nonumber
	\\
	&+5538121875 \pi ^4\bigg] \, ,
\\
\frac{\ell_{\rm Crease}^{(8)}}{M} =& \, \frac{\pi \chi^2}{8455716864} \bigg[3543750 \pi ^4 \left(134617 \chi ^2+4062456\right)+201600 \pi ^2
   \left(40 \chi ^4-22042188 \chi ^2\right. 
   \nonumber
	\\
	&\left.-139771467\right)
   +2048
   \left(846788 \chi ^6+2624154 \chi ^4-218175612 \chi
   ^2+6948967743\right)
   \nonumber
	\\
	&-1130234765625 \pi ^6\bigg] \, ,
\\
\frac{\ell_{\rm Crease}^{(9)}}{M} =& \,  \frac{5 \pi^2 \chi^2}{46179488366592} \bigg[-362880000 \pi ^4 \left(5826085 \chi ^2+123394599\right)+206438400
   \pi ^2 \left(55454 \chi^4\right.
   \nonumber
   \\
   &\left.+106672161 \chi
   ^2+651615615\right)-33554432 \left(29720 \chi ^6+828000 \chi
   ^4+313727196 \chi ^2\right.
   \nonumber
   \\
   &\left.+29072583\right)+3118705227421875 \pi^6\bigg] \, .
\end{align}
\endgroup
The fact that the caustics are displaced from $\varphi = \pi/2, 3\pi/2$ must be accounted for at $\mathcal{O}(\tau^8)$ and higher in the computation of the crease length. If this fact is neglected, one would naively find the length to diverge at that order. Note that to obtain this result it was necessary to invert the Boyer-Lindquist time function along the crease to obtain a relationship $\varepsilon = \varepsilon(\tau)$. This perturbative inversion is valid only for the large black hole horizon. As a result, the perturbative expansion for the length is not expected to be a good approximation in the vicinity of the moment of merger or near the small black hole horizon.


\section{Perturbative results for the false crease}
\label{app:false_crease_pert}
In Section~\ref{sec:perturbative}, we used perturbative techniques to obtain analytic formulae for the crease set valid far away from the small black hole. The same techniques can be used to obtain analytic expressions for the false crease. We have used these techniques, for example, in preparing Fig.~\ref{fig:crease_caustic_section}. Here we give a brief analysis of the false crease, omitting details and presenting the relevant expressions.

Recall that the pairs of geodesics that intersect at the false crease have impact parameters $(\alpha, \beta)$ and $(\alpha, -\beta)$. This fact makes the evaluation of the perturbative expressions much simpler. We work only to the order at which the false crease is first resolved. This yields the following Boyer-Lindquist coordinates along the false crease:
\begin{align}
    \frac{t^{(f)}_s}{M} =& \, -\frac{1}{4 \varepsilon^2} - \frac{15 \pi \varepsilon}{4} + \left(\frac{225 \pi^2}{256}-8 \right) \varepsilon^2 -\frac{15 \pi \left(375 \pi^{2}+1664\right)   \,\varepsilon^{3}}{8192} 
    \nonumber 
    \\
    &+ \frac{5 \left(3 \left(\mu_o^2-1 \right) \cos^{2} \varphi +7\right) \pi  \,\varepsilon^{3} \chi^{2}}{32} + \mathcal{O}(\varepsilon^4) \, ,
    \\
    \mu^{(f)}_s =& \, - \mu_o  \, ,
    \\
    \phi^{(f)}_s =&\,  \pi \, {\rm sign} (\cos \varphi)  - 4 \chi  \,\varepsilon^{2} -\frac{5}{4} \pi  \chi  \,\varepsilon^{3} +\frac{15 \sqrt{-\mu_{o}^{2}+1}\, \pi  \cos \! \left(\varphi \right) \varepsilon^{4} \chi^{2}}{16}
    \nonumber
    \\
    &+\left(\frac{225 \pi^{2}}{128}  -16\right) \chi \varepsilon^{4}  + \mathcal{O}(\varepsilon^5) \, .
\end{align}
As before, we have introduced the parameters $(\varepsilon, \varphi)$ as coordinates along the false crease. As with the true crease, allowing $\varphi$ to range over the full interval $0$ to $2\pi$ yields a double covering of the false crease. Here $\varphi$ should take values in the interval $[0, \pi]$, with the $\varphi = 0, \pi$ corresponding to the $A_3$ caustics. Note also that the result for $\mu_s^{(f)}$ is exact, while the remaining two coordinates are valid to the specified orders.

In terms of the quasi-Cartesian coordinates~\eqref{Emp_Cart}, the false crease is parameterized as
\begin{align}
    \frac{x^{(f)}}{M} &= - \frac{1}{4 \varepsilon^2} -\frac{5 \pi  \chi^{2} \left(1-\mu_{o}^{2}\right)   \,\varepsilon^{3}}{4} + \mathcal{O}(\varepsilon^4) \, ,
    \\
    \frac{y^{(f)}}{M} &= \, \chi \sqrt{1-\mu_o^2} \bigg[ 1+\frac{5 \pi  \varepsilon}{16} + \left(-\frac{15 \pi  \chi  \sqrt{1-\mu_{o}^{2}}\,  \cos  \varphi }{64}-\frac{225 \pi^{2}}{512}+4\right) \varepsilon^{2} 
    \nonumber
    \\
    &+\left(\frac{15 \pi \chi^{2} \left(1-\mu_o^2 \right) \cos^{2} \varphi }{128}+\frac{23625 \pi^{3}}{32768}+\frac{21 \pi \chi^{2}}{128}-\frac{801 \pi}{256}\right) \varepsilon^{3}\bigg] + \mathcal{O}(\varepsilon^4)\, ,
    \\
    \frac{z^{(f)}}{M} &= -\frac{5 \pi \chi^2 \mu_{o}   \sqrt{1-\mu_{o}^{2}} \, \varepsilon^{3}}{4}  + \mathcal{O}(\varepsilon^4)\, .
\end{align}

\begin{figure}
    \centering
    \includegraphics[width=0.45\textwidth]{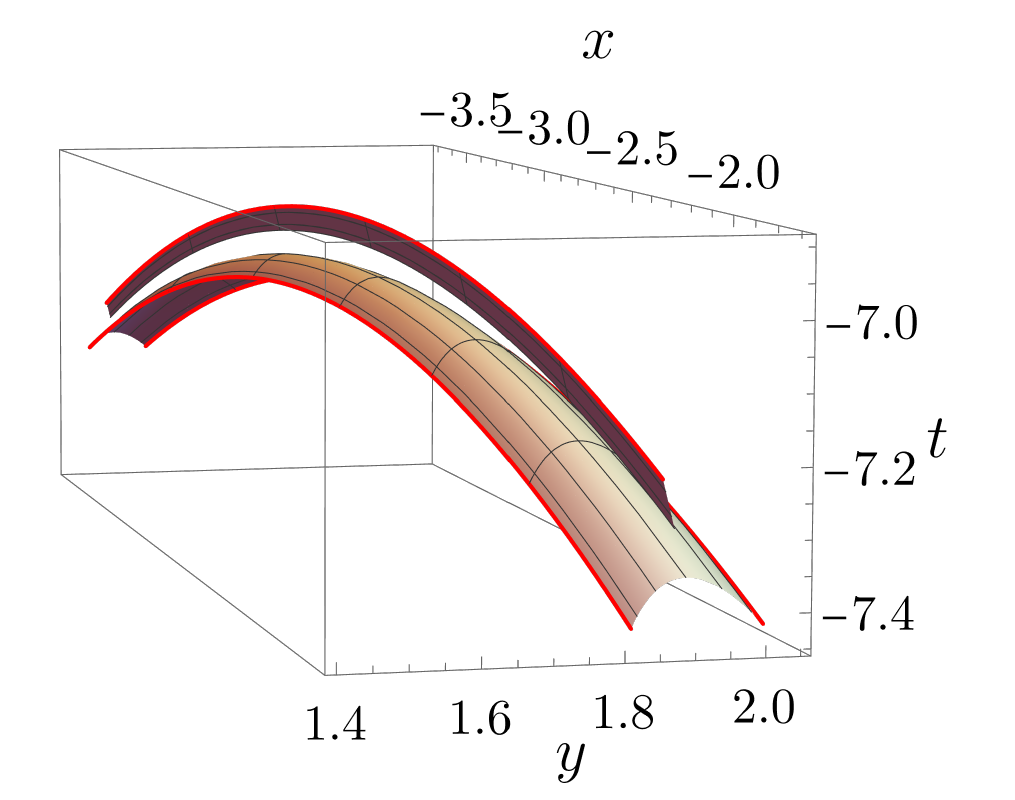}
    \caption{The false crease is shown for an orthogonal merger with an extremal small black hole. The red lines are lines of $A_3$ caustics. The false crease ``bows up'' away from the $A_3$ lines. The true crease is displayed as well and is located above the false crease. Since the $z$-direction is suppressed in this plot, the true crease appears as a thin strip. This figure is obtained from our perturbative results, which give a qualitatively correct depiction of the configuration.}
    \label{fig:false_crease}
\end{figure}
We illustrate the false crease in Fig.~\ref{fig:false_crease}, which can be compared with Fig.~14 of~\cite{Emparan:2017vyp}. The false crease ``bows up'' as one moves away from the lines of caustics. On the same plot we indicate the true crease, which appears above the false one. Here, as the $z$ direction is suppressed, the true crease appears as a thin vertical strip. In the figure we have chosen an orthogonal merger with an extremal small black hole. The result is qualitatively similar for other values of the spin or inclination angle. Decreasing the spin or increasing the inclination angle $\mu_o$ has the effect of shrinking the physical extent of the false crease, with it degenerating to a line for $a=0$ or $\mu_o = 1$. 

These analytic expressions are useful for simple consistency checks. We can compare the Boyer-Lindquist time function between the true and false creases,
\be 
t_s^{(t)} - t_s^{(f)} = \frac{15 \pi a^2 \left(1-\mu_o^2 \right) \varepsilon^3}{16} + \mathcal{O}(\varepsilon^4) \, ,
\ee
confirming the true crease occurs first (in the backward-in-time sense). It is also possible to compute geometric quantities for the false crease. For example, we find that the leading order contribution to its area is
\be 
A = \frac{15  \pi^2 a^2 (1-\mu_o^2) \varepsilon_{\rm Max}}{64} + \mathcal{O}(\varepsilon_{\rm Max}^2) \, ,
\ee
which is identical to the result for the true crease. (At higher orders in perturbation theory we find that the false crease has a larger area than the true crease. These differences appear at $\mathcal{O}(\varepsilon_{\rm max}^3)$.)

\bibliographystyle{JHEP}
\bibliography{creases.bib}

\end{document}